\documentclass[structabstract]{aa}  
\usepackage{graphicx}
\usepackage{txfonts}
\usepackage{natbib}
\bibpunct[]{(}{)}{;}{a}{}{,}
\begin{document}
   \title{The end of star formation in Chamaeleon~I\,?} 
   \subtitle{A LABOCA census of starless and protostellar cores
   \thanks{Based on observations carried out with the Atacama Pathfinder 
   Experiment telescope (APEX). APEX is a collaboration between the 
   Max-Planck Institut f\"ur Radioastronomie, the European Southern 
   Observatory, and the Onsala Space Observatory.}
   \thanks{Appendices A, B, and C are only available in electronic form via
   http://www.edpsciences.org.}
   \thanks{The image shown in Fig.~\ref{f:labocamap} and the content of 
   Table~\ref{t:id_gcl_simbad} are available in electronic form (FITS format 
   for the former) at the CDS via anonymous ftp to cdsarc.u-strasbg.fr 
   (130.79.128.5).}
   }

   \author{A. Belloche
          \inst{1}
          \and
          F. Schuller
          \inst{1}
          \and
          B. Parise
          \inst{1}
          \and
          Ph. Andr{\'e}
          \inst{2}
          \and
          J. Hatchell
          \inst{3}
          \and
          J.~K. J{\o}rgensen
          \inst{4}
          \and\\
          S. Bontemps
          \inst{5}
          \and
          A. Wei{\ss}
          \inst{1}
          \and
          K.~M. Menten
          \inst{1}
          \and
          D. Muders
          \inst{1}
          }

   \institute{Max-Planck Institut f\"ur Radioastronomie, Auf dem H\"ugel 69,
              53121 Bonn, Germany\\
              \email{belloche@mpifr-bonn.mpg.de}
         \and
             {Laboratoire AIM, CEA/DSM-CNRS-Universit{\'e} Paris Diderot, 
              IRFU/Service d'Astrophysique, CEA Saclay, 91191 Gif-sur-Yvette, 
              France}
         \and
             {School of Physics, University of Exeter, Stocker Road, Exeter 
              EX4 4QL, UK}
         \and
             {Centre for Star and Planet Formation, Natural History Museum of 
              Denmark, University of Copenhagen, {\O}ster Voldgade 5--7, 1350 
              Copenhagen K., Denmark}
         \and
             {Universit{\'e} de Bordeaux, Laboratoire d'Astrophysique 
              de Bordeaux, CNRS/INSU, UMR 5804, 
              BP 89, 33271 Floirac cedex, France}
             }

   \date{Received 10 September 2010; accepted 17 December 2010}

 
  \abstract
   {Chamaeleon I is the most active region in terms of star formation in the 
   Chamaeleon molecular cloud complex. Although it is one of the nearest 
   low-mass star forming regions, its population of prestellar and 
   protostellar cores is not known and a controversy exists concerning its
   history of star formation.}
   {Our goal is to search for prestellar and protostellar cores and 
   characterize the earliest stages of star formation in this cloud.}
   {We used the bolometer array LABOCA at the APEX telescope to map the 
   cloud in dust continuum emission at 870~$\mu$m with a high sensitivity. 
   This deep, unbiased survey was performed based on an extinction map derived 
   from 2MASS data. The 870~$\mu$m map is compared with the extinction map and 
   C$^{18}$O observations, and decomposed with a multiresolution 
   algorithm. The extracted sources are analysed by carefully taking into 
   account the spatial filtering inherent in the data reduction process. A 
   search for associations with young stellar objects is performed using 
   \textit{Spitzer} data and the SIMBAD database.}
   {Most of the detected 870~$\mu$m emission is distributed in five 
   filaments. We identify 59 starless cores, one candidate first 
   hydrostatic core, and 21 sources associated with more evolved young stellar 
   objects. The estimated 90$\%$ completeness limit of our survey 
   is 0.22~M$_\odot$ for the starless cores. The latter are only found above a 
   visual extinction threshold of 5 mag. They are less dense than those 
   detected in other nearby molecular clouds by a factor of a few on average,
   maybe because of the better sensitivity of our survey. 
   The core mass distribution is consistent with the IMF at the high-mass end 
   but is overpopulated at the low-mass end. In addition, at most 
   17$\%$ of 
   the cores have a mass larger than the critical Bonnor-Ebert mass. Both 
   results suggest that a large fraction of the starless cores may not be 
   prestellar in nature. Based on the census of prestellar cores, Class~0 
   protostars, and more evolved young stellar objects, we conclude that the 
   star formation rate has decreased with time in this cloud.}
   {The low fraction of candidate prestellar cores among the population of 
   starless cores, the small number of Class 0 protostars, the high 
   global star formation efficiency, the decrease of the star formation rate 
   with time, and the low mass per unit length of the detected filaments all 
   suggest that we may be witnessing the end of the star formation process in 
   Chamaeleon~I.}


   \maketitle
%

\section{Introduction}
\label{s:intro}

Large-scale, unbiased surveys in dust continuum emission have greatly 
improved our knowledge of the process of star formation in nearby, 
star-forming molecular clouds. During the past decade, (sub)mm dust continuum 
surveys established a close relationship between the prestellar core mass 
function (CMF) and the stellar initial mass function (IMF), suggesting that 
the IMF is already set at the early stage of cloud fragmentation
\citep[e.g.][]{Motte98,Testi98,Johnstone00}. In these submm surveys, 
prestellar cores are found only above a visual extinction of 5--7~mag, 
suggesting the existence of an extinction threshold for star formation to occur
\citep[see, e.g.,][, but also \citeauthor{Onishi98} 1998 based on molecular 
line observations]{Johnstone04,Enoch06,Kirk06}. 

In the mid-infrared range, five of these nearby molecular clouds 
-- Chamaeleon~II, Lupus, Perseus, Serpens, and Ophiuchus -- were mapped with 
good angular resolution and high sensitivity with the \textit{Spitzer} 
satellite in the frame of the c2d legacy project 
\citep[\textit{From molecular cores to planet forming disks}, see][]{Evans09}.
This extensive project produced a deep census of young stellar objects. It
delivered star formation efficiencies ranging from 3 to 6$\%$ and 
\citet{Evans09} estimate that these efficiencies could reach 15--30$\%$ if 
star formation continues at current rates in these clouds. Star 
formation is found to be slow compared to the free-fall time and highly 
concentrated to regions of high extinction \citep{Evans09}, consistent with 
the threshold mentioned above. 

Recently, the far-infrared/submm \textit{Herschel} Space Observatory revealed 
an impressive network of parsec-scale filamentary structures in nearby 
molecular clouds in the frame of the Gould Belt Survey \citep[][]{Andre10a}. 
This survey greatly increases the statistics of prestellar cores and confirms
the close relationship between the CMF and the IMF.
In the Aquila molecular cloud, several hundreds of prestellar cores are 
identified, and \citet[][]{Andre10a} show that most of these prestellar cores 
are located in supercritical filaments. They confirm the existence of an 
extinction threshold for star formation at $A_V \sim 7$~mag and propose that 
it may naturally come from the conditions required for a filament to become 
gravitationally unstable, collapse, and fragment into prestellar cores 
\citep[see also][]{Andre10b}. This finding will be further investigated in the 
frame of this \textit{Herschel} survey, but it can also be partly tested in 
other nearby clouds in the submm range, thanks to the advent of very sensitive 
cameras like LABOCA at the Atacama Pathfinder Experiment telescope (APEX).

Chamaeleon I (Cha~I) is one of the nearest, low-mass star forming regions in 
the southern sky \citep[150--160~pc,][; see Appendix~\ref{ss:distance} for more 
details]{Whittet97,Knude98}. Together with Cha~II and III, it belongs to the 
Chamaeleon cloud complex whose population of T-Tauri stars has been well 
studied \citep[see][ and references therein]{Luhman08c}. Cha~I contains nearly 
one order of magnitude more young stars than Cha~II, while Cha~III does not 
have any. Surveys performed in CO and its isotopologues showed 
that Cha~I contains a larger fraction of dense gas than Cha~II and III 
although the latter are somewhat more massive 
\citep[][]{Boulanger98,Mizuno99,Mizuno01}. Finally, several indications of 
jets and outflows were found in Cha I 
\citep[][]{Mattila89,Gomez04,Wang06,Belloche06}. 
Only three Herbig-Haro objects are known in Cha~II and none 
has been found in Cha~III \citep[][]{Schwartz77}. Cha~I is therefore much 
more actively forming stars than Cha~II and III. 

A controversy exists, however, concerning the history of star formation in 
Cha~I. Based on the H-R diagram of the young stellar population known 
at that time (47 members with known spectral type), \citet{Palla00} conclude
that star formation  in Cha~I began within the last 7 Myr and that its rate 
steadily increased until recently. They find similar results in several other 
clouds and draw the general conclusion that star formation in nearby 
associations and clusters started slowly some 10 Myr ago and
\textit{accelerated} until the present epoch. They speculate that this 
acceleration arises from contraction of the parent molecular cloud and results
from star formation being a critical phenomenon occuring above some 
threshold density. However, based on a larger sample of known members, 
\citet{Luhman07} recently came to the conclusion that star formation began 
3--6~Myr ago in Cha~I and has continued to the present time at a 
\textit{declining} rate. Since little is known about the earliest stages of 
star formation in Cha~I, investigating whether there is presently a population 
of condensations that are \textit{prestellar}, i.e. bound to form stars, 
should provide strong constraints on these two competing scenarii for the star 
formation history in Cha~I.

To unveil the present status of the earliest stages of star formation in 
Cha~I, we carried out a deep, unbiased dust continuum survey of this cloud at 
870~$\mu$m with the bolometer array LABOCA at APEX.
The observations and data reduction are described in Sects.~\ref{s:obs} and
\ref{s:reduction}, respectively. The maps are presented in 
Sect.~\ref{s:results}. Section~\ref{s:sourceid} explains the source extraction
and the sources are analysed in Sect.~\ref{s:analysis}. The implications are 
discussed in Sect.~\ref{s:discussion}. Section~\ref{s:conclusions} gives a 
summary of our results and conclusions.

%

\section{Observations}
\label{s:obs}

\subsection{870 $\mu$m continuum observations with APEX}
\label{ss:laboca}

\begin{figure}
\centerline{\resizebox{1.0\hsize}{!}{\includegraphics[angle=270]{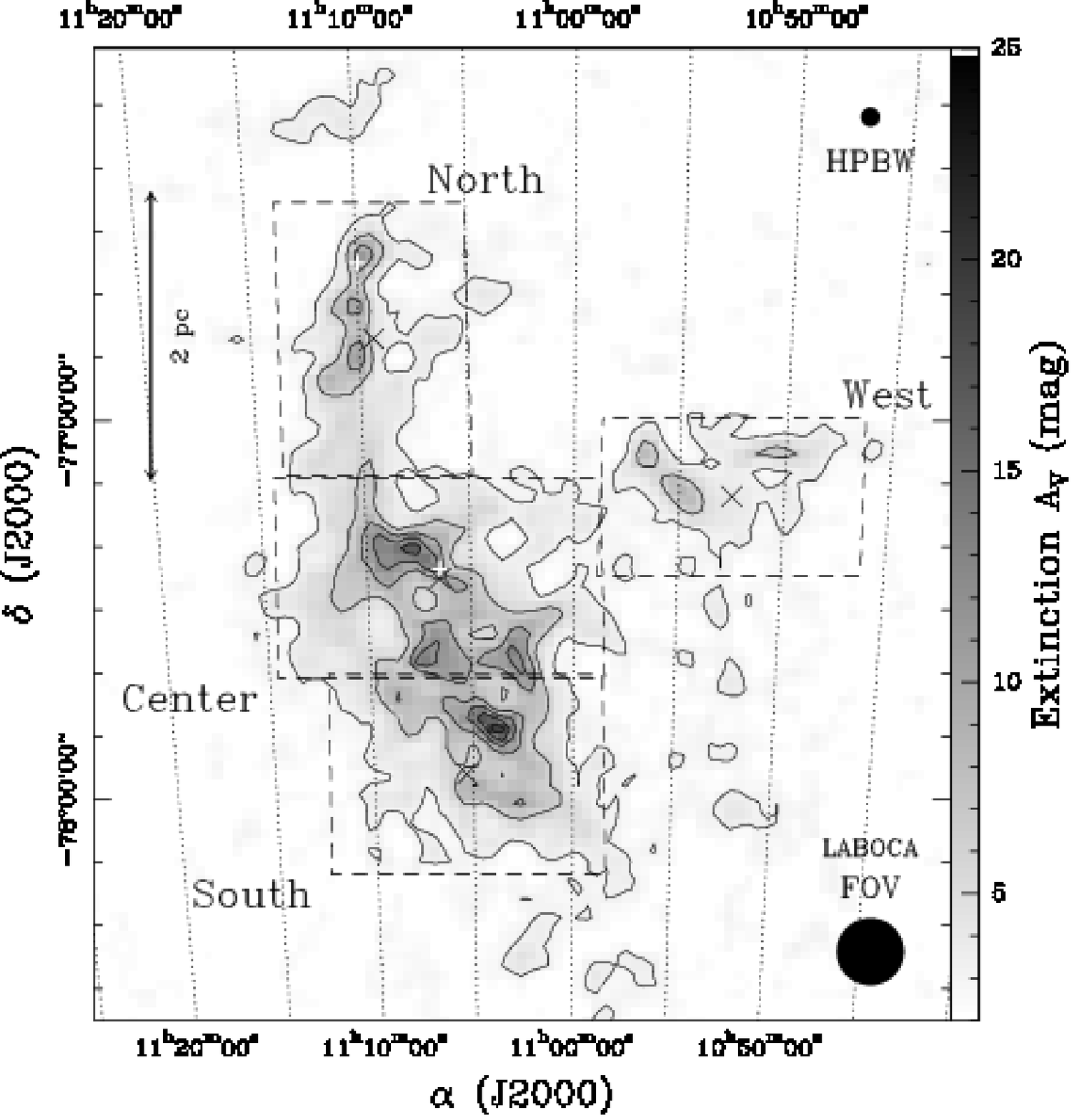}}}
\caption{Extinction map of Cha~I derived from 2MASS in radio projection.
The projection center is at ($\alpha, \delta$)$_{\mathrm{J2000}}$ =
($11^{\mathrm{h}}01^{\mathrm{m}}24^{\mathrm{s}}$, $-77^\circ15\arcmin00\arcsec$). 
The contours start at $A_V = 3$ mag and increase by step of 3 mag. The dotted 
lines are lines of constant right ascension. The angular resolution of the map 
($HPBW = 3\arcmin$) is shown in the upper right corner. 
The four fields selected for mapping with LABOCA are delimited with
dashed lines and their center is marked with a black cross. The white plus 
symbols in fields Cha-Center and Cha-North mark the positions of the dense 
cores Cha-MMS1 and Cha-MMS2, respectively, detected in dust 
continuum emission at 1.3~mm with the SEST \citep{Reipurth96}. The field of 
view of LABOCA is displayed in the lower right corner.}
\label{f:av}
\end{figure}

The region of Cha~I with a visual extinction higher than 3~mag was selected 
on the basis of an extinction map derived from 2MASS\footnote{The Two Micron 
All Sky Survey (2MASS) is a joint project of the University of Massachusetts 
and the Infrared Processing and Analysis Center/California Institute of 
Technology, funded by the National Aeronautics and Space Administration and 
the National Science Foundation.} (see Sect.~\ref{ss:2mass}). It was divided 
into four contiguous fields labeled Cha-North, Cha-Center, Cha-South, and 
Cha-West with a total angular area of 1.6 deg$^2$ (see Fig.~\ref{f:av}). 
The four fields were mapped in continuum emission with the Large APEX BOlometer 
CAmera \citep[LABOCA,][]{Siringo09} operating with about 250 working pixels in 
the 870~$\mu$m atmospheric window at the APEX 12~m 
submillimeter telescope \citep[][]{Guesten06}. The central frequency of 
LABOCA is 345~GHz and its angular resolution is 19.2$\arcsec$ ($HPBW$).
The observations were carried out for a total of 43 hours in August, October, 
November, and December 2007 (fields Cha-Center and Cha-South) and for 41 
hours in May 2008 (fields Cha-North and Cha-West), under excellent 
($\tau_{\mathrm{zenith}}^{\mathrm{870}\,\mu\mathrm{m}}$ = 0.13) to good 
($\tau_{\mathrm{zenith}}^{\mathrm{870}\,\mu\mathrm{m}}$ = 0.37) 
atmospheric conditions. The sky opacity was measured every 1 to 2 hours with
skydips. The focus was optimised on Mars, Saturn, or Venus at
least once per day/night. The pointing of the telescope was checked every hour 
on the nearby quasar \hbox{PKS1057-79} and was found to be accurate within
1.9$\arcsec$ for the 2007 data and 2.8$\arcsec$ for the 2008 data (rms). The 
calibration was performed with the secondary calibrators IRAS~13134-6264, 
IRAS~16293-2422, V883~Ori, or NGC 2071 that were observed every 1 to 2 hours 
\citep[see Table A.1 of][]{Siringo09}. Measurements on the primary calibrator 
Mars were also used.

The observations were performed on the fly, alternately with rectangular 
(``OTF'') and spiral scanning patterns \citep[see Sect.~8 of][]{Siringo09}.
The OTF maps were alternately scanned in right ascension and declination, 
with a random position angle between $-12^\circ$ and $+12^\circ$ to improve the 
sampling and reduce striping effects. The ``spiral'' maps were 
obtained 
using rasters of spirals (see Sect.~\ref{ss:redmethod} for the parameters), 
with a random offset around the field center (within 
$\pm 30\arcsec$ in each direction) to also improve the sampling.

\subsection{Extinction map from 2MASS}
\label{ss:2mass}

We derived an extinction map toward Cha~I from the publicly available 2MASS 
point source catalog (see Fig.~\ref{f:av}). We selected all sources detected 
in either both $J$ and $H$, or both $H$ and $K$, or in all three bands.
Before computing the extinction map, 223 young stellar objects (YSOs) 
that are known members of Cha~I \citep[see Table~1 of][]{Luhman08c} were
removed from the sample. The extinction was 
calculated from the average reddening of stars inside the elements of 
resolution of the map using the method described in \citet{Schneider10} that 
is adapted from \citet{Lada94}, \citet{Lombardi01}, and \citet{Cambresy02}. 
The extinction toward each star is obtained from the 
uncertainty-weighted combination of the [$J$-$H$] and [$H$-$K$] colors. The 
assumed average intrinsic colors are [$J$-$H$]$_0 = 0.45 \pm 0.15$, and 
[$H$-$K$]$_0 = 0.12 \pm 0.05$. These are 
derived from the typical dispersions for a population of galactic stars as 
measured using simulations with the Besan{\c c}on stellar population model 
\citep[][]{Robin03}\footnote{See also http://www.obs-besancon.fr/.}. The 
infrared color excess was converted to visual extinction assuming the standard 
extinction law of \citet{Rieke85}, with a ratio of total to selective 
extinction $R_V = 3.1$ (see Appendix~\ref{ss:avtoN}).
In addition, we use the Besan{\c c}on galactic model to derive
the predicted density of foreground stars in the 2MASS bands. For the distance 
and the direction of the Chameleon clouds, it is found to be less than 
0.01 star per arcmin$^2$. Therefore it is negligible and no filtering for 
foreground stars is applied. 

A Gaussian weight function for the local average of the individual 
extinctions defines the resolution of the final map, with a $FWHM$ of
3$\arcmin$ and a pixel size of 1.5$\arcmin$. The $FWHM$ is chosen so 
that most pixels (97$\%$) with $A_V < 6$~mag have at least 10 stars 
within a radius equal to $FWHM/2$. Most pixels (96$\%$) between 6 and 
9~mag and between 9 and 13~mag have at least 4 and 3 stars, respectively.
All pixels above 15~mag have between 2 and 4 stars. Only five pixels have no
background star within $FWHM/2$. Their extinction value is solely
estimated from the stars located between $FWHM/2$ and the truncation radius
($1.3\,FWHM$), which means a small local loss of resolution. Two of these 
pixels are located within the $A_V = 12$~mag contour of the Cha-MMS1 region.

The typical rms noise level in the outer parts of the map is 0.4~mag, i.e. 
a $3\sigma$ detection level of 1.2~mag for an $FWHM$ of 3$\arcmin$. This
rms noise level is however expected to increase towards the higher-extinction 
regions because of the decreasing number of stars per element of resolution.

Our map looks very similar to the extinction map of \citet{Kainulainen06} in 
terms of structure and sensitivity. Their map was also based on 2MASS 
data and a similar method was used to compute the extinction. The main 
differences are that they chose a higher angular resolution ($2\arcmin$) and 
used only the 2MASS sources with a signal-to-noise ratio larger than 10 in 
$J$, $H$, and $K$, at the cost of a few dozens of pixels being empty due to 
the lack of background stars. The extinction maps of 
\citet{Cambresy97} and \citet{Cambresy99}, based on star counts in the J and
R bands, respectively, are more sensitive in the low-extinction regions than
our map, but they miss the high-extinction regions 
\citep[see also discussion in][]{Cambresy02}.

\subsection{Archival C$^{18}$O 1--0 observations with SEST}
\label{ss:sest}

The eastern part of Cha~I was mapped with the Swedish ESO Submillimeter 
Telescope (SEST) in the C$^{18}$O 1--0 line by
\citet{Haikala05}. The angular resolution is 45$\arcsec$ but the map is 
highly undersampled since the observations were done with a step of 1$\arcmin$. 
The median rms noise level is 0.1~K for a channel width of 43~kHz 
(0.12~km~s$^{-1}$), but occasionally goes up to 0.22~K at certain positions.
We retrieved the publicly available FITS data cube from the A$\&$A website. We 
reprojected the cube from B1950 to J2000 equatorial coordinates in radio 
projection with a projection center at 
($\alpha, \delta$)$_{\mathrm{J2000}}$ =
($11^{\mathrm{h}}01^{\mathrm{m}}24^{\mathrm{s}}$, $-77^\circ15\arcmin00\arcsec$).
We applied a velocity correction of $+0.25$ km~s$^{-1}$ to all spectra 
(as recommended by L. Haikala, \textit{priv. comm.}).

%

\section{LABOCA data reduction}
\label{s:reduction}

\subsection{Reduction method}
\label{ss:redmethod}

The LABOCA data were reduced with the BoA software\footnote{see 
http://www.mpifr-bonn.mpg.de/div/submmtech/software/boa/ boa\_main.html.} 
following the 
procedures described in Sect.~10.2 of \citet{Siringo09} and Sect.~3.1 of 
\citet{Schuller09}. The removal of the correlated 
noise was done with the median noise method applied to all pixels with 5 
iterations and a relative gain of 0.9. Subsequently, the correlated noise 
computed separately for each group of pixels belonging to the same amplifier 
box was removed (BoA function \textit{correlbox} with 3 iterations and a 
relative gain of 0.95), as well as the correlated noise computed separately 
for each group of pixels connected to the same read-out cable (BoA function 
\textit{correlgroup} with 3 iterations and a relative gain of 0.8). 

The power at frequencies below $0.1$~Hz in the Fourier domain was partially 
filtered out to reduce the $1/f$ noise. It was replaced with the average power 
measured between 0.1 and 0.15~Hz (BoA function \textit{flattenFreq}). Since 
the OTF scans were performed with a mapping speed of \hbox{2~arcmin~s$^{-1}$}, 
the frequency 0.1~Hz corresponds to an angular scale of 
$20\arcmin$. The analysis is different for the spiral scans. An individual 
spiral subscan had a duration of 40~s, an angular speed of 60~deg~s$^{-1}$, 
and a radius linearly increasing from 2$\arcmin$ to 3$\arcmin$. A frequency of 
0.1~Hz corresponds to 10~s, i.e. about 1.7 turns. As a result, the typical 
angular scale associated with the cutoff frequency of 0.1~Hz is the spiral 
``diameter'' which varied between 4.5$\arcmin$ and 6$\arcmin$.
In addition to this low-frequency ``flattening'', a first order baseline was 
subtracted scan-wise. The gridding was done with a cell size of 6.1$\arcsec$ 
and the map was smoothed with a Gaussian kernel of size 9$\arcsec$ ($FWHM$). 
The angular resolution of the final map is 21.2$\arcsec$ ($HPBW$) and the rms 
noise level is 12~mJy/21.2$\arcsec$-beam (see Sect.~\ref{ss:labocamap}).

This whole process was performed 21 times in an iterative way. The pixel 
values in the map produced at iteration 0 (resp. 1) were set to zero below a 
signal-to-noise ratio of 4. The location of the remaining signal was used as 
an astronomical source model to \textit{mask} 
the raw data at the start of iteration 1 (resp. 2). The mask was removed at 
the end of iteration 1 (resp. 2) to compute the reduced map. Iterations 3 to 
20 were performed differently: the pixel values of the map resulting from the 
previous iteration were set to zero below a signal-to-noise ratio of 2.5. The 
remaining signal was 
\textit{subtracted} from the raw data before reduction and added back after 
reduction. Masking or subtracting the significant astronomical signal found 
after one iteration permits to protect these regions at the next iteration 
when removing the correlated signal, flattening the low-frequency part of the 
power spectrum, and subtracting the first-order baseline. In this way, 
negative bowls around strong sources are much reduced and more extended 
emission can be recovered.

\subsection{Spatial filtering and convergence}
\label{ss:summ_reduction_apendix}

Data reduction features dealing with spatial filtering and 
convergence are analysed in detail in Appendix~\ref{s:reduction_appendix}.
Here we give a brief summary. The correlated noise removal severely 
limits the extent of the 870~$\mu$m emission that can be recovered (see 
Fig.~\ref{f:filtering}). Elongated structures are better recovered than 
circular ones (see Table~\ref{t:filtering}). The recovery of extended 
structures is improved by increasing the number of iterations of the data 
reduction. With 21 iterations, convergence is reached for most relevant 
structures (see Figs.~\ref{f:total_flux} and \ref{f:convergence}).

%

\section{Basic results}
\label{s:results}

\begin{figure*}
\centerline{\resizebox{0.9\hsize}{!}{\includegraphics[angle=270]{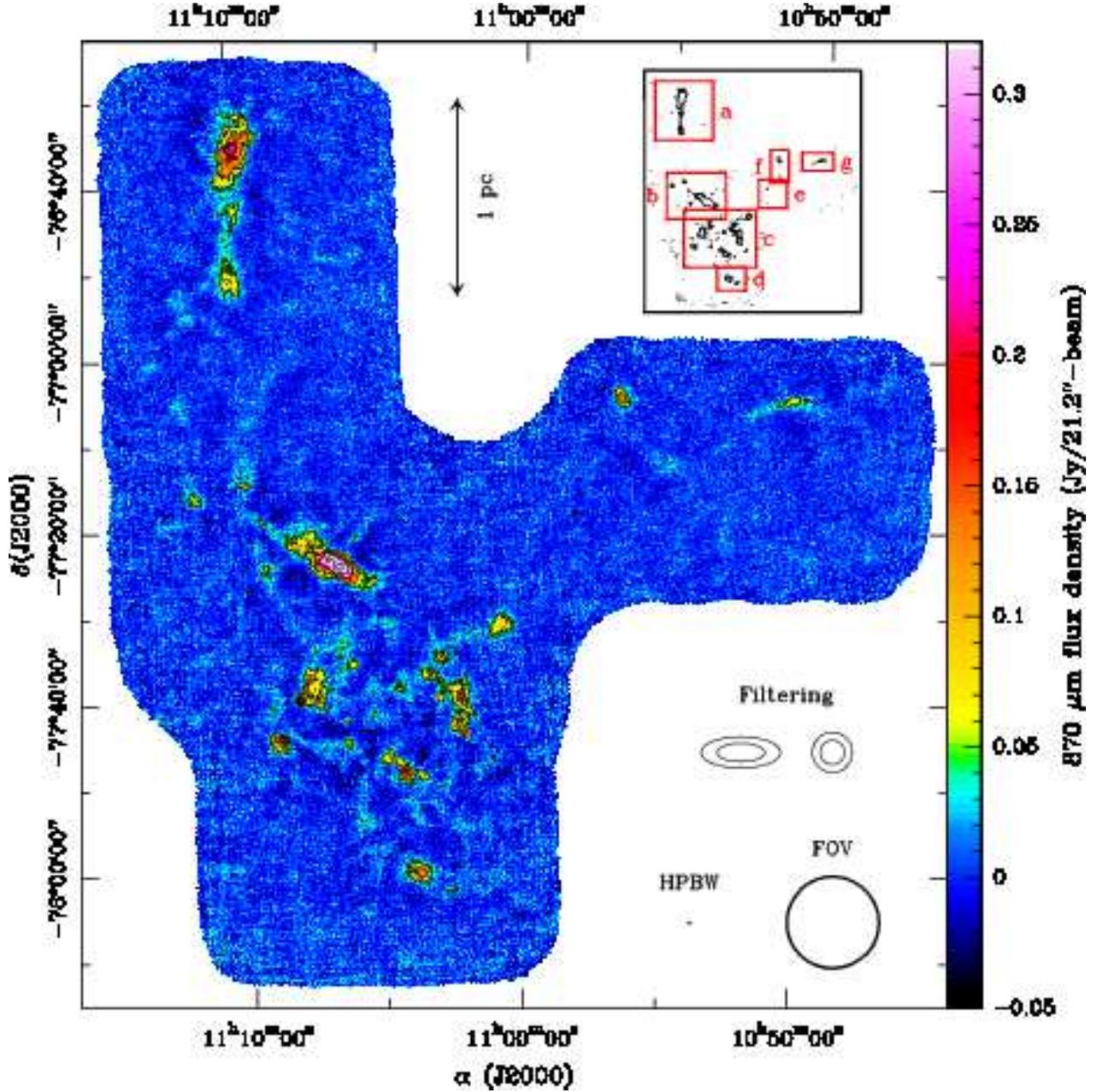}}}
\caption{870 $\mu$m continuum emission map of Cha~I obtained with LABOCA at 
APEX. The projection type and center are the same as in Fig~\ref{f:av}. The 
contour levels are $a$, $2a$, $4a$, $8a$, $16a$, and $32a$, with 
$a = 48$~mJy/21.2$\arcsec$-beam, i.e. about 4 times the rms noise level. The 
flux density color scale is shown on the right. 
The field of view of LABOCA (10.7$\arcmin$) and the angular resolution of the 
map ($HPBW = 21.2\arcsec$) are shown in the lower right corner. The typical 
sizes above which the filtering due to the data reduction becomes significant 
in terms of peak flux density are also displayed, for weak ($<150$~mJy/beam) 
and strong ($>150$~mJy/beam) sources with small and large symbols, 
respectively, and for elliptical sources with aspect ratio 2.5 and circular 
sources with ellipses and circles, respectively (see 
Appendix~\ref{ss:filtering} and Col.~2 of Table~\ref{t:filtering}). The pixel 
size is $6.1\arcsec$. The red boxes in the insert are labeled like 
Figs.~\ref{f:labocamapdet}a--g and show their limits overlaid on the first 
870 $\mu$m contour.
}
\label{f:labocamap}
\end{figure*}

The main assumptions made in this and the next sections to derive the physical 
properties of the detected sources are detailed in 
Appendix~\ref{s:assumptions}. These assumptions are not repeated in the 
following, except in the few cases where there could be an ambiguity.

\subsection{Maps of dust continuum emission in Cha~I}
\label{ss:labocamap}

\begin{figure}
\centerline{\resizebox{1.0\hsize}{!}{\includegraphics[angle=270]{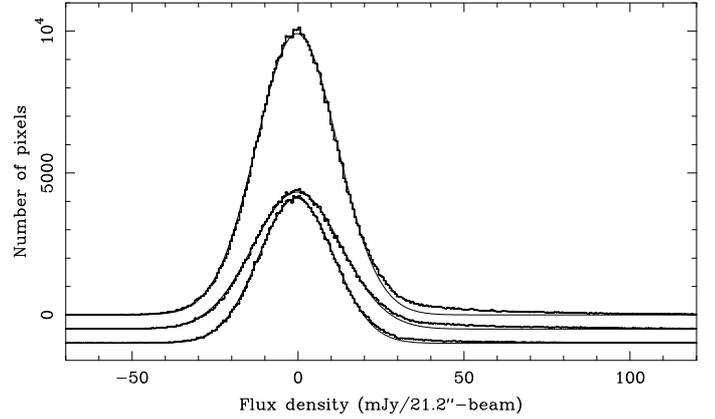}}}
\caption{Flux density distribution in the full map of Cha~I 
(\textit{upper} histogram), in fields Cha-Center and Cha-South 
(\textit{middle} histogram), and in fields Cha-North and Cha-West 
(\textit{lower} histogram). The middle and lower histograms were shifted 
vertically by -500 and -1000, respectively, for clarity. A Gaussian fit is
overlaid as a thin line on each histogram. The Gaussian standard deviation is 
12.2, 13.3, and 11.2 mJy/21.2$\arcsec$-beam for the full map, for fields 
Cha-Center and Cha-South, and for fields Cha-North and Cha-West, 
respectively.}
\label{f:pixelhisto}
\end{figure}

The final 870 $\mu$m continuum emission map of Cha~I obtained with LABOCA is 
shown in Fig.~\ref{f:labocamap}. Pixels with a number of independent 
measurements (``coverage'') smaller than 800 are masked. The resulting map 
contains 0.57 megapixels, corresponding to a total area of 1.6~deg$^2$ 
(11.0~pc$^2$). The mean and median coverages are 1472 and 1476 hits per pixel, 
respectively, with an integration time of 40~ms per hit. The noise 
distribution is fairly uniform and Gaussian (see Fig.~\ref{f:pixelhisto}). The 
average noise level is 12.2~mJy/21.2$\arcsec$-beam. It is slightly higher for 
fields Cha-Center and Cha-South (13.3~mJy/21.2$\arcsec$-beam) and slightly 
lower for fields Cha-North and Cha-West (11.2~mJy/21.2$\arcsec$-beam).
In the following, we will use 
12~mJy/21.2$\arcsec$-beam as the typical noise level. This translates into an
H$_2$ column density of $1.1 \times 10^{21}$~cm$^{-2}$ for a dust mass 
opacity of 0.01 cm$^{2}$~g$^{-1}$, and corresponds to a visual extinction 
$A_V \sim$ 1.2 mag with $R_V = 3.1$ (see the other assumptions in 
Appendix~\ref{s:assumptions}).

The dust continuum emission map of Cha~I reveals several very compact sources 
and many spatially resolved sources. Very faint filamentary structures are 
also present, especially in fields Cha-Center and Cha-North. 
Figure~\ref{f:labocamapdet} presents all the detected structures in more 
detail. The northern part (Cha-North) consists of a filamentary structure 
elongated along the south-north direction that ends with a prominent dense 
core dominated by a strong, compact source, \object{Cha-MMS2} (source S3, 
Fig.~\ref{f:labocamapdet}a). Three additional very compact sources with weak 
emission are also detected to the east and west of the filament (S16, S20, 
S21). The northern part of field Cha-Center features one filamentary structure 
with a position angle $PA \sim +60^\circ$ east of north dominated by a 
prominent dense core that contains two compact sources and \object{Cha-MMS1} 
(S4, S10, 
and C1, Fig.~\ref{f:labocamapdet}b). Three additional very compact sources are 
also detected outside the filament, in the south-east, north-west, and 
north-east, respectively (S9, S11, S19). The region covered by the southern 
part of field Cha-Center and the northern part of field Cha-South is very 
clumpy with five very compact sources (S1, S2, S6, S7, S12) and many extended 
structures (Fig.~\ref{f:labocamapdet}c). The southern part of field 
Cha-South contains a somewhat more isolated extended structure 
(Fig.~\ref{f:labocamapdet}d). Finally, field Cha-West features relatively 
isolated sources: four compact sources (S5, S8, S17, S18), one extended 
structure and one filamentary structure (Figs.~\ref{f:labocamapdet}e--g).

\begin{figure*}
\centerline{\resizebox{0.79\hsize}{!}{\includegraphics[angle=270]{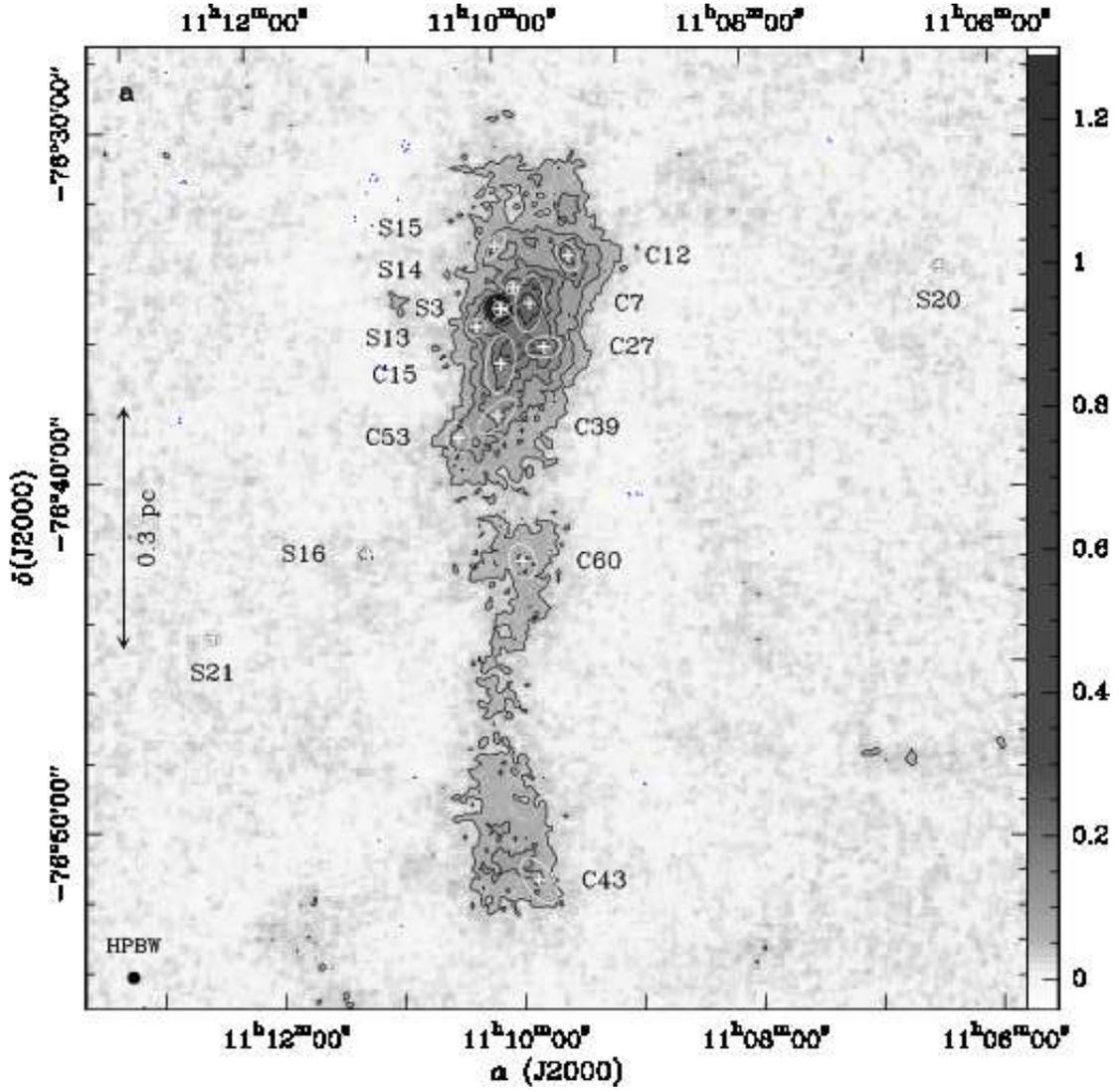}}}
\caption{Detailed 870 $\mu$m continuum emission maps of Cha~I extracted
from the map shown in Fig.~\ref{f:labocamap}. The flux density greyscale is 
shown on the right of each panel and labeled in Jy/21.2$\arcsec$-beam. It has 
been optimized to reveal the faint emission 
with a better contrast. The angular resolution of the map is shown in the 
lower left corner of each panel ($HPBW = 21.2\arcsec$). The white plus symbols 
and ellipses show the positions, sizes ($FWHM$), and orientations of the 
Gaussian sources extracted with \textit{Gaussclumps} or fitted with GAUSS\_2D 
in the filtered map shown in Fig.~\ref{f:mrmed}a. The sources are labeled like 
in the first column of Tables~\ref{t:starless}, \ref{t:ysos}, and 
\ref{t:motherysos}. ``C'' stands for starless core (or Class 0 protostellar 
core) and ``S'' for YSO (Class~I or more evolved). In the following, the 
contour levels are described with the parameter 
$b = 36$~mJy/21.2$\arcsec$-beam, i.e. about 3 times the rms noise level.
\textbf{a} Field Cha-North. The contour levels are $-b$ (in dotted 
blue), $b$, 
$2b$, $3b$, $4b$, $5b$, $6b$, $9b$, $12b$, $18b$, $24b$ and $30b$. The white, 
thin cross marks the SEST position of Cha-MMS2 \citep[][]{Reipurth96}.}
\label{f:labocamapdet}
\end{figure*}

\addtocounter{figure}{-1}
\begin{figure*}
\centerline{\resizebox{0.81\hsize}{!}{\includegraphics[angle=270]{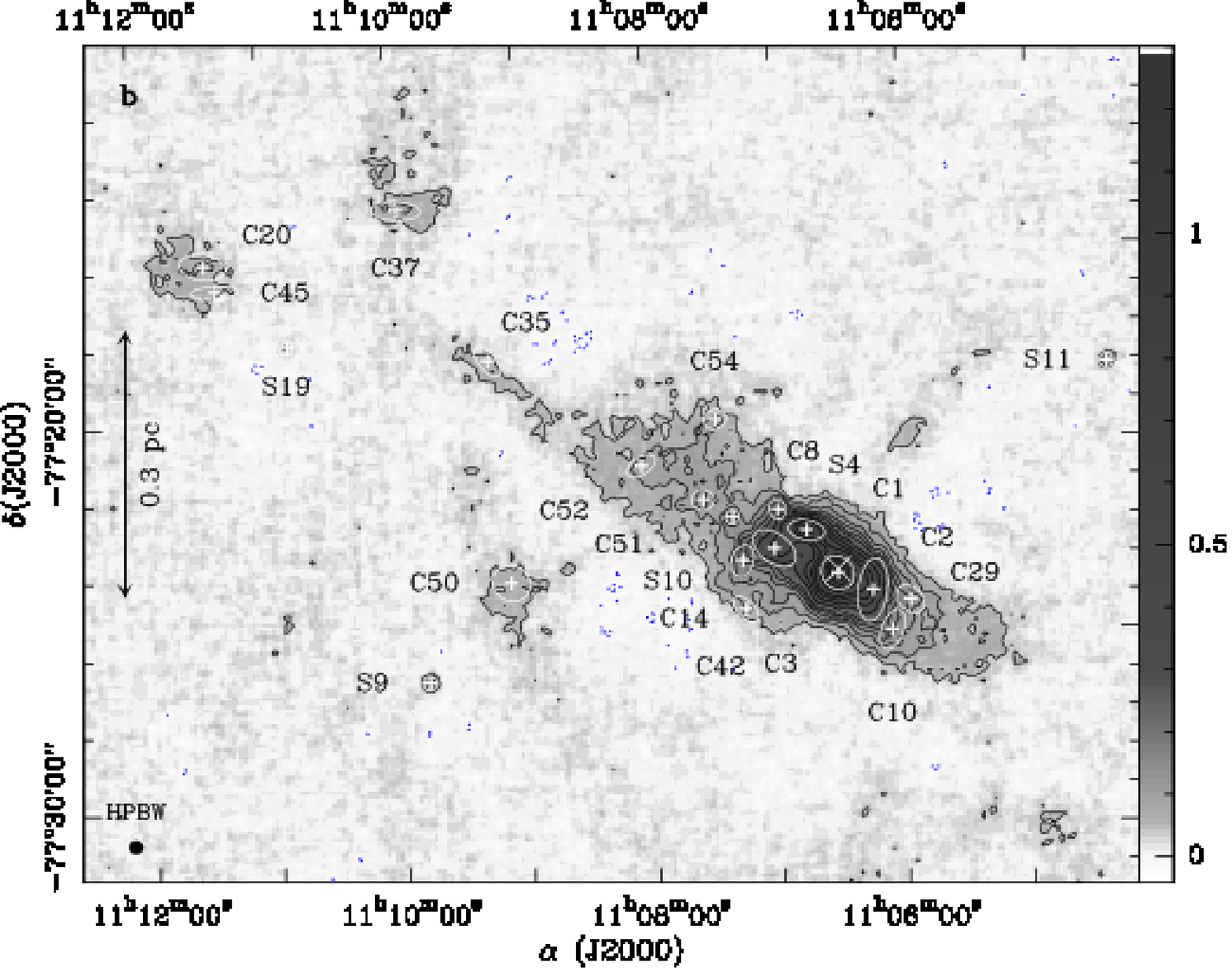}}}
\caption{(continued) \textbf{b} Northern part of field Cha-Center. The contour 
levels are $-b$ (in dotted blue), $b$, $2b$, $3b$, $4b$, $5b$, $6b$, 
$8b$, 
$10b$, $12b$, $14b$, $18b$, $22b$, $26b$, $30b$, and $34b$. The white, 
thin cross marks the SEST position of Cha-MMS1 \citep[][]{Reipurth96}.}
\end{figure*}

\addtocounter{figure}{-1}
\begin{figure*}
\centerline{\resizebox{0.96\hsize}{!}{\includegraphics[angle=270]{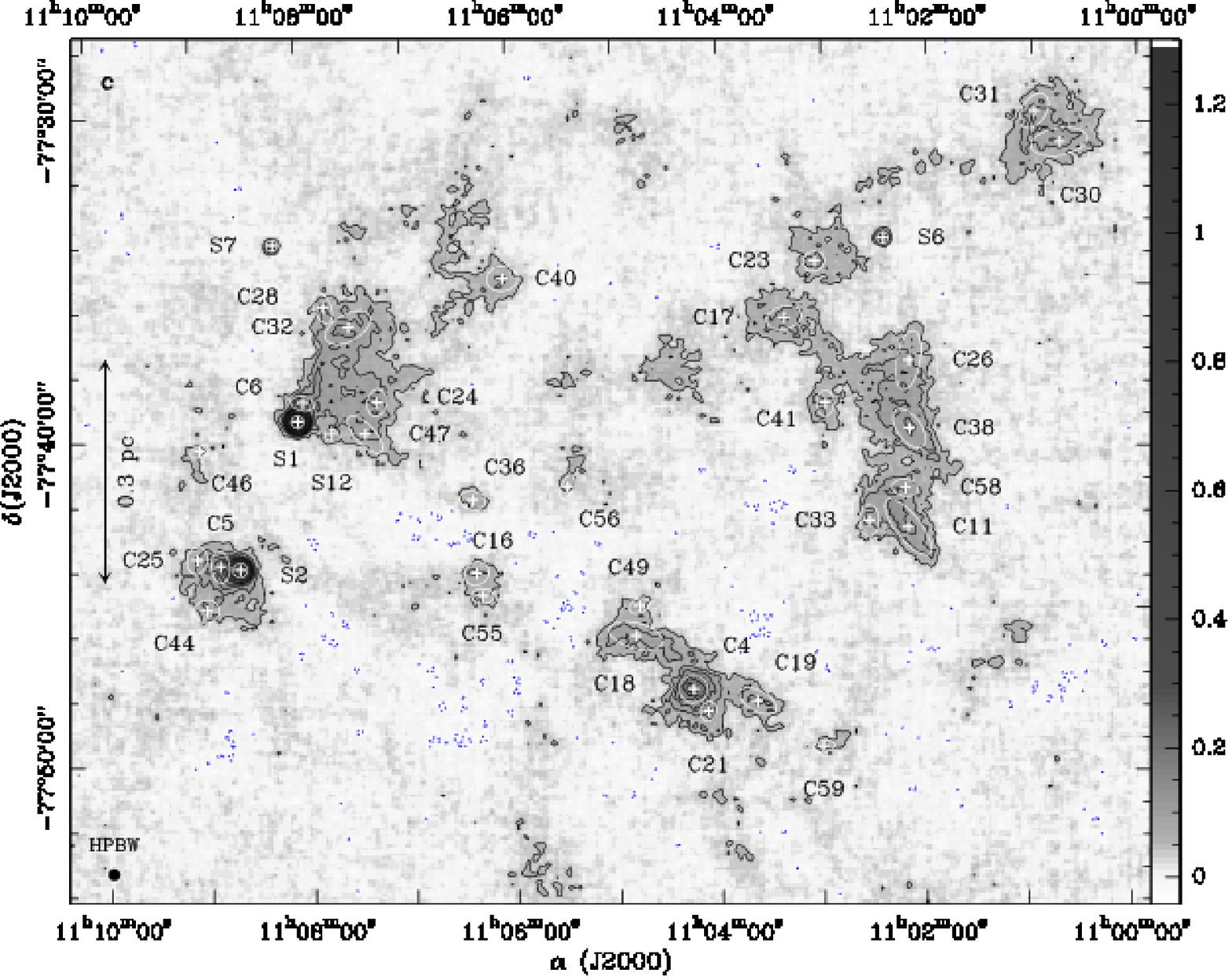}}}
\caption{(continued) \textbf{c} Southern part of field Cha-Center. The contour 
levels are $-b$ (in dotted blue), $b$, $2b$, $3b$, $4b$, $5b$, $6b$, 
$9b$, 
$15b$, $21b$, $32b$, $48b$, and $64b$.}
\end{figure*}

\addtocounter{figure}{-1}
\begin{figure}
\centerline{\resizebox{0.93\hsize}{!}{\includegraphics[angle=270]{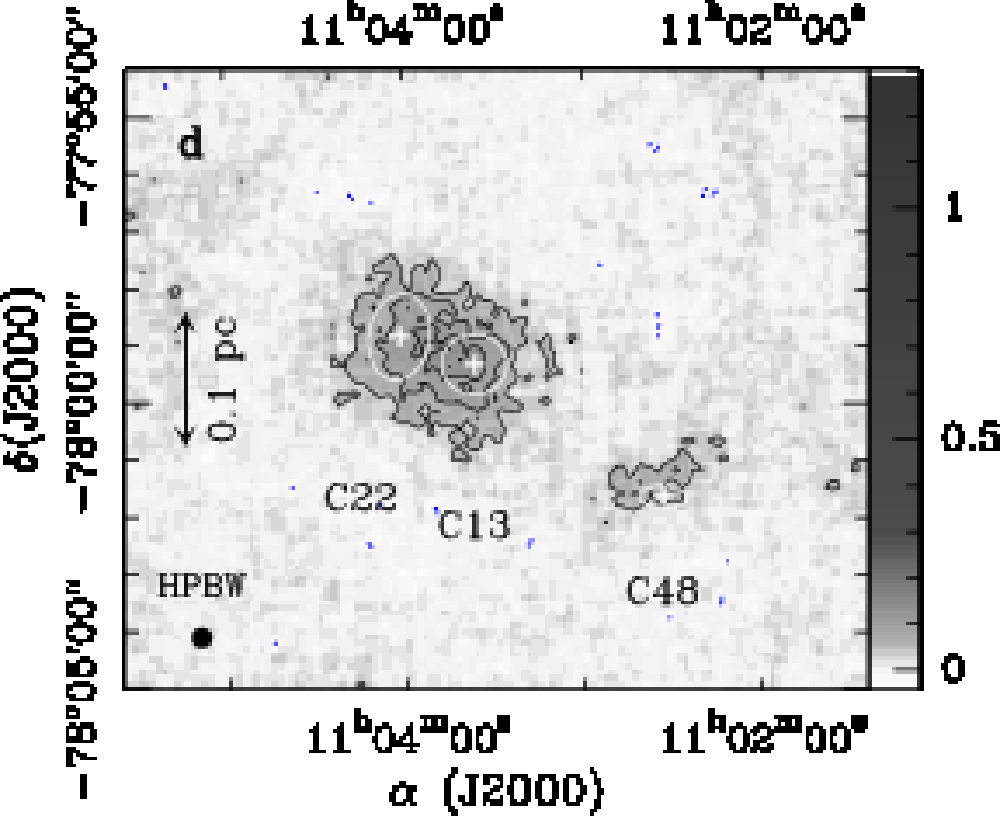}}}
\caption{(continued) \textbf{d} Field Cha-South. The contour levels are $-b$ 
(in dotted blue), $b$, $2b$, and $3b$.}
\end{figure}

\addtocounter{figure}{-1}
\begin{figure}
\centerline{\resizebox{0.96\hsize}{!}{\includegraphics[angle=270]{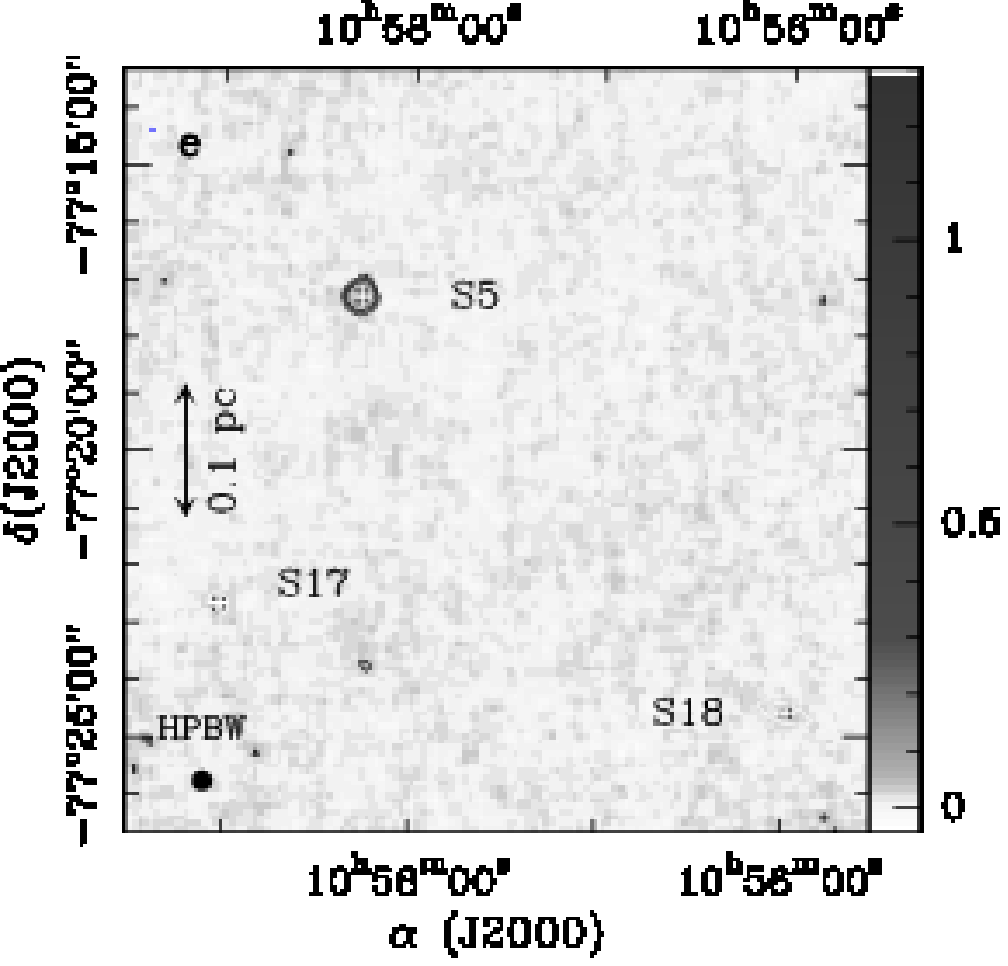}}}
\caption{(continued) \textbf{e} South-eastern part of field Cha-West. The 
contour levels are $-b$ (in dotted blue), $b$, $2b$, $4b$, and $7b$.}
\end{figure}

\addtocounter{figure}{-1}
\begin{figure}
\centerline{\resizebox{0.67\hsize}{!}{\includegraphics[angle=270]{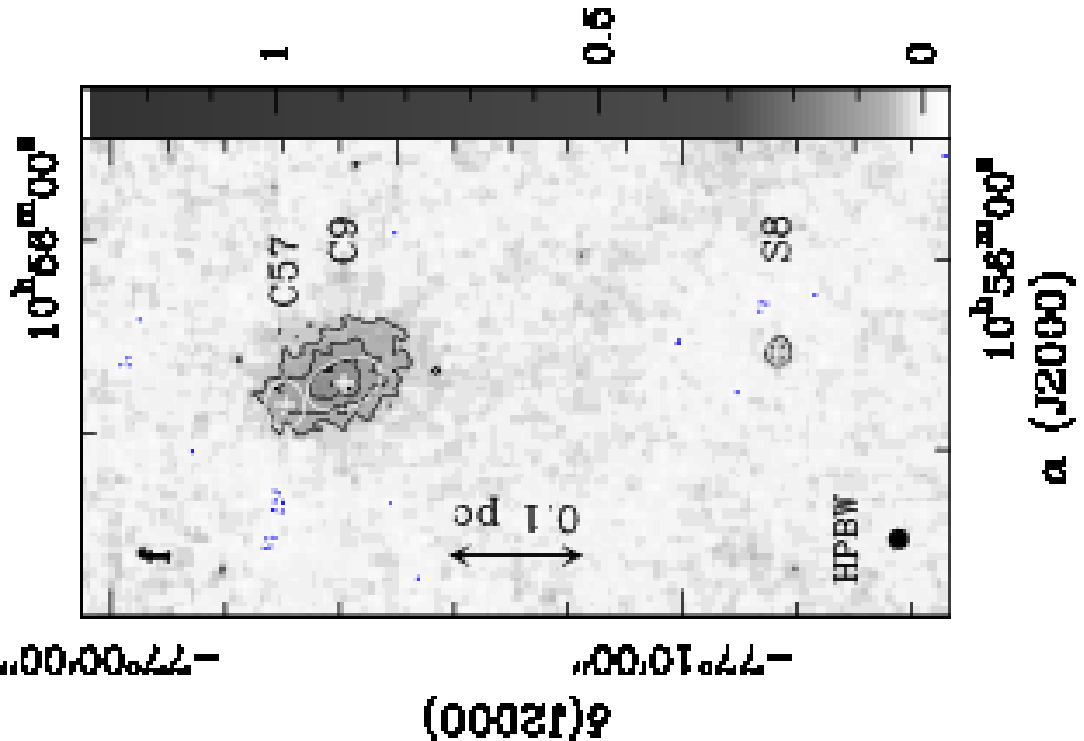}}}
\caption{(continued) \textbf{f} North-eastern part of field Cha-West. The 
contour levels are $-b$ (in dotted blue), $b$, $2b$, $3b$, and $4b$.}
\end{figure}

\addtocounter{figure}{-1}
\begin{figure}
\centerline{\resizebox{1.00\hsize}{!}{\includegraphics[angle=270]{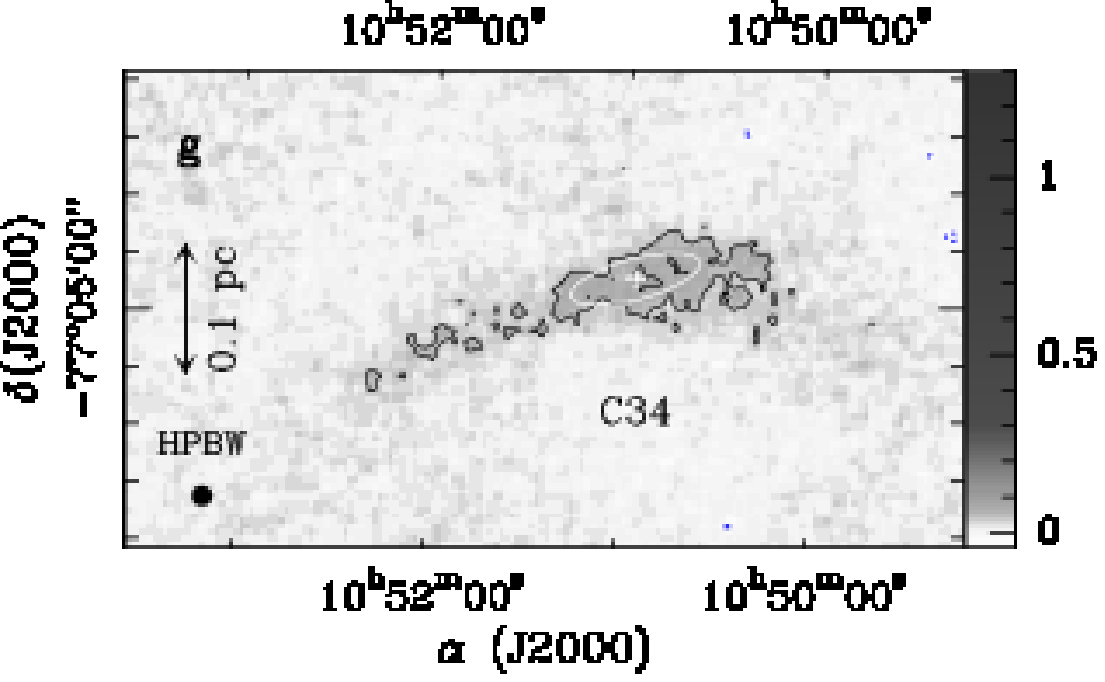}}}
\caption{(continued) \textbf{g} North-western part of field Cha-West. The 
contour levels are $-b$ (in dotted blue), $b$, and $2b$.}
\end{figure}

\subsection{Masses traced with LABOCA and 2MASS}
\label{ss:cha1masses}

The total 870~$\mu$m flux in the whole map of Cha I is about 115~Jy. 
Assuming a dust mass opacity $\kappa_{870} = 0.01$~cm$^2$~g$^{-1}$, this 
translates into a cloud mass of 61~M$_\odot$. It corresponds to 5.9$\%$ of the 
total mass traced by CO in Cha~I 
\citep[1030~M$_\odot$,][]{Mizuno01}, 7.7$\%$ of the mass traced by $^{13}$CO 
\citep[790~M$_\odot$,][]{Mizuno99}, and 27--32$\%$ of the mass traced by 
C$^{18}$O \citep[190--230~M$_\odot$,][]{Mizuno99,Haikala05}.

The extinction map shown in Fig.~\ref{f:av} traces larger scales than
the 870~$\mu$m dust emission map. The median and mean extinctions over the 
1.6~deg$^2$ covered with LABOCA are 3.3 and 4.1~mag, respectively. Assuming an 
extinction to H$_2$ column density 
conversion factor of $9.4 \times 10^{20}$~cm$^{-2}$~mag$^{-1}$ (for $R_V = 3.1$, 
see Appendix~\ref{ss:avtoN}), we derive a total gas+dust mass of 950~M$_\odot$.
However, only 62$\%$ of this mass, i.e. 590~M$_\odot$, is at $A_V < 6$~mag. 
With the appropriate conversion factor for $A_V > 6$~mag (see 
Appendix~\ref{ss:avtoN}), the remaining mass is reduced to 220~M$_\odot$, 
yielding a more accurate estimate of 810~M$_\odot$ for the total mass of Cha~I 
traced with the extinction. It is roughly consistent with the masses traced by 
CO and $^{13}$CO mentioned above. Since the latter masses may have been 
integrated on a somewhat different area, we consider the mass derived from the 
extinction map as the best estimate to compare with. Thus the mass traced with 
LABOCA represents about 7.5$\%$ of the cloud mass. Given that the median 
extinction is 3.3~mag in the extinction map, i.e. about 2.9 times the rms 
sensitivity achieved with LABOCA, the missing 92$\%$ were lost not only 
because of a lack of sensitivity but also because of the spatial filtering due 
to the correlated noise removal (see Appendix~\ref{ss:filtering}). 
Finally, we estimate the average density of free particles. We assume that the 
depth of the cloud along the line of sight is equal to the square root of its 
projected surface, i.e. 3.3~pc. This yields an average density of 
$\sim$~380~cm$^{-3}$. For the same depth, the 
density of free particles corresponding to the median 
visual extinction is estimated to be $\sim 350$~cm$^{-3}$.

\section{Source extraction and classification}
\label{s:sourceid}

\subsection{Multiresolution decomposition}
\label{ss:mrmed}

\citet{Motte98,Motte07} used a multiresolution program based on 
wavelet transforms to decompose their continuum maps into different scales
and better estimate the properties of condensations embedded in 
larger-scale structures. We follow the same strategy with a different filter. 
Appendix~\ref{s:mrmed} details our method.

\begin{table}
 \caption{Continuum flux distribution in Cha~I.}
 \label{t:flux_mrmed}
 \centering
 \begin{tabular}{ccccc}
  \hline\hline
  \multicolumn{1}{c}{Scale} & \multicolumn{1}{c}{Typical size} & \multicolumn{1}{c}{Flux} & \multicolumn{1}{c}{$\%$ of total flux} & \multicolumn{1}{c}{Mass} \\
   \multicolumn{1}{c}{} &   &  \multicolumn{1}{c}{\scriptsize (Jy)} & &  \multicolumn{1}{c}{\scriptsize (M$_\odot$)}   \\
  \multicolumn{1}{c}{(1)} & \multicolumn{1}{c}{(2)} & \multicolumn{1}{c}{(3)} & \multicolumn{1}{c}{(4)} & \multicolumn{1}{c}{(5)} \\
  \hline
  3 & $<60\arcsec$  &  12.3 &  11 &  6.5 \\
  4 & $<120\arcsec$ &  25.8 &  22 & 13.6 \\
  5 & $<200\arcsec$ &  56.3 &  49 & 29.8 \\
  6 & $<300\arcsec$ &  94.4 &  82 & 49.9 \\
  7 & $\sim$ all    & 113.9 &  99 & 60.3 \\ 
  -- & all           & 114.9 & 100 & 60.8 \\
  \hline
 \end{tabular}
 \tablefoot{The mass is computed assuming a dust mass opacity 
$\kappa_{870} = 0.01$~cm$^2$~g$^{-1}$. The last row corresponds to 
the full map, while rows 1 to 5 correspond to the sum of the filtered maps up 
to scale $i$ listed in the first column (i.e. the \textit{sum} map at scale 
$i$). Column 2 gives the range of sizes of the sources that contribute 
significantly to the emission with more than $40\%$ of their peak flux density 
(see Col.~2 of Table~\ref{t:mrmed}).}
\end{table}

\begin{figure*}
\centerline{\resizebox{1.00\hsize}{!}{\includegraphics[angle=270]{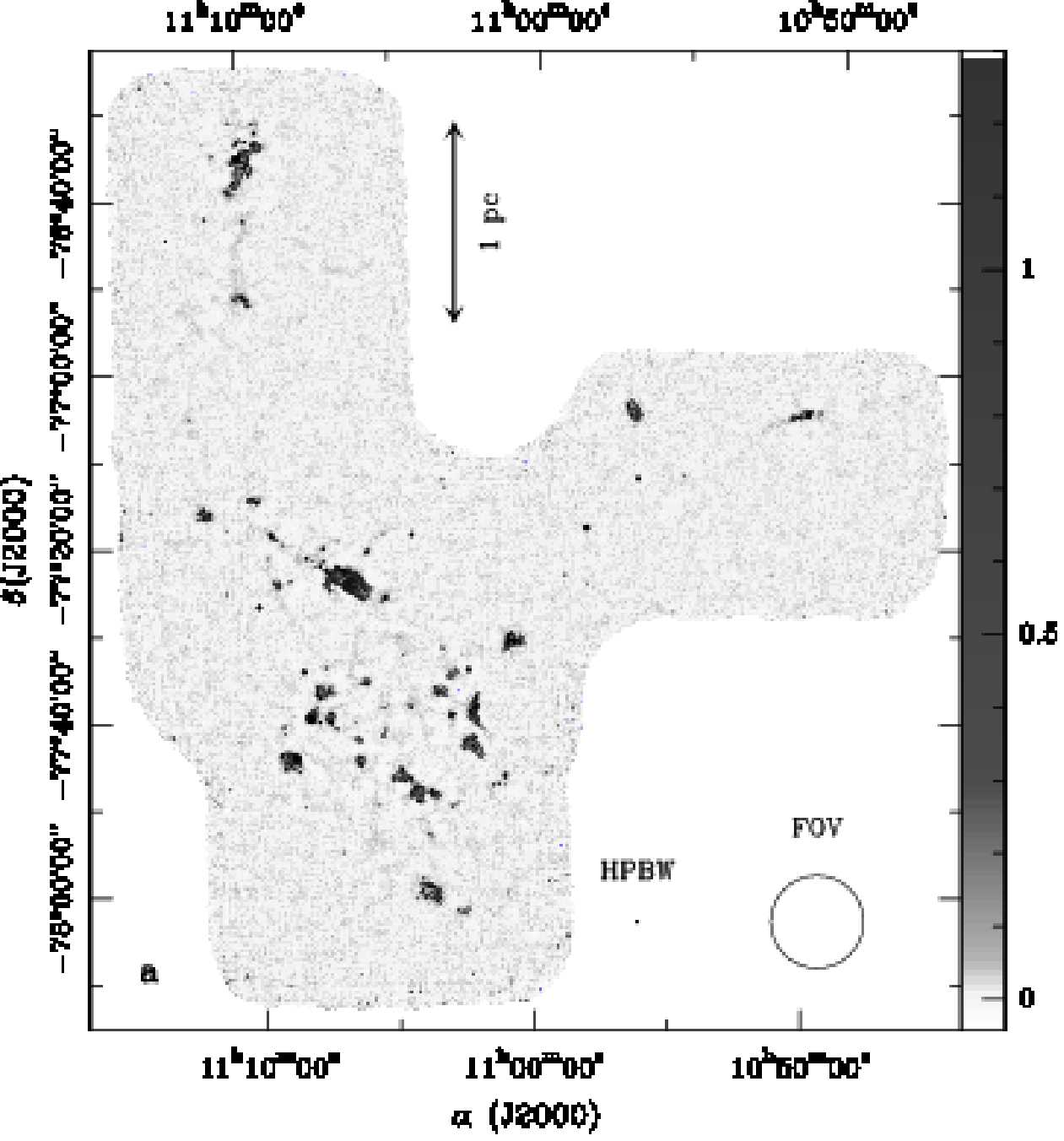}\hspace*{3ex}\includegraphics[angle=270]{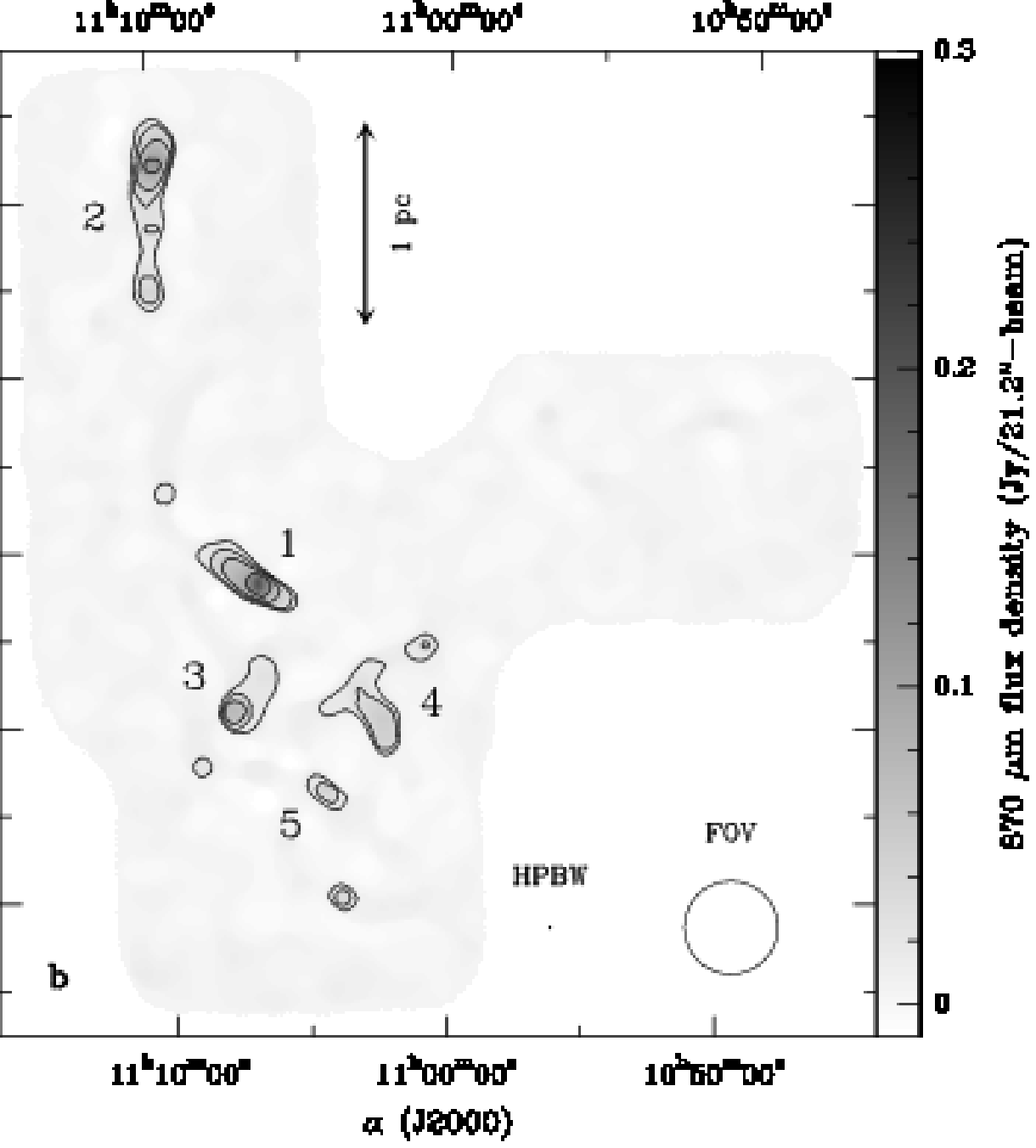}}}
\caption{\textbf{a} 870 $\mu$m continuum emission \textit{sum} map of Cha~I at 
scale 5 (see Sect.~\ref{ss:mrmed}). The contour levels are $-a$ (in dashed 
blue), $a$, $2a$, $4a$, $8a$, $16a$, and $32a$, with 
$a = 48$~mJy/21.2$\arcsec$-beam, i.e. about 4 times the rms noise level. 
\textbf{b} Smoothed map, i.e. residuals, at scale 5. The contour levels are 
$-c$ (in dashed blue), $c$, $2c$, $4c$, and $8c$, with 
$c = 13.5$~mJy/21.2$\arcsec$-beam, i.e. about 4.5 times the rms noise level in 
this map. The labels indicate the filaments listed in Table~\ref{t:clf_ls}.
The greyscales of both maps are different. The sum of these two 
maps is strictly equal to the original map (Fig.~\ref{f:labocamap}).}
\label{f:mrmed}
\end{figure*}

This multiresolution decomposition was performed on the continuum map of 
Cha~I. The total fluxes measured in the \textit{sum} maps at scales 3 to 7 
(see definition in Appendix~\ref{s:mrmed}) are listed in 
Table~\ref{t:flux_mrmed}, as well as the corresponding masses. About half of 
the total flux is emitted by structures smaller than $\sim 200\arcsec$ 
($FWHM$), and only 11$\%$ by structures smaller than $\sim 60\arcsec$. The 
\textit{sum} map at scale 5 and its associated smoothed map are shown in 
Fig.~\ref{f:mrmed}. The sum of these two maps is strictly equal to the 
original map shown in Fig.~\ref{f:labocamap}.

\subsection{Source extraction with \textit{Gaussclumps}}
\label{ss:gaussclumps}

\textit{Gaussclumps} \citep[][]{Stutzki90,Kramer98} and \textit{Clumpfind} 
\citep[][]{Williams94} are two of the most popular numerical codes used to 
extract sources from large-scale molecular line data cubes and dust continuum 
maps. \citet{Mookerjea04} found 
similar core mass distributions in the massive star forming region RCW~106 
with both algorithms, even if there was no one-to-one correspondence between 
source positions and masses. On the other hand, depending on the source 
extraction algorithm, \citet{Curtis10} recently obtained significant 
differences between the core mass distributions derived for the Perseus 
molecular cloud, as well as opposite conclusions concerning the size of 
protostellar versus starless cores. They mentioned in their conclusions that 
\textit{Gaussclumps} may be more reliable than \textit{Clumpfind} to 
disentangle blended sources in highly-clustered regions. Here, we decided to 
use \textit{Gaussclumps} to identify and extract sources in the continuum map 
of Cha~I.

We set all three stiffness parameters to 1, as recommended by 
\citet{Kramer98}. The initial guesses for the aperture cutoff and aperture 
$FWHM$ were set to 8 and 3 times the angular resolution ($HPBW$), 
respectively. The initial guess for the source $FWHM$ was set to 
$1.5 \times HPBW$, and the peak flux density threshold to 
60~mJy/$21.2\arcsec$-beam, i.e. $5\sigma$ to secure the detections.

\begin{table*}
 \caption{
 Sources extracted with \textit{Gaussclumps} in the 870~$\mu$m continuum \textit{sum} map of Cha~I at scale 5, and possible associations found in the SIMBAD database.
 }
 \label{t:id_gcl_simbad}
 \centering
 \begin{tabular}{cccccccccclc}
 \hline\hline
 \multicolumn{1}{c}{$N_{\mathrm{gcl}}$} & \multicolumn{1}{c}{R.A.} & \multicolumn{1}{c}{Decl.} & \multicolumn{1}{c}{${f_{\mathrm{peak}}}$\tablefootmark{a}} & \multicolumn{1}{c}{${f_{\mathrm{tot}}}$\tablefootmark{a}} & \multicolumn{1}{c}{maj.\tablefootmark{a}} & \multicolumn{1}{c}{min.\tablefootmark{a}} & \multicolumn{1}{c}{P.A.\tablefootmark{a}} & \multicolumn{1}{c}{$S$\tablefootmark{b}} & \multicolumn{1}{c}{Typ.\tablefootmark{c}} & \multicolumn{1}{c}{\hspace*{-1.5ex} SIMBAD possible} & \multicolumn{1}{c}{Dist.\tablefootmark{d}} \\ 
  & \multicolumn{1}{c}{\scriptsize (J2000)} & \multicolumn{1}{c}{\scriptsize (J2000)} & \multicolumn{1}{c}{\scriptsize (Jy/beam)} & \multicolumn{1}{c}{\scriptsize (Jy)} & \multicolumn{1}{c}{\scriptsize ($\arcsec$)} & \multicolumn{1}{c}{\scriptsize ($\arcsec$)} & \multicolumn{1}{c}{\scriptsize ($^\circ$)} & \multicolumn{1}{c}{\scriptsize ($\arcsec$)} & & \multicolumn{1}{c}{\hspace*{-1.5ex} associations\tablefootmark{d}} & \multicolumn{1}{c}{\scriptsize ($\arcsec$)} \\ 
 \multicolumn{1}{c}{(1)} & \multicolumn{1}{c}{(2)} & \multicolumn{1}{c}{(3)} & \multicolumn{1}{c}{(4)} & \multicolumn{1}{c}{(5)} & \multicolumn{1}{c}{(6)} & \multicolumn{1}{c}{(7)} & \multicolumn{1}{c}{(8)} & \multicolumn{1}{c}{(9)} & \multicolumn{1}{c}{(10)} & \multicolumn{1}{c}{\hspace*{-1.5ex} (11)} & \multicolumn{1}{c}{(12)} \\ 
 \hline
  1 & 11:08:03.22 & -77:39:17.9 &  2.580 &    3.093 &  23.5 &  22.9 &  50.0 &  23.2 & S & \hspace*{-1.5ex} V* CU Cha                      (Or*) &   0.5 \\ 
 & & & & & & & & & & \hspace*{-1.5ex} BRAN 341B                      (BNe) &   1.3 \\ 
  2 & 11:08:38.96 & -77:43:52.9 &  1.206 &    1.893 &  26.6 &  26.6 & -33.2 &  26.6 & S & \hspace*{-1.5ex} Cha IR NEBULA             (TT*) &   6.2 \\ 
  3 & 11:10:00.13 & -76:34:59.1 &  1.169 &    1.501 &  25.1 &  23.0 &  61.3 &  24.0 & S & \hspace*{-1.5ex} V* WW Cha                      (Or*) &   1.2 \\ 
 & & & & & & & & & & \hspace*{-1.5ex} CHXR 44                        (X) &   6.6 \\ 
 & & & & & & & & & & \hspace*{-1.5ex} [RNC96] Cha-MMS 2              (mm) &   7.5 \\ 
  4 & 11:06:31.94 & -77:23:38.9 &  1.053 &    6.350 &  55.1 &  49.3 &  41.5 &  52.1 & C & \hspace*{-1.5ex} [RNC96] Cha-MMS 1a             (mm) &   6.9 \\ 
  5 & 11:06:46.44 & -77:22:32.2 &  0.607 &    2.418 &  57.8 &  31.0 &  80.9 &  42.3 & S & \hspace*{-1.5ex} [PMG2001] NIR 69               (IR) &   2.2 \\ 
 & & & & & & & & & & \hspace*{-1.5ex} IRAS 11051-7706                (Y*O) &   2.9 \\ 
 & & & & & & & & & & \hspace*{-1.5ex} [PMG2001] NIR 72               (IR) &   5.9 \\ 
 & & & & & & & & & & \hspace*{-1.5ex} [PMG2001] NIR 74               (IR) &   6.4 \\ 
 & & & & & & & & & & \hspace*{-1.5ex} [PMG2001] NIR 80               (IR) &  12.8 \\ 
  6 & 11:06:15.51 & -77:24:04.9 &  0.306 &    3.067 &  98.2 &  45.9 &  -7.0 &  67.2 & C & \hspace*{-1.5ex} [PMG2001] NIR 16               (IR) &  38.2 \\ 
  7 & 10:58:16.92 & -77:17:18.3 &  0.293 &    0.314 &  22.7 &  21.3 & -34.0 &  21.9 & S & \hspace*{-1.5ex} V* SZ Cha                      (Or*) &   1.3 \\ 
  8 & 11:07:01.85 & -77:23:00.8 &  0.283 &    2.197 &  76.1 &  45.9 &  53.7 &  59.1 & C & \hspace*{-1.5ex} ISO-ChaI 86                    (Y*O) &  14.6 \\ 
 & & & & & & & & & & \hspace*{-1.5ex} IRAS 11057-7706                (Y*O) &  26.4 \\ 
 & & & & & & & & & & \hspace*{-1.5ex} [PCW91] Ced 110 IRS 6A         (IR) &  27.0 \\ 
  9 & 11:04:16.41 & -77:47:32.1 &  0.209 &    1.249 &  53.4 &  50.4 & -78.1 &  51.9 & C & & \\ 
 10 & 11:02:24.98 & -77:33:36.3 &  0.195 &    0.197 &  21.4 &  21.2 & -63.0 &  21.3 & S & \hspace*{-1.5ex} V* CS Cha                      (Or*) &   0.6 \\ 
 11 & 11:08:50.40 & -77:43:46.7 &  0.167 &    0.773 &  60.3 &  34.5 &   5.2 &  45.6 & C & \hspace*{-1.5ex} HJM C 9-1                      (IR) &   8.2 \\ 
 12 & 11:08:15.28 & -77:33:51.9 &  0.168 &    0.173 &  21.9 &  21.2 & -63.0 &  21.6 & S & \hspace*{-1.5ex} Glass Ia                  (TT*) &   1.2 \\ 
 & & & & & & & & & & \hspace*{-1.5ex} SSTc2d J110815.3-773353        (*) &   1.4 \\ 
 13 & 11:08:00.07 & -77:38:43.8 &  0.161 &    0.412 &  44.3 &  26.0 &  79.7 &  33.9 & C & \hspace*{-1.5ex} HH 920B                        (HH) &   6.8 \\ 
 & & & & & & & & & & \hspace*{-1.5ex} 2M J11075792-7738449        (TT*) &   7.0 \\ 
 & & & & & & & & & & \hspace*{-1.5ex} 2M J11080297-7738425        (Y*O) &   9.4 \\ 
 & & & & & & & & & & \hspace*{-1.5ex} [HTN2003] HD 97048 2           (IR) &  11.5 \\ 
 & & & & & & & & & & \hspace*{-1.5ex} ESO-HA 562                     (Y*O) &  12.3 \\ 
 14 & 11:09:45.85 & -76:34:49.7 &  0.148 &    1.552 &  98.3 &  48.1 &  -7.9 &  68.8 & C & \hspace*{-1.5ex} 2M J11094621-7634463        (TT*) &   3.5 \\ 
 & & & & & & & & & & \hspace*{-1.5ex} Hn 10W                         (Em*) &   4.5 \\ 
 & & & & & & & & & & \hspace*{-1.5ex} [ACA2003] Cha I4               (Em*) &   7.3 \\ 
 & & & & & & & & & & \hspace*{-1.5ex} OTS 20                         (*) &  13.0 \\ 
 & & & & & & & & & & \hspace*{-1.5ex} [ACA2003] Cha I3               (pr*) &  16.2 \\ 
 15 & 11:06:59.93 & -77:22:00.6 &  0.154 &    0.276 &  31.0 &  26.1 &  15.0 &  28.4 & C & \hspace*{-1.5ex} [PMG2001] NIR 103              (IR) &   7.5 \\ 
 & & & & & & & & & & \hspace*{-1.5ex} [PMG2001] NIR 96               (IR) &   7.9 \\ 
 & & & & & & & & & & \hspace*{-1.5ex} [PMG2001] NIR 101              (IR) &  10.8 \\ 
 & & & & & & & & & & \hspace*{-1.5ex} [PMG2001] NIR 93               (IR) &  11.9 \\ 
 16 & 10:56:42.95 & -77:04:05.1 &  0.134 &    1.278 &  82.6 &  51.8 &  23.0 &  65.4 & C & & \\ 
 17 & 10:56:30.56 & -77:11:39.8 &  0.137 &    0.142 &  22.0 &  21.2 & -88.6 &  21.6 & S & \hspace*{-1.5ex} V* SY Cha                      (Or*) &   0.6 \\ 
 18 & 11:09:47.42 & -77:26:31.5 &  0.129 &    0.132 &  21.8 &  21.2 & -10.2 &  21.5 & S & \hspace*{-1.5ex} 2M J11094742-7726290        (Y*O) &   2.4 \\ 
 19 & 11:06:39.88 & -77:24:18.1 &  0.130 &    0.521 &  68.1 &  26.5 &  53.7 &  42.4 & R & & \\ 
 20 & 11:07:21.34 & -77:22:12.1 &  0.119 &    0.121 &  21.6 &  21.2 &  -5.7 &  21.4 & S & \hspace*{-1.5ex} DENIS-P J1107.4-7722           (Y*O) &   1.3 \\ 
 & & & & & & & & & & \hspace*{-1.5ex} [PMG2001] NIR 137              (IR) &   3.8 \\ 
 & & & & & & & & & & \hspace*{-1.5ex} [PMG2001] NIR 140              (IR) &   4.4 \\ 
 21 & 11:06:06.48 & -77:25:06.7 &  0.116 &    0.583 &  65.9 &  34.3 & -20.9 &  47.5 & C & & \\ 
 22 & 11:04:23.35 & -77:18:07.8 &  0.114 &    0.115 &  21.4 &  21.2 & -19.0 &  21.3 & S & \hspace*{-1.5ex} HH 48 IRS                 (HH) &   2.0 \\ 
 & & & & & & & & & & \hspace*{-1.5ex} HH 48E                         (HH) &   7.5 \\ 
 23 & 11:02:10.85 & -77:42:31.8 &  0.111 &    1.434 & 126.7 &  45.8 &  38.6 &  76.2 & C & & \\ 
 24 & 11:09:26.24 & -76:33:28.3 &  0.111 &    0.455 &  56.3 &  32.7 &  27.1 &  42.9 & C & \hspace*{-1.5ex} DENIS-P J1109.4-7633           (Y*O) &   5.2 \\ 
 & & & & & & & & & & \hspace*{-1.5ex} [ACA2003] Cha I2               (pr*) &   8.1 \\ 
 25 & 11:03:36.59 & -77:59:19.3 &  0.104 &    1.023 &  69.4 &  63.7 &  85.8 &  66.5 & C & & \\ 
 26 & 11:07:17.13 & -77:23:20.2 &  0.109 &    0.403 &  49.8 &  33.6 & -17.3 &  40.9 & C & \hspace*{-1.5ex} [PMG2001] NIR 134              (IR) &   4.8 \\ 
 & & & & & & & & & & \hspace*{-1.5ex} DENIS-P J1107.3-7723           (Y*O) &  11.9 \\ 
 & & & & & & & & & & \hspace*{-1.5ex} [PMG2001] NIR 135              (IR) &  15.4 \\ 
 & & & & & & & & & & \hspace*{-1.5ex} [PMG2001] NIR 142              (IR) &  16.7 \\ 
 27 & 11:06:35.35 & -77:22:52.8 &  0.106 &    0.263 &  52.3 &  21.4 & -74.9 &  33.5 & R & \hspace*{-1.5ex} 1AXG J110700-7728              (X) &  25.3 \\ 
 28 & 11:07:43.89 & -77:39:41.6 &  0.093 &    0.094 &  21.4 &  21.2 & -35.9 &  21.3 & S & \hspace*{-1.5ex} Ass Cha T 2-28                 (TT*) &   0.8 \\ 
 29 & 11:10:01.05 & -76:36:32.9 &  0.090 &    0.999 & 102.4 &  48.6 &  -8.0 &  70.5 & C & \hspace*{-1.5ex} OTS 37                         (*) &  28.8 \\ 
 30 & 11:06:21.67 & -77:43:58.4 &  0.089 &    0.392 &  47.4 &  41.7 & -60.9 &  44.5 & C & & \\ 
 31 & 11:10:12.26 & -76:35:30.1 &  0.093 &    0.211 &  35.8 &  28.5 & -79.5 &  32.0 & Sc & \hspace*{-1.5ex} 2M J11101141-7635292        (pr*) &   3.0 \\ 
 32 & 11:03:21.97 & -77:36:04.4 &  0.088 &    0.554 &  67.3 &  42.4 & -77.0 &  53.4 & C & & \\ 
 33 & 11:04:49.58 & -77:45:53.5 &  0.086 &    0.926 & 102.1 &  47.7 &  75.1 &  69.8 & C & & \\ 
 \hline
 \end{tabular}
 \end{table*}
\begin{table*}
 \addtocounter{table}{-1}
 \caption{
continued.
 }
 \centering
 \begin{tabular}{cccccccccclc}
 \hline\hline
 \multicolumn{1}{c}{$N_{\mathrm{gcl}}$} & \multicolumn{1}{c}{R.A.} & \multicolumn{1}{c}{Decl.} & \multicolumn{1}{c}{${f_{\mathrm{peak}}}$\tablefootmark{a}} & \multicolumn{1}{c}{${f_{\mathrm{tot}}}$\tablefootmark{a}} & \multicolumn{1}{c}{maj.\tablefootmark{a}} & \multicolumn{1}{c}{min.\tablefootmark{a}} & \multicolumn{1}{c}{P.A.\tablefootmark{a}} & \multicolumn{1}{c}{$S$\tablefootmark{b}} & \multicolumn{1}{c}{Typ.\tablefootmark{c}} & \multicolumn{1}{c}{\hspace*{-1.5ex} SIMBAD possible} & \multicolumn{1}{c}{Dist.\tablefootmark{d}} \\ 
  & \multicolumn{1}{c}{\scriptsize (J2000)} & \multicolumn{1}{c}{\scriptsize (J2000)} & \multicolumn{1}{c}{\scriptsize (Jy/beam)} & \multicolumn{1}{c}{\scriptsize (Jy)} & \multicolumn{1}{c}{\scriptsize ($\arcsec$)} & \multicolumn{1}{c}{\scriptsize ($\arcsec$)} & \multicolumn{1}{c}{\scriptsize ($^\circ$)} & \multicolumn{1}{c}{\scriptsize ($\arcsec$)} & & \multicolumn{1}{c}{\hspace*{-1.5ex} associations\tablefootmark{d}} & \multicolumn{1}{c}{\scriptsize ($\arcsec$)} \\ 
 \multicolumn{1}{c}{(1)} & \multicolumn{1}{c}{(2)} & \multicolumn{1}{c}{(3)} & \multicolumn{1}{c}{(4)} & \multicolumn{1}{c}{(5)} & \multicolumn{1}{c}{(6)} & \multicolumn{1}{c}{(7)} & \multicolumn{1}{c}{(8)} & \multicolumn{1}{c}{(9)} & \multicolumn{1}{c}{(10)} & \multicolumn{1}{c}{\hspace*{-1.5ex} (11)} & \multicolumn{1}{c}{(12)} \\ 
 \hline
 34 & 11:03:39.07 & -77:47:54.0 &  0.083 &    0.529 &  76.3 &  37.5 &  60.7 &  53.5 & C & & \\ 
 35 & 11:11:27.78 & -77:15:45.2 &  0.078 &    0.507 &  73.5 &  39.9 &  83.3 &  54.1 & C & & \\ 
 36 & 11:04:07.89 & -77:48:13.1 &  0.082 &    0.197 &  37.4 &  29.0 & -12.9 &  32.9 & C & & \\ 
 37 & 11:04:01.79 & -77:58:50.4 &  0.078 &    1.016 &  90.7 &  64.3 &  -8.9 &  76.3 & C & & \\ 
 38 & 11:03:04.43 & -77:34:19.4 &  0.079 &    0.262 &  40.5 &  37.0 & -62.7 &  38.7 & C & & \\ 
 39 & 11:07:17.30 & -77:38:42.3 &  0.077 &    0.307 &  50.3 &  35.8 & -10.8 &  42.5 & C & \hspace*{-1.5ex} HH 927                         (HH) &  12.3 \\ 
 40 & 11:09:03.86 & -77:43:34.6 &  0.081 &    0.346 &  51.4 &  37.2 &  -9.1 &  43.8 & C & & \\ 
 41 & 11:02:10.46 & -77:37:22.9 &  0.077 &    0.830 & 111.5 &  43.7 & -10.3 &  69.8 & C & & \\ 
 42 & 11:09:39.48 & -76:36:04.6 &  0.080 &    0.400 &  61.1 &  36.7 & -83.9 &  47.3 & C & & \\ 
 43 & 11:07:47.07 & -77:35:46.8 &  0.075 &    0.319 &  59.6 &  32.1 &  22.0 &  43.7 & C & & \\ 
 44 & 11:05:57.64 & -77:24:18.9 &  0.082 &    0.323 &  49.3 &  36.1 &  44.8 &  42.2 & C & & \\ 
 45 & 11:00:43.98 & -77:30:38.5 &  0.072 &    0.862 & 105.2 &  51.2 & -80.5 &  73.4 & C & & \\ 
 46 & 11:00:58.29 & -77:29:41.2 &  0.074 &    0.528 &  71.6 &  44.6 & -33.4 &  56.5 & C & & \\ 
 47 & 11:07:32.72 & -77:36:23.0 &  0.075 &    0.752 &  91.5 &  49.5 & -61.2 &  67.3 & C & & \\ 
 48 & 11:10:01.60 & -76:34:54.3 &  0.077 &    0.078 &  21.4 &  21.2 &  -6.3 &  21.3 & R & \hspace*{-1.5ex} CHXR 44                        (X) &   1.8 \\ 
 & & & & & & & & & & \hspace*{-1.5ex} V* WW Cha                      (Or*) &   6.3 \\ 
 49 & 11:02:33.40 & -77:42:20.1 &  0.075 &    0.348 &  58.2 &  36.0 &   0.0 &  45.8 & C & & \\ 
 50 & 10:50:56.54 & -77:04:29.1 &  0.072 &    0.952 & 139.3 &  42.7 & -75.1 &  77.1 & C & & \\ 
 51 & 11:08:12.67 & -77:39:12.1 &  0.080 &    0.130 &  34.6 &  21.2 &   0.3 &  27.1 & R & & \\ 
 52 & 11:09:15.17 & -77:18:10.7 &  0.071 &    0.235 &  55.8 &  26.6 &  38.4 &  38.5 & C & & \\ 
 53 & 11:06:23.17 & -77:41:42.7 &  0.074 &    0.271 &  49.1 &  33.7 &  79.1 &  40.7 & C & & \\ 
 54 & 11:09:56.09 & -77:14:16.7 &  0.070 &    0.365 &  77.6 &  30.1 &  86.9 &  48.3 & C & & \\ 
 55 & 11:02:10.12 & -77:39:29.2 &  0.071 &    0.587 &  83.4 &  44.6 &  36.2 &  61.0 & C & & \\ 
 56 & 11:10:03.64 & -76:38:02.6 &  0.068 &    0.445 &  82.3 &  35.8 & -46.8 &  54.2 & C & & \\ 
 57 & 11:06:03.63 & -77:34:52.7 &  0.070 &    0.426 &  58.7 &  46.6 & -81.7 &  52.3 & C & & \\ 
 58 & 11:02:58.44 & -77:38:42.5 &  0.070 &    0.241 &  48.3 &  31.9 &  -0.7 &  39.3 & C & & \\ 
 59 & 11:07:15.99 & -77:24:35.6 &  0.068 &    0.192 &  48.7 &  26.0 &  44.0 &  35.6 & C & \hspace*{-1.5ex} [PMG2001] NIR 129              (IR) &   8.7 \\ 
 & & & & & & & & & & \hspace*{-1.5ex} [PMG2001] NIR 130              (IR) &   8.8 \\ 
 & & & & & & & & & & \hspace*{-1.5ex} [PMG2001] NIR 126              (IR) &  12.8 \\ 
 & & & & & & & & & & \hspace*{-1.5ex} [PMG2001] NIR 136              (IR) &  14.0 \\ 
 60 & 11:09:51.10 & -76:51:19.2 &  0.066 &    0.598 &  90.0 &  45.6 &  34.2 &  64.0 & C & \hspace*{-1.5ex} ISO-ChaI 206                   (IR) &  11.7 \\ 
 61 & 11:08:59.13 & -77:45:08.6 &  0.067 &    0.292 &  46.5 &  42.4 & -76.4 &  44.4 & C & & \\ 
 62 & 11:09:53.92 & -76:34:23.2 &  0.070 &    0.071 &  21.5 &  21.2 &  10.2 &  21.3 & Sc & \hspace*{-1.5ex} HH 914                         (HH) &   1.8 \\ 
 & & & & & & & & & & \hspace*{-1.5ex} CED 112 IRS 4             (TT*) &   2.9 \\ 
 63 & 11:10:01.87 & -76:33:11.5 &  0.068 &    0.199 &  54.5 &  24.3 & -33.5 &  36.4 & Sc & \hspace*{-1.5ex} 2M J11100336-7633111        (Y*O) &   5.3 \\ 
 & & & & & & & & & & \hspace*{-1.5ex} [FL2004] 55                    (pr*) &  18.8 \\ 
 64 & 11:11:22.58 & -77:16:24.4 &  0.066 &    0.229 &  73.2 &  21.4 & -87.1 &  39.6 & C & & \\ 
 65 & 11:09:00.36 & -77:40:12.6 &  0.065 &    0.112 &  36.6 &  21.2 & -66.1 &  27.9 & C & & \\ 
 66 & 11:05:38.45 & -77:49:12.5 &  0.065 &    0.069 &  22.4 &  21.4 &  69.9 &  21.9 & A & & \\ 
 67 & 11:07:24.50 & -77:39:40.6 &  0.065 &    0.459 &  87.6 &  36.2 &  42.5 &  56.3 & C & & \\ 
 68 & 11:02:33.46 & -78:01:38.2 &  0.063 &    0.170 &  50.7 &  24.0 &  87.8 &  34.9 & C & & \\ 
 69 & 11:06:24.83 & -77:22:58.5 &  0.070 &    0.112 &  33.8 &  21.2 &  53.0 &  26.8 & R & \hspace*{-1.5ex} [PMG2001] NIR 35               (IR) &   5.3 \\ 
 70 & 11:04:47.42 & -77:44:58.1 &  0.064 &    0.154 &  40.2 &  26.9 & -63.6 &  32.9 & C & & \\ 
 71 & 11:09:07.16 & -77:23:55.4 &  0.063 &    0.457 &  60.6 &  54.3 &  52.1 &  57.4 & C & & \\ 
 72 & 11:07:35.16 & -77:21:45.8 &  0.064 &    0.158 &  34.9 &  31.9 & -86.8 &  33.4 & C & & \\ 
 73 & 11:08:03.98 & -77:20:53.1 &  0.061 &    0.188 &  50.5 &  27.8 & -55.9 &  37.4 & C & & \\ 
 74 & 11:10:23.04 & -76:38:40.4 &  0.063 &    0.295 &  52.8 &  40.1 &  13.1 &  46.0 & C & & \\ 
 75 & 11:06:30.97 & -77:24:25.2 &  0.068 &    0.090 &  27.7 &  21.4 &  66.2 &  24.4 & R & & \\ 
 76 & 11:08:45.48 & -77:44:22.7 &  0.063 &    0.080 &  27.1 &  21.2 &  68.3 &  24.0 & R & & \\ 
 77 & 11:11:09.52 & -76:42:00.2 &  0.062 &    0.072 &  24.8 &  21.2 &  29.5 &  22.9 & S & \hspace*{-1.5ex} 2M J11111083-7641574        (Y*O) &   5.3 \\ 
 78 & 11:07:28.55 & -77:19:40.6 &  0.060 &    0.145 &  36.6 &  29.9 & -70.1 &  33.1 & C & & \\ 
 79 & 11:06:18.59 & -77:44:40.1 &  0.061 &    0.159 &  40.7 &  28.8 &  87.1 &  34.2 & C & & \\ 
 80 & 11:05:28.19 & -77:41:18.1 &  0.060 &    0.137 &  46.3 &  22.4 & -23.4 &  32.2 & C & & \\ 
 81 & 10:56:49.93 & -77:03:02.8 &  0.061 &    0.280 &  51.1 &  40.1 & -86.3 &  45.3 & C & & \\ 
 82 & 11:02:12.48 & -77:41:19.1 &  0.063 &    0.214 &  48.5 &  31.6 &  82.2 &  39.2 & C & & \\ 
 83 & 11:03:00.75 & -77:49:18.2 &  0.061 &    0.118 &  41.0 &  21.4 & -82.5 &  29.6 & C & & \\ 
 84 & 11:09:53.67 & -76:42:12.0 &  0.058 &    0.350 &  61.6 &  43.7 &  23.5 &  51.9 & C & & \\ 
 \hline
 \end{tabular}
 \tablefoot{
 \tablefoottext{a}{Peak flux density (in Jy/21.2$\arcsec$-beam), total flux, $FWHM$ along the major and minor axes, and position angle (east from north) of the fitted Gaussian.}
 \tablefoottext{b}{Mean source size, equal to the geometrical mean of the major and minor $FWHM$.}
 \tablefoottext{c}{Type of source based on the associations found in the SIMBAD database (see Table~\ref{t:classes}).}
 \tablefoottext{d}{Source found in the SIMBAD database within the $FWHM$ ellipse, its type in parentheses, and its distance to the peak position. 2MASS was abbreviated to 2M. The abbreviated references in square brackets and the source types in parentheses are defined in SIMBAD.}
 }
\end{table*}

\textit{Gaussclumps} was applied to all \textit{sum} maps. The \textit{sum} 
maps at scales 1 to 7 were decomposed into 10, 18, 28, 42, 84, 103, and 114 
Gaussian sources, respectively, while the full map was decomposed into 116 
Gaussian sources. These counts do not include the sources found too close to 
the noisier map edges (coverage $<$ 1250), which we consider as artefacts. 

\subsection{Source classification}
\label{ss:classification}

We now consider the results obtained with \textit{Gaussclumps} for the 
\textit{sum} map at scale 5 (i.e. the map shown in Fig.~\ref{f:mrmed}a), which 
is a good scale to characterize sources with $FWHM < 120\arcsec$ as shown in 
Appendix~\ref{s:mrmed} and Table~\ref{t:mrmed}. The positions, sizes,
orientations, and indexes of the 84 extracted Gaussian sources are listed in 
Table~\ref{t:id_gcl_simbad} in the order in which \textit{Gaussclumps} found 
them. We looked for associations with sources in the SIMBAD astronomical 
database that provides basic data, cross-identifications, bibliography and 
measurements for astronomical objects outside the solar system. 
We used SIMBAD4 (release 1.148) as of April 19th, 2010, and removed the 
objects that did not correspond to young stellar objects or stars. The 
SIMBAD sources that are located within the $FWHM$ ellipse of each 
\textit{Gaussclumps} source after this filtering are listed in
Table~\ref{t:id_gcl_simbad}, along with their type and their distance to the 
fitted peak position. Based on these possible associations, we 
classify the sources found with \textit{Gaussclumps} into five categories 
(Col.~10 of Table~\ref{t:id_gcl_simbad}): 
\begin{itemize}
\item S: very compact or unresolved source associated with a young stellar 
object (Class I or more evolved) the position of which agrees within a few 
arcsec (13 sources),
\item Sc: tentative association like S, but with some uncertainty due to the 
somewhat extended 870~$\mu$m emission (3 sources),
\item R: source that looks like a residual (departure from Gaussianity) of a 
stronger nearby source and may therefore not be real (7 sources),
\item A: likely artefact, i.e. unresolved source with no SIMBAD association (1 
source),
\item C: source that looks like a \textit{bona-fide} starless core or Class 0 
protostar (60 sources).
\end{itemize} 
This classification is summarized in Table~\ref{t:classes}.

\begin{table}
 \caption{
 Classification of the sources extracted with \textit{Gaussclumps} from the 870~$\mu$m continuum \textit{sum} map of Cha~I at scale 5.
 }
 \label{t:classes}
 \centering
 \begin{tabular}{ccl}
 \hline\hline
 \multicolumn{1}{c}{Category} & \multicolumn{1}{c}{\hspace*{-1ex} Number} & \multicolumn{1}{c}{\hspace*{-1ex} Definition} \\ 
 \multicolumn{1}{c}{(1)} & \multicolumn{1}{c}{\hspace*{-2ex} (2)} & \multicolumn{1}{c}{\hspace*{-3ex} (3)} \\ 
 \hline
C & \hspace*{-2ex} 60 & \hspace*{-3ex} starless core or Class 0 protostar \\ 
S & \hspace*{-2ex} 13 & \hspace*{-3ex} young stellar object (Class I or more evolved) \\ 
Sc & \hspace*{-2ex} 3 & \hspace*{-3ex} candidate association with a YSO \\ 
R & \hspace*{-2ex} 7 & \hspace*{-3ex} residual of a stronger nearby source \\ 
A & \hspace*{-2ex} 1 & \hspace*{-3ex} artefact \\ 
 \hline
 \end{tabular} \\[0.5ex] 
\end{table}

Twelve\footnote{We do not count Cha-MMS1a since it is only a 1.3~mm source in 
SIMBAD.} LABOCA sources classified as \textit{bona-fide} starless cores (C) in 
Table~\ref{t:id_gcl_simbad} have a possible SIMBAD association within their 
$FWHM$ ellipse ($N_{\mathrm{gcl}} = 6$, 8, 11, 13 14, 15, 24, 26, 29, 39, 59, 
and 60). They are still classified as starless 
cores because either they are extended and do not show any evidence for a 
compact structure at the position of the possibly associated SIMBAD source, or 
their peak position is significantly offset from it. We consider that the 
SIMBAD source is either not physically associated with the dense core
(chance association, especially in the large dense core in Cha-Center where an 
IR cluster overlaps with its outskirts) or it is embedded in the
dense core but the latter may still form an additional star. However, we 
cannot exclude that a fraction of these twelve dense cores are remnant of a 
previous episode of star formation and will never form new stars.

Cha~I was observed with the \textit{Spitzer} instruments IRAC and MIPS in 
the framework of the Gould's Belt legacy program \citep[][]{Allen06}. The maps
are currently being analysed (J{\o}rgensen et al., \textit{in prep.}) and
were used to search for possible associations with the LABOCA sources.
Among the 60 LABOCA sources classified as \textit{bona-fide} starless cores 
(C), 7 have a compact counterpart in emission in the 24 or 70~$\mu$m 
\textit{Spitzer} MIPS maps within a radius of 15$\arcsec$. 
These \textit{Spitzer} sources  are listed in Table~\ref{t:starless_spitzer}. 
The first one ($N_{\mathrm{sp}} = 1$) is associated with source 
$N_{\mathrm{gcl}} = 4$ (Cha-MMS1) and was already reported by 
\citet{Belloche06}, who argued that it is a very young Class 0 protostar, or 
maybe even at the stage of the first hydrostatic core. At this stage, the mass 
in the protostellar envelope completely dominates the mass of the stellar 
embryo. Therefore it makes sense to compare its properties with those of 
starless cores and we keep it in category C. Note that the peak of the 
870~$\mu$m emission is actually a bit shifted compared to the position fitted 
by \textit{Gaussclumps} and reported in Table~\ref{t:id_gcl_simbad}. The 
deviation comes from the asymmetry of the core. By eye, we measure a peak 
position  ($\alpha, \delta$)$_{\mathrm{J2000}}$ =
($11^{\mathrm{h}}06^{\mathrm{m}}32\fs94$, 
$-77^\circ23\arcmin35\farcs2$), i.e. much closer to the \textit{Spitzer} 
position with a relative offset (-0.6$\arcsec$,-1.1$\arcsec$), which 
strengthens the association of the \textit{Spitzer} source with 
the 870~$\mu$m source $N_{\mathrm{gcl}} = 4$ even further.

The second and third \textit{Spitzer} sources 
($N_{\mathrm{sp}} = 2$ and 3) are coincident within 0.3$\arcsec$ with SIMBAD 
YSOs already listed in Table~\ref{t:id_gcl_simbad}. Their mid-IR slope 
indicates that they are Class~II objects (J{\o}rgensen et al., 
\textit{in prep.}). The fourth and fifth \textit{Spitzer} sources 
($N_{\mathrm{sp}} = 4$ and 5) are also coincident with SIMBAD objects, but only 
within 3$\arcsec$, which makes the association less secure. Their mid-IR slope 
indicates that they are Class~II and flat-spectrum objects, respectively. The 
four 870~$\mu$m sources $N_{\mathrm{gcl}} = 13$, 14, 24, and 26 possibly 
associated with the \textit{Spitzer} sources $N_{\mathrm{sp}} = 2$ to 5 
are extended. Since they do not show evidence for a compact structure toward 
the \textit{Spitzer} source, we still classify them as \textit{bona-fide} 
starless cores (C). The last two \textit{Spitzer} sources 
($N_{\mathrm{sp}} = 6$ and 7) 
are Class~I and II objects, respectively, according to their mid-IR slope, but 
they do not have any SIMBAD counterpart. They are located $\sim 6\arcsec$ and 
$\sim 7\arcsec$ from the peak of the 870~$\mu$m sources $N_{\mathrm{gcl}} = 37$ 
and 39, respectively. Since the latter are also extended and do not show 
evidence for a compact structure, we also still classify them as 
\textit{bona-fide} starless cores (C). 

\begin{table*}
 \caption{
 Properties of the \textit{Spitzer} sources possibly associated with a 870~$\mu$m source classified C in Table~\ref{t:id_gcl_simbad}.
 }
 \label{t:starless_spitzer}
 \centering
 \begin{tabular}{cccccclc}
 \hline\hline
 \multicolumn{1}{c}{$N_{\mathrm{sp}}$} & \multicolumn{1}{c}{R.A.} & \multicolumn{1}{c}{Decl.} & \multicolumn{1}{c}{$\alpha$\tablefootmark{a}} & \multicolumn{1}{c}{$N_{\mathrm{gcl}}$\tablefootmark{b}} & \multicolumn{1}{c}{Dist.\tablefootmark{c}} & \multicolumn{1}{c}{SIMBAD possible} & \multicolumn{1}{c}{Dist.\tablefootmark{d}} \\ 
  & \multicolumn{1}{c}{\scriptsize (J2000)} & \multicolumn{1}{c}{\scriptsize (J2000)} & & & \multicolumn{1}{c}{\scriptsize ($\arcsec$)} & \multicolumn{1}{c}{associations\tablefootmark{d}} & \multicolumn{1}{c}{\scriptsize ($\arcsec$)} \\ 
 \multicolumn{1}{c}{(1)} & \multicolumn{1}{c}{(2)} & \multicolumn{1}{c}{(3)} & \multicolumn{1}{c}{(4)} & \multicolumn{1}{c}{(5)} & \multicolumn{1}{c}{(6)} & \multicolumn{1}{c}{(7)} & \multicolumn{1}{c}{(8)} \\ 
 \hline
 1 & 11:06:33.13 & -77:23:35.1  & -- & 4 &   5.5 & & \\ 
 2 & 11:07:57.96 & -77:38:45.2  & $-0.3$ & 13 &   7.0 & 2M J11075792-7738449        (TT*) &   0.3 \\ 
 3 & 11:09:46.19 & -76:34:46.6  & $-0.4$ & 14 &   3.3 & 2M J11094621-7634463        (TT*) &   0.2 \\ 
 & & & & & & Hn 10W                         (Em*) &   1.6 \\ 
 & & & & & & [ACA2003] Cha I4               (Em*) &   4.8 \\ 
 4 & 11:09:26.09 & -76:33:33.9  & $-1.0$ & 24 &   5.7 & DENIS-P J1109.4-7633           (Y*O) &   3.0 \\ 
 5 & 11:07:16.19 & -77:23:06.9  & $-0.1$ & 26 &  13.5 & DENIS-P J1107.3-7723           (Y*O) &   3.0 \\ 
 6 & 11:04:01.15 & -77:58:44.7  & $1.0$ & 37 &   6.1 & & \\ 
 7 & 11:07:16.32 & -77:38:48.3  & $-0.3$ & 39 &   6.9 & & \\ 
 \hline
 \end{tabular}
 \tablefoot{
 \tablefoottext{a}{Mid-IR spectral slope (J{\o}rgensen et al., \textit{in prep.}). A dash means that it is not detected shortward of 24~$\mu$m.}
 \tablefoottext{b}{Possibly associated \textit{Gaussclumps} 870~$\mu$m source (same number as in Table~\ref{t:id_gcl_simbad}).}
 \tablefoottext{c}{Distance to the peak of the \textit{Gaussclumps} 870~$\mu$m source.}
 \tablefoottext{d}{Source found in the SIMBAD database, its type in parentheses, and its distance to the \textit{Spitzer} source. Only sources with a distance smaller than $5\arcsec$ are listed. 2MASS was abbreviated to 2M.}
 }
\end{table*}

In summary, out of 84 sources found with \textit{Gaussclumps} in the 
870~$\mu$m map of Cha~I, 76 look like real sources. Sixteen of these sources 
($21 \%$) are associated with a young stellar object (Class I or more evolved) 
and most likely trace a circumstellar disk and/or a residual circumstellar 
envelope. The remaining 60 sources ($79 \%$) appear to be starless based on
\textit{Spitzer} or contain an embedded Class 0 protostar or first hydrostatic 
core (one case). The position, sizes, and orientation of these 76 sources, 
plus the additional sources mentioned in Sect.~\ref{ss:additional} below, 
are shown in Figs.~\ref{f:labocamapdet}a to g. The labels correspond to the 
first column of Tables~\ref{t:starless}, \ref{t:ysos}, and \ref{t:motherysos}. 
``C'' stands for starless core (or Class 0 protostellar core) and ``S'' for 
YSO (Class~I or more evolved).

\subsection{Additional sources}
\label{ss:additional}

Since the detection threshold for \textit{Gaussclumps} was set to $5\sigma$, 
compact sources with a SIMBAD counterpart and a 870~$\mu$m detection 
between $3.5\sigma$ and 
$\sim 5\sigma$ that may be trusted based on the association were missed by 
\textit{Gaussclumps}. Therefore, we also looked for 870~$\mu$m emission above 
$3.5\sigma$ in the \textit{sum} map at scale 5 at the position of each source 
in the SIMBAD database. Clear associations between a SIMBAD source and
a compact 870~$\mu$m emission were visually selected and all SIMBAD 
sources associated with extended emission but no 
clear peak at 870~$\mu$m were discarded, in particular those in the prominent 
dense cores around Cha-MMS1 and Cha-MMS2. As a result, there are five 
additional compact 870~$\mu$m sources with a likely SIMBAD association and a 
formal peak flux density above $3.5\sigma$. They are listed in 
Table~\ref{t:otherysos}, along with the parameters derived from a Gaussian fit 
performed with the task GAUSS\_2D in GREG\footnote{GREG is part of the 
collection of softwares GILDAS (see http://www.iram.fr/IRAMFR/GILDAS).}. The 
fitted peak flux density is 
below $3.5\sigma$  in one case. Given the number of SIMBAD sources in the 
Cha~I field (1045 objects over 1.6 deg$^2$), the probability of chance 
association within a radius of $5\arcsec$ is 0.4$\%$ only, so we are confident 
that these five additional SIMBAD associations are real.

\begin{table*}
 \caption{
 Additional compact sources with SIMBAD association with a formal peak flux density above 3.5$\sigma$ in the 870~$\mu$m continuum map of Cha~I filtered up to scale 5.
 }
 \label{t:otherysos}
 \centering
 \begin{tabular}{ccccccccclc}
 \hline\hline
  \multicolumn{1}{c}{$N_{\mathrm{add}}$} & \multicolumn{1}{c}{\hspace*{-2ex} R.A.} & \multicolumn{1}{c}{\hspace*{-1ex} Decl.} & \multicolumn{1}{c}{\hspace*{-2ex} ${f_{\mathrm{peak}}}$\tablefootmark{a}} & \multicolumn{1}{c}{\hspace*{-2ex} ${f_{\mathrm{tot}}}$\tablefootmark{a}} & \multicolumn{1}{c}{maj.\tablefootmark{a}} & \multicolumn{1}{c}{\hspace*{-1ex} min.\tablefootmark{a}} & \multicolumn{1}{c}{\hspace*{-1ex} P.A.\tablefootmark{a}} & \multicolumn{1}{c}{$S$\tablefootmark{b}} & \multicolumn{1}{c}{\hspace*{-1ex} SIMBAD possible} & \multicolumn{1}{c}{\hspace*{-2ex} Dist.\tablefootmark{c}} \\ 
  & \multicolumn{1}{c}{\hspace*{-2ex} \scriptsize (J2000)} & \multicolumn{1}{c}{\hspace*{-1ex} \scriptsize (J2000)} & \multicolumn{1}{c}{\hspace*{-2ex} \scriptsize (Jy/beam)} & \multicolumn{1}{c}{\hspace*{-2ex} \scriptsize (Jy)} & \multicolumn{1}{c}{\scriptsize ($\arcsec$)} & \multicolumn{1}{c}{\hspace*{-1ex} \scriptsize ($\arcsec$)} & \multicolumn{1}{c}{\hspace*{-1ex} \scriptsize ($^\circ$)} & \multicolumn{1}{c}{\scriptsize ($\arcsec$)} & \multicolumn{1}{c}{\hspace*{-1ex} associations\tablefootmark{c}} & \multicolumn{1}{c}{\hspace*{-2ex} \scriptsize ($\arcsec$)} \\ 
 \multicolumn{1}{c}{(1)} & \multicolumn{1}{c}{\hspace*{-2ex} (2)} & \multicolumn{1}{c}{\hspace*{-1ex} (3)} & \multicolumn{1}{c}{\hspace*{-2ex} (4)} & \multicolumn{1}{c}{\hspace*{-2ex} (5)} & \multicolumn{1}{c}{(6)} & \multicolumn{1}{c}{\hspace*{-1ex} (7)} & \multicolumn{1}{c}{\hspace*{-1ex} (8)} & \multicolumn{1}{c}{(9)} & \multicolumn{1}{c}{\hspace*{-1ex} (10)} & \multicolumn{1}{c}{\hspace*{-2ex} (11)} \\ 
 \hline
   1 &  \hspace*{-2ex} 10:59:01.28 &  \hspace*{-1ex} -77:22:40.4 &  \hspace*{-2ex}  0.055 &  \hspace*{-2ex}    0.062 &  24.4 & \hspace*{-1ex}  20.7 & \hspace*{-1ex}  17.9 &  22.5  & \hspace*{-1ex} V* TW Cha                      (Or*) & \hspace*{-2ex}   0.7 \\ 
   2 &  \hspace*{-2ex} 10:55:59.55 &  \hspace*{-1ex} -77:24:35.9 &  \hspace*{-2ex}  0.032 &  \hspace*{-2ex}    0.075 &  56.9 & \hspace*{-1ex}  18.6 & \hspace*{-1ex}  69.5 &  32.5  & \hspace*{-1ex} Ass Cha T 2-3 B                (Y*O) & \hspace*{-2ex}   4.1 \\ 
 & & & & & & & & & \hspace*{-1ex} 2M J10555973-7724399        (Y*O) & \hspace*{-2ex}   4.1 \\ 
 & & & & & & & & & \hspace*{-1ex} V* SX Cha                      (Or*) & \hspace*{-2ex}   4.2 \\ 
   3 &  \hspace*{-2ex} 11:10:49.76 &  \hspace*{-1ex} -77:17:51.0 &  \hspace*{-2ex}  0.047 &  \hspace*{-2ex}    0.045 &  23.2 & \hspace*{-1ex}  18.4 & \hspace*{-1ex} -22.5 &  20.7  & \hspace*{-1ex} HBC 584                        (Em*) & \hspace*{-2ex}   0.9 \\ 
   4 &  \hspace*{-2ex} 11:06:25.43 &  \hspace*{-1ex} -76:33:44.8 &  \hspace*{-2ex}  0.059 &  \hspace*{-2ex}    0.044 &  20.4 & \hspace*{-1ex}  16.6 & \hspace*{-1ex} -88.3 &  18.4  & \hspace*{-1ex} 2M J11062554-7633418        (Y*O) & \hspace*{-2ex}   3.0 \\ 
   5 &  \hspace*{-2ex} 11:12:28.26 &  \hspace*{-1ex} -76:44:24.2 &  \hspace*{-2ex}  0.061 &  \hspace*{-2ex}    0.041 &  18.6 & \hspace*{-1ex}  16.3 & \hspace*{-1ex}  87.2 &  17.4  & \hspace*{-1ex} V* CV Cha                      (Or*) & \hspace*{-2ex}   2.7 \\ 
 & & & & & & & & & \hspace*{-1ex} BRAN 341D                      (RNe) & \hspace*{-2ex}   4.0 \\ 
 & & & & & & & & & \hspace*{-1ex} 2E 1110.9-7628                 (X) & \hspace*{-2ex}   5.5 \\ 
 \hline
 \end{tabular}
 \tablefoot{
 \tablefoottext{a}{Peak flux density (in Jy/21.2$\arcsec$-beam), total flux, $FWHM$ along the major and minor axes, and position angle (east from north) of the fitted Gaussian. The fitted peak flux density may slightly depart from the formal peak flux density measured directly in the map.}
 \tablefoottext{b}{Mean source size, equal to the geometrical mean of the major and minor $FWHM$.}
 \tablefoottext{c}{Source found in the SIMBAD database within the $FWHM$ ellipse, its type in parentheses, and its distance to the peak position. 2MASS was abbreviated to 2M.}
 }
\end{table*}

\section{Analysis}
\label{s:analysis}

\subsection{Starless cores}
\label{ss:starless}

\begin{table*}
 \caption{
 Characteristics of starless (or Class 0) sources extracted with \textit{Gaussclumps} in the 870~$\mu$m continuum map of Cha~I filtered up to scale 5.
 }
 \label{t:starless}
 \centering
 \begin{tabular}{cccccccccccccc}
 \hline\hline
  \multicolumn{1}{c}{Name} & \multicolumn{1}{c}{\hspace*{-2ex} ${N_{\mathrm{gcl}}}$\tablefootmark{a}} & \multicolumn{1}{c}{\hspace*{-2ex} $FWHM$\tablefootmark{b}} & \multicolumn{1}{c}{\hspace*{-2ex} $FWHM$\tablefootmark{b}} & \multicolumn{1}{c}{${R_{\mathrm{a}}}$\tablefootmark{b}} & \multicolumn{1}{c}{\hspace*{-3ex} ${N_{\mathrm{peak}}}$\tablefootmark{c}} & \multicolumn{1}{c}{\hspace*{-2.5ex} ${A_V}$\tablefootmark{d}} & \multicolumn{1}{c}{\hspace*{-1ex} ${M_{\mathrm{peak}}}$\tablefootmark{e}} & \multicolumn{1}{c}{${M_{\mathrm{tot}}}$\tablefootmark{e}} & \multicolumn{1}{c}{${M_{50\arcsec}}$\tablefootmark{e}} & \multicolumn{1}{c}{${C_M}$\tablefootmark{f}} & \multicolumn{1}{c}{\hspace*{-3ex} ${n_{\mathrm{peak}}}$\tablefootmark{g}} & \multicolumn{1}{c}{\hspace*{-3ex} ${n_{50\arcsec}}$\tablefootmark{g}} & \multicolumn{1}{c}{\hspace*{-3ex} ${c_n}$\tablefootmark{h}} \\ 
  & \hspace*{-2ex} & \multicolumn{1}{c}{\hspace*{-2ex} \scriptsize ($\arcsec \times \arcsec$)} & \multicolumn{1}{c}{\hspace*{-2ex} \scriptsize (1000 AU)$^2$} & & \multicolumn{1}{c}{\hspace*{-3ex} \scriptsize ($10^{21}$ cm$^{-2}$)} & \multicolumn{1}{c}{\hspace*{-2.5ex} \scriptsize (mag)} & \multicolumn{1}{c}{\hspace*{-1ex} \scriptsize (M$_\odot$)} & \multicolumn{1}{c}{\scriptsize (M$_\odot$)} & \multicolumn{1}{c}{\scriptsize (M$_\odot$)} &  \multicolumn{1}{c}{\scriptsize ($\%$)} & \multicolumn{1}{c}{\hspace*{-3ex} \scriptsize ($10^{5}$ cm$^{-3}$)} & \multicolumn{1}{c}{\hspace*{-3ex} \scriptsize ($10^{4}$ cm$^{-3}$)} & \hspace*{-3ex} \\ 
 \multicolumn{1}{c}{(1)} & \multicolumn{1}{c}{\hspace*{-2ex} (2)} & \multicolumn{1}{c}{\hspace*{-2ex} (3)} & \multicolumn{1}{c}{\hspace*{-2ex} (4)} & \multicolumn{1}{c}{(5)} & \multicolumn{1}{c}{\hspace*{-3ex} (6)} & \multicolumn{1}{c}{\hspace*{-2.5ex} (7)} & \multicolumn{1}{c}{\hspace*{-1ex} (8)} & \multicolumn{1}{c}{(9)} & \multicolumn{1}{c}{(10)} & \multicolumn{1}{c}{(11)} & \multicolumn{1}{c}{\hspace*{-3ex} (12)} & \multicolumn{1}{c}{\hspace*{-3ex} (13)} & \multicolumn{1}{c}{\hspace*{-3ex} (14)} \\ 
 \hline
 Cha1-C1 & \hspace*{-2ex} 4 & \hspace*{-2ex} 55.1 $\times$ 49.3 & \hspace*{-2ex}  7.6 $\times$  6.7 &  1.1 & \hspace*{-3ex}  92 & \hspace*{-2.5ex}   14 & \hspace*{-1ex}  0.56 &  3.36 &  1.44 &  39( 1) & \hspace*{-3ex} 49.5 & \hspace*{-3ex} 97.7 & \hspace*{-3ex} 5.1(0.1) \\  Cha1-C2 & \hspace*{-2ex} 6 & \hspace*{-2ex} 98.2 $\times$ 45.9 & \hspace*{-2ex} 14.4 $\times$  6.1 &  2.4 & \hspace*{-3ex}  27 & \hspace*{-2.5ex}    9 & \hspace*{-1ex}  0.16 &  1.62 &  0.58 &  28( 1) & \hspace*{-3ex} 14.4 & \hspace*{-3ex} 39.4 & \hspace*{-3ex} 3.7(0.2) \\  Cha1-C3 & \hspace*{-2ex} 8 & \hspace*{-2ex} 76.1 $\times$ 45.9 & \hspace*{-2ex} 11.0 $\times$  6.1 &  1.8 & \hspace*{-3ex}  25 & \hspace*{-2.5ex}   13 & \hspace*{-1ex}  0.15 &  1.16 &  0.45 &  33( 2) & \hspace*{-3ex} 13.3 & \hspace*{-3ex} 30.7 & \hspace*{-3ex} 4.3(0.2) \\  Cha1-C4 & \hspace*{-2ex} 9 & \hspace*{-2ex} 53.4 $\times$ 50.4 & \hspace*{-2ex}  7.3 $\times$  6.9 &  1.1 & \hspace*{-3ex}  18 & \hspace*{-2.5ex}   17 & \hspace*{-1ex}  0.11 &  0.66 &  0.29 &  38( 3) & \hspace*{-3ex}  9.8 & \hspace*{-3ex} 19.9 & \hspace*{-3ex} 4.9(0.3) \\  Cha1-C5 & \hspace*{-2ex} 11 & \hspace*{-2ex} 60.3 $\times$ 34.5 & \hspace*{-2ex}  8.5 $\times$  4.1 &  2.1 & \hspace*{-3ex}  15 & \hspace*{-2.5ex}    9 & \hspace*{-1ex}  0.088 &  0.41 &  0.32 &  28( 2) & \hspace*{-3ex}  7.9 & \hspace*{-3ex} 21.4 & \hspace*{-3ex} 3.7(0.3) \\  Cha1-C6 & \hspace*{-2ex} 13 & \hspace*{-2ex} 44.3 $\times$ 26.0 & \hspace*{-2ex}  5.8 $\times$  2.2 &  2.6 & \hspace*{-3ex}  14 & \hspace*{-2.5ex}    8 & \hspace*{-1ex}  0.085 &  0.22 &  0.33 &  26( 2) & \hspace*{-3ex}  7.6 & \hspace*{-3ex} 22.3 & \hspace*{-3ex} 3.4(0.3) \\  Cha1-C7 & \hspace*{-2ex} 14 & \hspace*{-2ex} 98.3 $\times$ 48.1 & \hspace*{-2ex} 14.4 $\times$  6.5 &  2.2 & \hspace*{-3ex}  13 & \hspace*{-2.5ex}   11 & \hspace*{-1ex}  0.078 &  0.82 &  0.23 &  34( 3) & \hspace*{-3ex}  6.9 & \hspace*{-3ex} 15.4 & \hspace*{-3ex} 4.5(0.4) \\  Cha1-C8 & \hspace*{-2ex} 15 & \hspace*{-2ex} 31.0 $\times$ 26.1 & \hspace*{-2ex}  3.4 $\times$  2.3 &  1.5 & \hspace*{-3ex}  13 & \hspace*{-2.5ex}   14 & \hspace*{-1ex}  0.081 &  0.15 &  0.22 &  36( 3) & \hspace*{-3ex}  7.2 & \hspace*{-3ex} 15.2 & \hspace*{-3ex} 4.8(0.4) \\  Cha1-C9 & \hspace*{-2ex} 16 & \hspace*{-2ex} 82.6 $\times$ 51.8 & \hspace*{-2ex} 12.0 $\times$  7.1 &  1.7 & \hspace*{-3ex}  12 & \hspace*{-2.5ex}    7 & \hspace*{-1ex}  0.071 &  0.68 &  0.20 &  36( 4) & \hspace*{-3ex}  6.3 & \hspace*{-3ex} 13.4 & \hspace*{-3ex} 4.7(0.5) \\  Cha1-C10 & \hspace*{-2ex} 21 & \hspace*{-2ex} 65.9 $\times$ 34.3 & \hspace*{-2ex}  9.4 $\times$  4.0 &  2.3 & \hspace*{-3ex}  10 & \hspace*{-2.5ex}    9 & \hspace*{-1ex}  0.061 &  0.31 &  0.19 &  33( 4) & \hspace*{-3ex}  5.5 & \hspace*{-3ex} 12.8 & \hspace*{-3ex} 4.3(0.5) \\  Cha1-C11 & \hspace*{-2ex} 23 & \hspace*{-2ex} 126.7 $\times$ 45.8 & \hspace*{-2ex} 18.7 $\times$  6.1 &  3.1 & \hspace*{-3ex} 9.7 & \hspace*{-2.5ex}    9 & \hspace*{-1ex}  0.059 &  0.76 &  0.16 &  36( 5) & \hspace*{-3ex}  5.2 & \hspace*{-3ex} 11.1 & \hspace*{-3ex} 4.7(0.6) \\  Cha1-C12 & \hspace*{-2ex} 24 & \hspace*{-2ex} 56.3 $\times$ 32.7 & \hspace*{-2ex}  7.8 $\times$  3.7 &  2.1 & \hspace*{-3ex} 9.7 & \hspace*{-2.5ex}   10 & \hspace*{-1ex}  0.059 &  0.24 &  0.14 &  42( 5) & \hspace*{-3ex}  5.2 & \hspace*{-3ex}  9.6 & \hspace*{-3ex} 5.5(0.7) \\  Cha1-C13 & \hspace*{-2ex} 25 & \hspace*{-2ex} 69.4 $\times$ 63.7 & \hspace*{-2ex}  9.9 $\times$  9.0 &  1.1 & \hspace*{-3ex} 9.1 & \hspace*{-2.5ex}    9 & \hspace*{-1ex}  0.055 &  0.54 &  0.16 &  35( 5) & \hspace*{-3ex}  4.9 & \hspace*{-3ex} 10.5 & \hspace*{-3ex} 4.7(0.6) \\  Cha1-C14 & \hspace*{-2ex} 26 & \hspace*{-2ex} 49.8 $\times$ 33.6 & \hspace*{-2ex}  6.8 $\times$  3.9 &  1.7 & \hspace*{-3ex} 9.5 & \hspace*{-2.5ex}   13 & \hspace*{-1ex}  0.057 &  0.21 &  0.17 &  34( 4) & \hspace*{-3ex}  5.1 & \hspace*{-3ex} 11.6 & \hspace*{-3ex} 4.4(0.6) \\  Cha1-C15 & \hspace*{-2ex} 29 & \hspace*{-2ex} 102.4 $\times$ 48.6 & \hspace*{-2ex} 15.0 $\times$  6.6 &  2.3 & \hspace*{-3ex} 7.9 & \hspace*{-2.5ex}    8 & \hspace*{-1ex}  0.048 &  0.53 &  0.13 &  37( 6) & \hspace*{-3ex}  4.2 & \hspace*{-3ex}  8.8 & \hspace*{-3ex} 4.8(0.8) \\  Cha1-C16 & \hspace*{-2ex} 30 & \hspace*{-2ex} 47.4 $\times$ 41.7 & \hspace*{-2ex}  6.4 $\times$  5.4 &  1.2 & \hspace*{-3ex} 7.8 & \hspace*{-2.5ex}   10 & \hspace*{-1ex}  0.047 &  0.21 &  0.11 &  44( 7) & \hspace*{-3ex}  4.2 & \hspace*{-3ex}  7.3 & \hspace*{-3ex} 5.8(1.0) \\  Cha1-C17 & \hspace*{-2ex} 32 & \hspace*{-2ex} 67.3 $\times$ 42.4 & \hspace*{-2ex}  9.6 $\times$  5.5 &  1.7 & \hspace*{-3ex} 7.7 & \hspace*{-2.5ex}   11 & \hspace*{-1ex}  0.046 &  0.29 &  0.11 &  41( 7) & \hspace*{-3ex}  4.1 & \hspace*{-3ex}  7.7 & \hspace*{-3ex} 5.3(0.9) \\  Cha1-C18 & \hspace*{-2ex} 33 & \hspace*{-2ex} 102.1 $\times$ 47.7 & \hspace*{-2ex} 15.0 $\times$  6.4 &  2.3 & \hspace*{-3ex} 7.5 & \hspace*{-2.5ex}   13 & \hspace*{-1ex}  0.045 &  0.49 &  0.12 &  38( 6) & \hspace*{-3ex}  4.0 & \hspace*{-3ex}  8.0 & \hspace*{-3ex} 5.0(0.8) \\  Cha1-C19 & \hspace*{-2ex} 34 & \hspace*{-2ex} 76.3 $\times$ 37.5 & \hspace*{-2ex} 11.0 $\times$  4.6 &  2.4 & \hspace*{-3ex} 7.3 & \hspace*{-2.5ex}   14 & \hspace*{-1ex}  0.044 &  0.28 &  0.098 &  45( 8) & \hspace*{-3ex}  3.9 & \hspace*{-3ex}  6.7 & \hspace*{-3ex} 5.9(1.1) \\  Cha1-C20 & \hspace*{-2ex} 35 & \hspace*{-2ex} 73.5 $\times$ 39.9 & \hspace*{-2ex} 10.6 $\times$  5.1 &  2.1 & \hspace*{-3ex} 6.8 & \hspace*{-2.5ex}  5.2 & \hspace*{-1ex}  0.041 &  0.27 &  0.095 &  43( 8) & \hspace*{-3ex}  3.7 & \hspace*{-3ex}  6.5 & \hspace*{-3ex} 5.7(1.1) \\  Cha1-C21 & \hspace*{-2ex} 36 & \hspace*{-2ex} 37.4 $\times$ 29.0 & \hspace*{-2ex}  4.6 $\times$  3.0 &  1.6 & \hspace*{-3ex} 7.1 & \hspace*{-2.5ex}   19 & \hspace*{-1ex}  0.043 &  0.10 &  0.11 &  38( 7) & \hspace*{-3ex}  3.8 & \hspace*{-3ex}  7.7 & \hspace*{-3ex} 5.0(0.9) \\  Cha1-C22 & \hspace*{-2ex} 37 & \hspace*{-2ex} 90.7 $\times$ 64.3 & \hspace*{-2ex} 13.2 $\times$  9.1 &  1.5 & \hspace*{-3ex} 6.9 & \hspace*{-2.5ex}    9 & \hspace*{-1ex}  0.041 &  0.54 &  0.12 &  35( 6) & \hspace*{-3ex}  3.7 & \hspace*{-3ex}  8.1 & \hspace*{-3ex} 4.5(0.8) \\  Cha1-C23 & \hspace*{-2ex} 38 & \hspace*{-2ex} 40.5 $\times$ 37.0 & \hspace*{-2ex}  5.2 $\times$  4.5 &  1.1 & \hspace*{-3ex} 6.9 & \hspace*{-2.5ex}   10 & \hspace*{-1ex}  0.042 &  0.14 &  0.087 &  48( 9) & \hspace*{-3ex}  3.7 & \hspace*{-3ex}  5.9 & \hspace*{-3ex} 6.3(1.2) \\  Cha1-C24 & \hspace*{-2ex} 39 & \hspace*{-2ex} 50.3 $\times$ 35.8 & \hspace*{-2ex}  6.8 $\times$  4.3 &  1.6 & \hspace*{-3ex} 6.7 & \hspace*{-2.5ex}   14 & \hspace*{-1ex}  0.041 &  0.16 &  0.081 &  50(10) & \hspace*{-3ex}  3.6 & \hspace*{-3ex}  5.5 & \hspace*{-3ex} 6.6(1.3) \\  Cha1-C25 & \hspace*{-2ex} 40 & \hspace*{-2ex} 51.4 $\times$ 37.2 & \hspace*{-2ex}  7.0 $\times$  4.6 &  1.5 & \hspace*{-3ex} 7.1 & \hspace*{-2.5ex}    8 & \hspace*{-1ex}  0.043 &  0.18 &  0.12 &  35( 6) & \hspace*{-3ex}  3.8 & \hspace*{-3ex}  8.3 & \hspace*{-3ex} 4.6(0.8) \\  Cha1-C26 & \hspace*{-2ex} 41 & \hspace*{-2ex} 111.5 $\times$ 43.7 & \hspace*{-2ex} 16.4 $\times$  5.7 &  2.9 & \hspace*{-3ex} 6.7 & \hspace*{-2.5ex}    7 & \hspace*{-1ex}  0.041 &  0.44 &  0.10 &  40( 8) & \hspace*{-3ex}  3.6 & \hspace*{-3ex}  6.9 & \hspace*{-3ex} 5.3(1.0) \\  Cha1-C27 & \hspace*{-2ex} 42 & \hspace*{-2ex} 61.1 $\times$ 36.7 & \hspace*{-2ex}  8.6 $\times$  4.5 &  1.9 & \hspace*{-3ex} 7.0 & \hspace*{-2.5ex}    8 & \hspace*{-1ex}  0.042 &  0.21 &  0.13 &  32( 5) & \hspace*{-3ex}  3.8 & \hspace*{-3ex}  9.1 & \hspace*{-3ex} 4.1(0.7) \\  Cha1-C28 & \hspace*{-2ex} 43 & \hspace*{-2ex} 59.6 $\times$ 32.1 & \hspace*{-2ex}  8.4 $\times$  3.6 &  2.3 & \hspace*{-3ex} 6.6 & \hspace*{-2.5ex}    8 & \hspace*{-1ex}  0.040 &  0.17 &  0.087 &  46( 9) & \hspace*{-3ex}  3.5 & \hspace*{-3ex}  5.9 & \hspace*{-3ex} 6.0(1.2) \\  Cha1-C29 & \hspace*{-2ex} 44 & \hspace*{-2ex} 49.3 $\times$ 36.1 & \hspace*{-2ex}  6.7 $\times$  4.4 &  1.5 & \hspace*{-3ex} 7.1 & \hspace*{-2.5ex}    9 & \hspace*{-1ex}  0.043 &  0.17 &  0.13 &  33( 6) & \hspace*{-3ex}  3.8 & \hspace*{-3ex}  8.8 & \hspace*{-3ex} 4.3(0.7) \\  Cha1-C30 & \hspace*{-2ex} 45 & \hspace*{-2ex} 105.2 $\times$ 51.2 & \hspace*{-2ex} 15.5 $\times$  7.0 &  2.2 & \hspace*{-3ex} 6.3 & \hspace*{-2.5ex}    7 & \hspace*{-1ex}  0.038 &  0.46 &  0.099 &  39( 8) & \hspace*{-3ex}  3.4 & \hspace*{-3ex}  6.7 & \hspace*{-3ex} 5.1(1.0) \\  Cha1-C31 & \hspace*{-2ex} 46 & \hspace*{-2ex} 71.6 $\times$ 44.6 & \hspace*{-2ex} 10.3 $\times$  5.9 &  1.7 & \hspace*{-3ex} 6.5 & \hspace*{-2.5ex}  5.7 & \hspace*{-1ex}  0.039 &  0.28 &  0.10 &  39( 7) & \hspace*{-3ex}  3.5 & \hspace*{-3ex}  6.9 & \hspace*{-3ex} 5.1(1.0) \\  Cha1-C32 & \hspace*{-2ex} 47 & \hspace*{-2ex} 91.5 $\times$ 49.5 & \hspace*{-2ex} 13.3 $\times$  6.7 &  2.0 & \hspace*{-3ex} 6.5 & \hspace*{-2.5ex}   11 & \hspace*{-1ex}  0.039 &  0.40 &  0.10 &  38( 7) & \hspace*{-3ex}  3.5 & \hspace*{-3ex}  7.1 & \hspace*{-3ex} 4.9(0.9) \\  Cha1-C33 & \hspace*{-2ex} 49 & \hspace*{-2ex} 58.2 $\times$ 36.0 & \hspace*{-2ex}  8.1 $\times$  4.4 &  1.9 & \hspace*{-3ex} 6.5 & \hspace*{-2.5ex}    8 & \hspace*{-1ex}  0.039 &  0.18 &  0.10 &  38( 7) & \hspace*{-3ex}  3.5 & \hspace*{-3ex}  7.0 & \hspace*{-3ex} 5.0(1.0) \\  Cha1-C34 & \hspace*{-2ex} 50 & \hspace*{-2ex} 139.3 $\times$ 42.7 & \hspace*{-2ex} 20.7 $\times$  5.6 &  3.7 & \hspace*{-3ex} 6.3 & \hspace*{-2.5ex}  5.5 & \hspace*{-1ex}  0.038 &  0.50 &  0.10 &  37( 7) & \hspace*{-3ex}  3.4 & \hspace*{-3ex}  6.9 & \hspace*{-3ex} 4.9(1.0) \\  Cha1-C35 & \hspace*{-2ex} 52 & \hspace*{-2ex} 55.8 $\times$ 26.6 & \hspace*{-2ex}  7.7 $\times$  2.4 &  3.2 & \hspace*{-3ex} 6.2 & \hspace*{-2.5ex}   13 & \hspace*{-1ex}  0.038 &  0.12 &  0.076 &  49(11) & \hspace*{-3ex}  3.4 & \hspace*{-3ex}  5.2 & \hspace*{-3ex} 6.5(1.4) \\  \hline
 \end{tabular}
 \end{table*}
\begin{table*}
 \addtocounter{table}{-1}
 \caption{
continued.
 }
 \centering
 \begin{tabular}{cccccccccccccc}
 \hline\hline
  \multicolumn{1}{c}{Name} & \multicolumn{1}{c}{\hspace*{-2ex} ${N_{\mathrm{gcl}}}$\tablefootmark{a}} & \multicolumn{1}{c}{\hspace*{-2ex} $FWHM$\tablefootmark{b}} & \multicolumn{1}{c}{\hspace*{-2ex} $FWHM$\tablefootmark{b}} & \multicolumn{1}{c}{${R_{\mathrm{a}}}$\tablefootmark{b}} & \multicolumn{1}{c}{\hspace*{-3ex} ${N_{\mathrm{peak}}}$\tablefootmark{c}} & \multicolumn{1}{c}{\hspace*{-2.5ex} ${A_V}$\tablefootmark{d}} & \multicolumn{1}{c}{\hspace*{-1ex} ${M_{\mathrm{peak}}}$\tablefootmark{e}} & \multicolumn{1}{c}{${M_{\mathrm{tot}}}$\tablefootmark{e}} & \multicolumn{1}{c}{${M_{50\arcsec}}$\tablefootmark{e}} & \multicolumn{1}{c}{${C_M}$\tablefootmark{f}} & \multicolumn{1}{c}{\hspace*{-3ex} ${n_{\mathrm{peak}}}$\tablefootmark{g}} & \multicolumn{1}{c}{\hspace*{-3ex} ${n_{50\arcsec}}$\tablefootmark{g}} & \multicolumn{1}{c}{\hspace*{-3ex} ${c_n}$\tablefootmark{h}} \\ 
  & \hspace*{-2ex} & \multicolumn{1}{c}{\hspace*{-2ex} \scriptsize ($\arcsec \times \arcsec$)} & \multicolumn{1}{c}{\hspace*{-2ex} \scriptsize (1000 AU)$^2$} & & \multicolumn{1}{c}{\hspace*{-3ex} \scriptsize ($10^{21}$ cm$^{-2}$)} & \multicolumn{1}{c}{\hspace*{-2.5ex} \scriptsize (mag)} & \multicolumn{1}{c}{\hspace*{-1ex} \scriptsize (M$_\odot$)} & \multicolumn{1}{c}{\scriptsize (M$_\odot$)} & \multicolumn{1}{c}{\scriptsize (M$_\odot$)} &  \multicolumn{1}{c}{\scriptsize ($\%$)} & \multicolumn{1}{c}{\hspace*{-3ex} \scriptsize ($10^{5}$ cm$^{-3}$)} & \multicolumn{1}{c}{\hspace*{-3ex} \scriptsize ($10^{4}$ cm$^{-3}$)} & \hspace*{-3ex} \\ 
 \multicolumn{1}{c}{(1)} & \multicolumn{1}{c}{\hspace*{-2ex} (2)} & \multicolumn{1}{c}{\hspace*{-2ex} (3)} & \multicolumn{1}{c}{\hspace*{-2ex} (4)} & \multicolumn{1}{c}{(5)} & \multicolumn{1}{c}{\hspace*{-3ex} (6)} & \multicolumn{1}{c}{\hspace*{-2.5ex} (7)} & \multicolumn{1}{c}{\hspace*{-1ex} (8)} & \multicolumn{1}{c}{(9)} & \multicolumn{1}{c}{(10)} & \multicolumn{1}{c}{(11)} & \multicolumn{1}{c}{\hspace*{-3ex} (12)} & \multicolumn{1}{c}{\hspace*{-3ex} (13)} & \multicolumn{1}{c}{\hspace*{-3ex} (14)} \\ 
 \hline
 Cha1-C36 & \hspace*{-2ex} 53 & \hspace*{-2ex} 49.1 $\times$ 33.7 & \hspace*{-2ex}  6.6 $\times$  3.9 &  1.7 & \hspace*{-3ex} 6.5 & \hspace*{-2.5ex}    9 & \hspace*{-1ex}  0.039 &  0.14 &  0.092 &  42( 8) & \hspace*{-3ex}  3.5 & \hspace*{-3ex}  6.3 & \hspace*{-3ex} 5.5(1.1) \\  Cha1-C37 & \hspace*{-2ex} 54 & \hspace*{-2ex} 77.6 $\times$ 30.1 & \hspace*{-2ex} 11.2 $\times$  3.2 &  3.5 & \hspace*{-3ex} 6.1 & \hspace*{-2.5ex}    7 & \hspace*{-1ex}  0.037 &  0.19 &  0.091 &  41( 8) & \hspace*{-3ex}  3.3 & \hspace*{-3ex}  6.2 & \hspace*{-3ex} 5.4(1.1) \\  Cha1-C38 & \hspace*{-2ex} 55 & \hspace*{-2ex} 83.4 $\times$ 44.6 & \hspace*{-2ex} 12.1 $\times$  5.9 &  2.1 & \hspace*{-3ex} 6.2 & \hspace*{-2.5ex}   11 & \hspace*{-1ex}  0.038 &  0.31 &  0.097 &  39( 8) & \hspace*{-3ex}  3.3 & \hspace*{-3ex}  6.6 & \hspace*{-3ex} 5.1(1.0) \\  Cha1-C39 & \hspace*{-2ex} 56 & \hspace*{-2ex} 82.3 $\times$ 35.8 & \hspace*{-2ex} 11.9 $\times$  4.3 &  2.8 & \hspace*{-3ex} 6.0 & \hspace*{-2.5ex}    6 & \hspace*{-1ex}  0.036 &  0.24 &  0.098 &  37( 8) & \hspace*{-3ex}  3.2 & \hspace*{-3ex}  6.7 & \hspace*{-3ex} 4.8(1.0) \\  Cha1-C40 & \hspace*{-2ex} 57 & \hspace*{-2ex} 58.7 $\times$ 46.6 & \hspace*{-2ex}  8.2 $\times$  6.2 &  1.3 & \hspace*{-3ex} 6.1 & \hspace*{-2.5ex}   12 & \hspace*{-1ex}  0.037 &  0.23 &  0.095 &  39( 8) & \hspace*{-3ex}  3.3 & \hspace*{-3ex}  6.4 & \hspace*{-3ex} 5.1(1.0) \\  Cha1-C41 & \hspace*{-2ex} 58 & \hspace*{-2ex} 48.3 $\times$ 31.9 & \hspace*{-2ex}  6.5 $\times$  3.6 &  1.8 & \hspace*{-3ex} 6.1 & \hspace*{-2.5ex}   12 & \hspace*{-1ex}  0.037 &  0.13 &  0.083 &  45(10) & \hspace*{-3ex}  3.3 & \hspace*{-3ex}  5.6 & \hspace*{-3ex} 5.9(1.3) \\  Cha1-C42 & \hspace*{-2ex} 59 & \hspace*{-2ex} 48.7 $\times$ 26.0 & \hspace*{-2ex}  6.6 $\times$  2.3 &  2.9 & \hspace*{-3ex} 6.0 & \hspace*{-2.5ex}    8 & \hspace*{-1ex}  0.036 &  0.10 &  0.064 &  57(14) & \hspace*{-3ex}  3.2 & \hspace*{-3ex}  4.3 & \hspace*{-3ex} 7.4(1.8) \\  Cha1-C43 & \hspace*{-2ex} 60 & \hspace*{-2ex} 90.0 $\times$ 45.6 & \hspace*{-2ex} 13.1 $\times$  6.0 &  2.2 & \hspace*{-3ex} 5.7 & \hspace*{-2.5ex}    9 & \hspace*{-1ex}  0.035 &  0.32 &  0.084 &  41( 9) & \hspace*{-3ex}  3.1 & \hspace*{-3ex}  5.7 & \hspace*{-3ex} 5.4(1.2) \\  Cha1-C44 & \hspace*{-2ex} 61 & \hspace*{-2ex} 46.5 $\times$ 42.4 & \hspace*{-2ex}  6.2 $\times$  5.5 &  1.1 & \hspace*{-3ex} 5.8 & \hspace*{-2.5ex}  5.9 & \hspace*{-1ex}  0.035 &  0.15 &  0.074 &  48(11) & \hspace*{-3ex}  3.1 & \hspace*{-3ex}  5.0 & \hspace*{-3ex} 6.3(1.4) \\  Cha1-C45 & \hspace*{-2ex} 64 & \hspace*{-2ex} 73.2 $\times$ 21.4 & \hspace*{-2ex} 10.5 $\times$  2.1 &  5.0 & \hspace*{-3ex} 5.7 & \hspace*{-2.5ex}  5.1 & \hspace*{-1ex}  0.035 &  0.12 &  0.074 &  47(11) & \hspace*{-3ex}  3.1 & \hspace*{-3ex}  5.0 & \hspace*{-3ex} 6.1(1.4) \\  Cha1-C46 & \hspace*{-2ex} 65 & \hspace*{-2ex} 36.6 $\times$ 21.2 & \hspace*{-2ex}  4.5 $\times$  2.1 &  2.1 & \hspace*{-3ex} 5.7 & \hspace*{-2.5ex}    7 & \hspace*{-1ex}  0.034 &  0.059 &  0.051 &  67(19) & \hspace*{-3ex}  3.0 & \hspace*{-3ex}  3.5 & \hspace*{-3ex} 8.8(2.4) \\  Cha1-C47 & \hspace*{-2ex} 67 & \hspace*{-2ex} 87.6 $\times$ 36.2 & \hspace*{-2ex} 12.8 $\times$  4.4 &  2.9 & \hspace*{-3ex} 5.7 & \hspace*{-2.5ex}    9 & \hspace*{-1ex}  0.034 &  0.24 &  0.090 &  38( 8) & \hspace*{-3ex}  3.1 & \hspace*{-3ex}  6.1 & \hspace*{-3ex} 5.0(1.1) \\  Cha1-C48 & \hspace*{-2ex} 68 & \hspace*{-2ex} 50.7 $\times$ 24.0 & \hspace*{-2ex}  6.9 $\times$  2.1 &  3.3 & \hspace*{-3ex} 5.5 & \hspace*{-2.5ex}    7 & \hspace*{-1ex}  0.033 &  0.090 &  0.059 &  56(15) & \hspace*{-3ex}  2.9 & \hspace*{-3ex}  4.0 & \hspace*{-3ex} 7.4(1.9) \\  Cha1-C49 & \hspace*{-2ex} 70 & \hspace*{-2ex} 40.2 $\times$ 26.9 & \hspace*{-2ex}  5.1 $\times$  2.5 &  2.1 & \hspace*{-3ex} 5.6 & \hspace*{-2.5ex}    9 & \hspace*{-1ex}  0.034 &  0.082 &  0.072 &  47(11) & \hspace*{-3ex}  3.0 & \hspace*{-3ex}  4.9 & \hspace*{-3ex} 6.2(1.5) \\  Cha1-C50 & \hspace*{-2ex} 71 & \hspace*{-2ex} 60.6 $\times$ 54.3 & \hspace*{-2ex}  8.5 $\times$  7.5 &  1.1 & \hspace*{-3ex} 5.5 & \hspace*{-2.5ex}   12 & \hspace*{-1ex}  0.033 &  0.24 &  0.086 &  39( 9) & \hspace*{-3ex}  2.9 & \hspace*{-3ex}  5.8 & \hspace*{-3ex} 5.1(1.2) \\  Cha1-C51 & \hspace*{-2ex} 72 & \hspace*{-2ex} 34.9 $\times$ 31.9 & \hspace*{-2ex}  4.2 $\times$  3.6 &  1.2 & \hspace*{-3ex} 5.6 & \hspace*{-2.5ex}   14 & \hspace*{-1ex}  0.034 &  0.084 &  0.065 &  52(13) & \hspace*{-3ex}  3.0 & \hspace*{-3ex}  4.4 & \hspace*{-3ex} 6.8(1.7) \\  Cha1-C52 & \hspace*{-2ex} 73 & \hspace*{-2ex} 50.5 $\times$ 27.8 & \hspace*{-2ex}  6.9 $\times$  2.7 &  2.6 & \hspace*{-3ex} 5.3 & \hspace*{-2.5ex}   17 & \hspace*{-1ex}  0.032 &  0.100 &  0.065 &  49(13) & \hspace*{-3ex}  2.9 & \hspace*{-3ex}  4.4 & \hspace*{-3ex} 6.5(1.7) \\  Cha1-C53 & \hspace*{-2ex} 74 & \hspace*{-2ex} 52.8 $\times$ 40.1 & \hspace*{-2ex}  7.2 $\times$  5.1 &  1.4 & \hspace*{-3ex} 5.5 & \hspace*{-2.5ex}    6 & \hspace*{-1ex}  0.033 &  0.16 &  0.084 &  40( 9) & \hspace*{-3ex}  2.9 & \hspace*{-3ex}  5.7 & \hspace*{-3ex} 5.2(1.2) \\  Cha1-C54 & \hspace*{-2ex} 78 & \hspace*{-2ex} 36.6 $\times$ 29.9 & \hspace*{-2ex}  4.5 $\times$  3.2 &  1.4 & \hspace*{-3ex} 5.2 & \hspace*{-2.5ex}   17 & \hspace*{-1ex}  0.032 &  0.077 &  0.060 &  53(14) & \hspace*{-3ex}  2.8 & \hspace*{-3ex}  4.1 & \hspace*{-3ex} 6.9(1.8) \\  Cha1-C55 & \hspace*{-2ex} 79 & \hspace*{-2ex} 40.7 $\times$ 28.8 & \hspace*{-2ex}  5.2 $\times$  2.9 &  1.8 & \hspace*{-3ex} 5.3 & \hspace*{-2.5ex}   10 & \hspace*{-1ex}  0.032 &  0.084 &  0.077 &  42(10) & \hspace*{-3ex}  2.9 & \hspace*{-3ex}  5.2 & \hspace*{-3ex} 5.5(1.3) \\  Cha1-C56 & \hspace*{-2ex} 80 & \hspace*{-2ex} 46.3 $\times$ 22.4 & \hspace*{-2ex}  6.2 $\times$  2.1 &  2.9 & \hspace*{-3ex} 5.2 & \hspace*{-2.5ex}    9 & \hspace*{-1ex}  0.031 &  0.073 &  0.056 &  57(16) & \hspace*{-3ex}  2.8 & \hspace*{-3ex}  3.8 & \hspace*{-3ex} 7.4(2.1) \\  Cha1-C57 & \hspace*{-2ex} 81 & \hspace*{-2ex} 51.1 $\times$ 40.1 & \hspace*{-2ex}  7.0 $\times$  5.1 &  1.4 & \hspace*{-3ex} 5.4 & \hspace*{-2.5ex}    7 & \hspace*{-1ex}  0.032 &  0.15 &  0.12 &  27( 6) & \hspace*{-3ex}  2.9 & \hspace*{-3ex}  8.1 & \hspace*{-3ex} 3.6(0.8) \\  Cha1-C58 & \hspace*{-2ex} 82 & \hspace*{-2ex} 48.5 $\times$ 31.6 & \hspace*{-2ex}  6.5 $\times$  3.5 &  1.9 & \hspace*{-3ex} 5.5 & \hspace*{-2.5ex}   11 & \hspace*{-1ex}  0.033 &  0.11 &  0.088 &  38( 9) & \hspace*{-3ex}  3.0 & \hspace*{-3ex}  6.0 & \hspace*{-3ex} 5.0(1.1) \\  Cha1-C59 & \hspace*{-2ex} 83 & \hspace*{-2ex} 41.0 $\times$ 21.4 & \hspace*{-2ex}  5.3 $\times$  2.1 &  2.5 & \hspace*{-3ex} 5.3 & \hspace*{-2.5ex}   14 & \hspace*{-1ex}  0.032 &  0.063 &  0.052 &  61(17) & \hspace*{-3ex}  2.9 & \hspace*{-3ex}  3.5 & \hspace*{-3ex} 8.0(2.3) \\  Cha1-C60 & \hspace*{-2ex} 84 & \hspace*{-2ex} 61.6 $\times$ 43.7 & \hspace*{-2ex}  8.7 $\times$  5.7 &  1.5 & \hspace*{-3ex} 5.1 & \hspace*{-2.5ex}    9 & \hspace*{-1ex}  0.031 &  0.18 &  0.071 &  43(11) & \hspace*{-3ex}  2.8 & \hspace*{-3ex}  4.8 & \hspace*{-3ex} 5.7(1.4) \\  \hline
 \end{tabular}
 \tablefoot{
 \tablefoottext{a}{Numbering of \textit{Gaussclumps} sources like in Table~\ref{t:id_gcl_simbad}.}
 \tablefoottext{b}{Size of the fitted Gaussian, deconvolved physical source size, and aspect ratio. Sizes smaller than 25.4$\arcsec$ were set to 25.4$\arcsec$ to compute the deconvolved sizes, in order to account for a fit inaccuracy corresponding to a 5$\sigma$ detection in peak flux density. As a result, the minimum size that can be measured is about 2100 AU. The aspect ratio is the ratio of the deconvolved sizes along the major and minor axes.}
 \tablefoottext{c}{Peak H$_2$ column density computed assuming a dust opacity of 0.01~cm$^2$~g$^{-1}$. The statistical rms uncertainty is $1.1 \times 10^{21}$~cm$^{-2}$.}
 \tablefoottext{d}{Visual extinction derived from 2MASS.}
 \tablefoottext{e}{Mass in the central beam ($HPBW = 21.2\arcsec$), total mass derived from the Gaussian fit, and mass computed from the flux measured in an aperture of 50$\arcsec$ in diameter. The statistical rms uncertainties of $M_{\mathrm{peak}}$ and $M_{50\arcsec}$ are 0.006 and 0.011 M$_\odot$, respectively.}
 \tablefoottext{f}{Mass concentration $m_{\mathrm{peak}}/m_{50\arcsec}$. The statistical rms uncertainty is given in parentheses.}
 \tablefoottext{g}{Beam-averaged free-particle density within the central beam and mean free-particle density computed for the mass $M_{50\arcsec}$ in the aperture of diameter 50\arcsec. The statistical rms uncertainties are $5.6 \times 10^{4}$ and $7.2 \times 10^{3}$ cm$^{-3}$, respectively.}
 \tablefoottext{h}{Density contrast $n_{\mathrm{peak}}/n_{50\arcsec}$. The statistical rms uncertainty is given in parentheses.}
 }
\end{table*}

\begin{figure*}
\centerline{\resizebox{0.85\hsize}{!}{\includegraphics[angle=270]{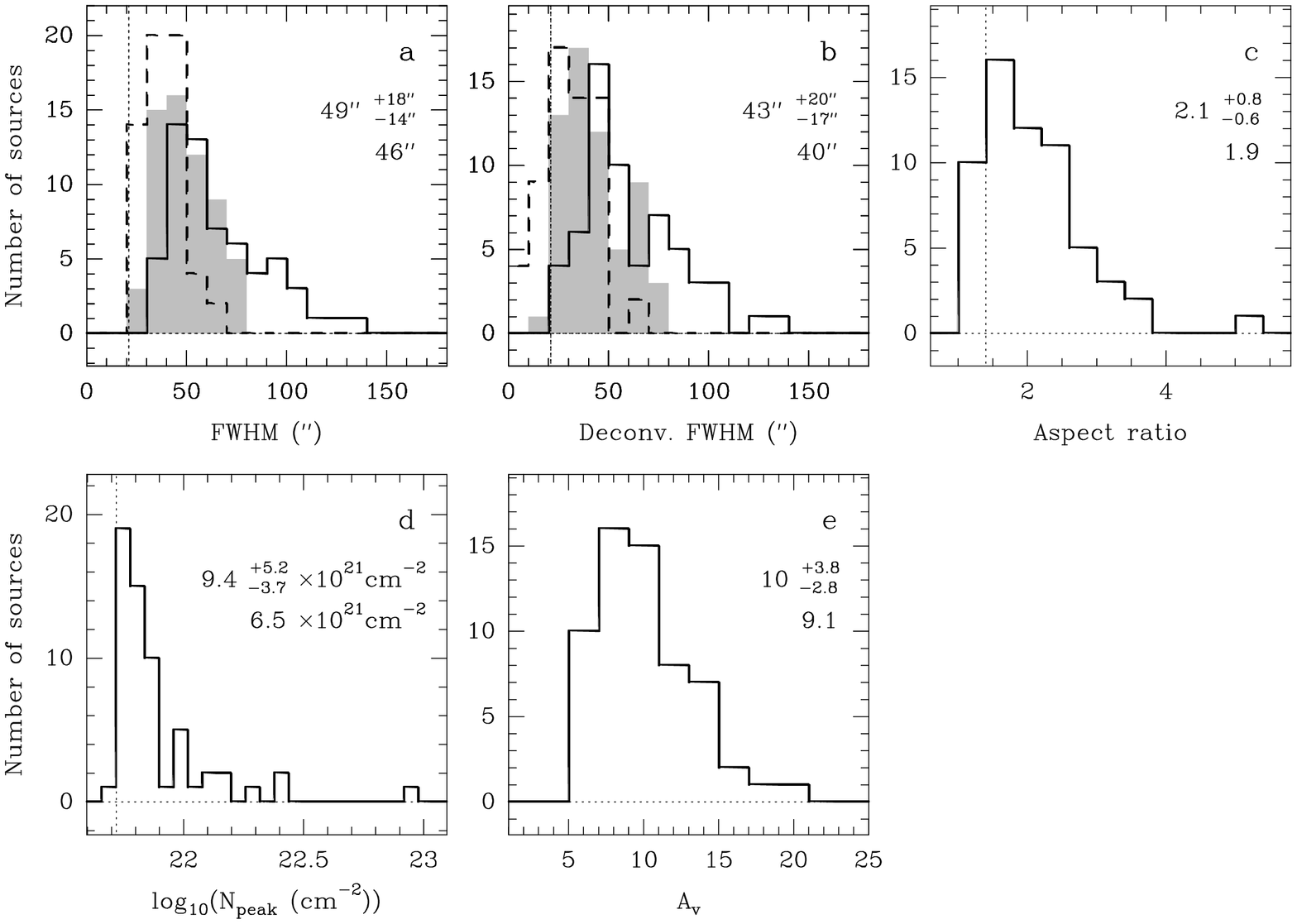}}}
\caption{Distribution of physical properties obtained for the 60 starless (or 
Class 0) sources found with \textit{Gaussclumps} in the \textit{sum} map of 
Cha~I at scale 5. The mean, standard deviation, and median of the 
distribution are given in each panel. The asymmetric standard deviation defines
the range containing 68$\%$ of the sample.
\textbf{a} $FWHM$ sizes along the major (solid line) and minor (dashed 
line) axes. The filled histogram shows the distribution of geometrical mean 
of major and minor sizes. The mean and median values refer to the filled
histogram. The dotted line indicates the angular resolution 
(21.2$\arcsec$). 
\textbf{b} Same as \textbf{a} but for the deconvolved sizes.
\textbf{c} Aspect ratios computed with the deconvolved sizes. The dotted line 
at 1.4 shows the threshold above which the deviation from 1 (elongation) can 
be considered as significant.
\textbf{d} Peak H$_2$ column density. The dotted line at 
$5.25 \times 10^{21}$~cm$^{-2}$ is the 5$\sigma$ sensitivity limit.
\textbf{e} Visual extinction derived from 2MASS.}
\label{f:histo}
\end{figure*}

The properties of the 60 starless (or Class 0) sources are listed in 
Table~\ref{t:starless} and their distribution is shown in Figs.~\ref{f:histo} 
and \ref{f:histom}. The column density (Col.~6) and masses (Cols.~8, 
9, and 10) are computed with the fluxes fitted with \textit{Gaussclumps} or 
directly measured in the \textit{sum} map at scale 5, assuming a dust mass 
opacity $\kappa_{870} = 0.01$~cm$^2$~g$^{-1}$. As a caveat, we remind that the 
assumption of a uniform temperature may be inaccurate and bias the 
measurements of the masses and column densities, as well as the mass 
concentration (or equivalently the density contrast). Since only one of the 60
sources has an embedded YSO (as traced at 24~$\mu$m with \textit{Spitzer}, see 
Sect.~\ref{ss:classification}), the other 59 have no central heating and their 
temperature should be rather uniform. However a dust temperature drop toward 
the center of starless dense cores is possible (see 
Appendix~\ref{ss:temperature}).

\subsubsection{Extinction}
\label{sss:starless_extinction}

The visual extinctions listed in Table~\ref{t:starless} are extracted from the 
extinction map derived from 2MASS (see Sect.~\ref{ss:2mass}). Given the lower 
resolution of this map ($HPBW = 3\arcmin$) compared to the 870~$\mu$m map, it 
provides an estimate of the extinction of the environment in which the 
870~$\mu$m sources are embedded.

The 870~$\mu$m sources are found down to a visual extinction $A_V \sim 5$~mag
(as traced with 2MASS at low angular resolution), but not below. A similar 
result was obtained by \citet{Johnstone04} in Ophiuchus (threshold at 
$A_V = 7$~mag), by \citet{Enoch06} and \citet{Kirk06} in Perseus 
(threshold at $A_V = 5$~mag), and by \citet{Andre10a} in Aquila 
\citep[threshold at $A_V = 5$~mag, see Fig.~5 of][, but see below for the peak
of the distribution]{Andre10b}. In Taurus, \citet{Goldsmith08} also found a 
threshold at $A_V = 6$~mag based on the distribution of YSOs, somewhat lower 
than the earlier findings of \citet{Onishi98} who proposed a threshold at 
$N_{{\mathrm{H}}_2} = 8 \times 10^{21}$~cm$^{-2}$ ($A_V \sim 9$~mag) based on 
C$^{18}$O cores and IRAS sources. In both latter cases, however, this 
threshold is not sharp but rather indicates a significant increase in the 
probability for star formation to occur.

The extinction toward the only confirmed young protostar in Cha~I (Cha-MMS1 or 
Cha1-C1 in Table~\ref{t:starless}) is high, at $A_V = 14$~mag. The 
distribution of extinctions of the environment in which the starless cores are 
embedded has a median of 9.1~mag and presents a sharp peak at $A_V \sim$ 
8--9~mag (see Fig.~\ref{f:histo}e). This is very similar in shape to the 
distribution of extinctions found for Aquila with \textit{Herschel}, which 
peaks at $A_V \sim 7.5$~mag \citep[see Fig.~5 of][]{Andre10b}.

\begin{figure*}
\centerline{\hspace*{-0.03\hsize}\resizebox{0.80\hsize}{!}{\includegraphics[angle=0,trim=15pt 0pt 10pt 105pt,clip=true]{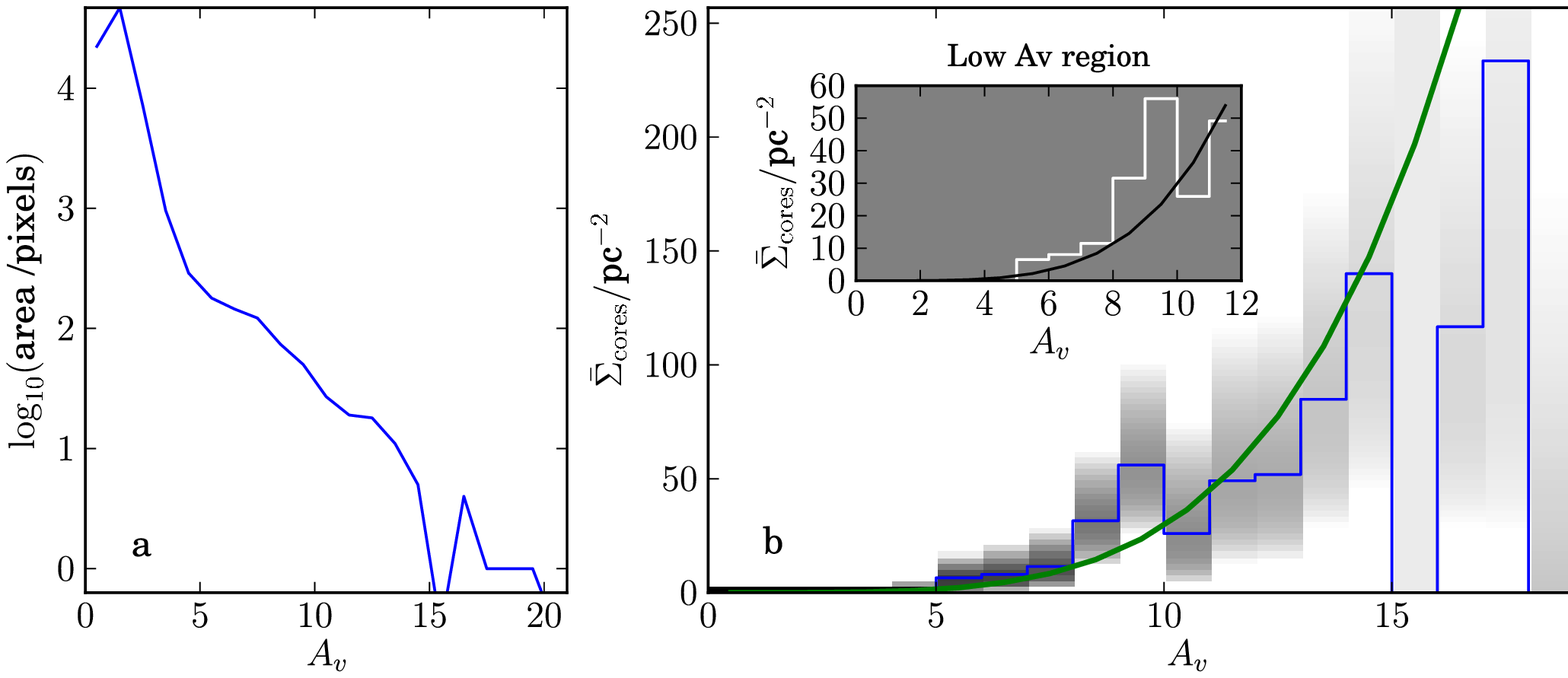}}}
\caption{\textbf{a} Distribution of extinction across the map. The area for 
$A_V<3$ regions is limited by the map boundary.  
\textbf{b}
Mean surface density of starless (+ Class 0) cores as a function of gas
column density as measured by $A_V$. For each $A_V$ value, the 
probability distribution for the true value of the mean surface density is 
displayed in greyscale. The corresponding colorbar on the right is labeled in 
natural logarithm. The insert is a zoom on the low-extinction range. The 
best-fit power-law curve to the mean surface density is shown in green/black 
(see Sect.~\ref{sss:starless_extinction} for details).
}
\label{f:jenny}
\end{figure*}

Star formation is more likely at high extinction, as discussed above,
and in Cha I the areas of high extinction are small. We can further quantify 
the relationship between the extinction and the location of the dense
cores using the distribution of extinctions across the map 
(Fig.~\ref{f:jenny}a). We calculate the average surface density of cores 
$\bar\Sigma_\mathrm{ cores}$ in each extinction bin by dividing the number of 
starless (+ Class 0) cores in that bin by the total area of the map 
(in pc$^2$) in that extinction range. The resulting average surface density of 
cores is plotted as a function of gas column density $\Sigma_\mathrm{gas}$ (as 
measured by $A_v$) as a blue histogram in Fig.~\ref{f:jenny}b.

If we assume that each core will ultimately form a star, then we could adopt 
the extragalactic concept of the Kennicutt-Schmidt law and convert the density 
of cores to a star formation rate $\Sigma_\mathrm{SF}$ in 
M$_\odot$~yr$^{-1}$~kpc$^{-2}$. Assuming a mean lifetime in the detectable 
prestellar (+ Class 0) stage of 0.5~Myr (see Sect.~\ref{ss:lifetimes}) and a 
mean final stellar mass of 0.5~M$_\odot$ would give a conversion factor from 
$\Sigma_\mathrm{SF}$ in M$_\odot$~yr$^{-1}$~kpc$^{-2}$ to 
$\bar\Sigma_\mathrm{cores}$ in pc$^{-2}$ equal to 1.

To quantify the relationship between star formation and gas column 
density in this local cloud we fit a power law of the form 
$\bar\Sigma_\mathrm{cores} = k\Sigma_\mathrm{{gas}}^\alpha$. One has to be very 
careful with the uncertainty budget when dividing two quantities each of which 
may lie close to zero, particularly if this applies for the 
denominator, and here the area of the 
cloud becomes small at high extinction. We assume the uncertainties are 
dominated by Poisson ($\sqrt{N}$) counting statistics in 
each extinction bin and use the Bayesian method as laid out in 
\citet[][, Appendix A]{Hatchell05} to calculate the resulting uncertainties on 
the ratio. The greyscale in Fig.~\ref{f:jenny}b gives the probability 
distribution for the true value of the mean surface density of cores for each 
$A_V$ value, given the measured counts and assuming Poisson statistics. At high 
extinction values, where the number of pixels is close to zero, the mean 
surface density of cores is very uncertain and this shows as a broad spread in 
the greyscale. Using these uncertainties, we perform a minimisation to 
calculate the power-law index $\alpha$ and its associated uncertainties. 
Fitting over the full range ($A_v = 0\hbox{--}20$) gives a steep 
power-law index $\alpha=4.34^{+0.30}_{-0.18}$ (68.3\% confidence 
limits). This is even steeper than the power law found in 
Perseus \citep[$\alpha=3.0\pm 0.2$,][]{Hatchell05}.
The power-law fit suggests an alternative interpretation to 
that of an $A_V$ threshold, which is that star formation is simply unusual at 
low column densities, and a wide area search is required to find a dense core 
which is not embedded in an extended area of raised extinction 
\citep{Hatchell05}. 

However, we note that the previous fit tends to 
underestimate the mean surface density of cores in the range 8--10~mag
(see insert of Fig.~\ref{f:jenny}b), where the peak of the distribution in 
number is located (Fig.~\ref{f:histo}e). This suggests that the peak at 
$A_V \sim$8--9~mag may have a specific physical signification, as 
proposed by \citet{Andre10a} (see Sect.~\ref{s:intro}).

\subsubsection{Sizes}
\label{sss:starless_sizes}

The source sizes along the major and minor axes before and after deconvolution 
are listed in Cols.~3 and 4 of Table~\ref{t:starless}, and their 
distribution is shown in Fig.~\ref{f:histo}a and b, respectively, along with 
the distribution of mean size (geometrical mean of major and minor sizes,
i.e. $\sqrt{FWHM_{\mathrm{maj}} \times FWHM_{\mathrm{min}}}$). 
The average major, minor, and mean sizes are 
$65\,^{+26}_{-18}\,\arcsec$, 
$38\,^{+10}_{-9}\,\arcsec$, and $49\,^{+18}_{-14}\,\arcsec$, respectively.
Only three sources have a major $FWHM$ size larger than 110$\arcsec$, and no 
source has a minor or mean $FWHM$ size larger than 80$\arcsec$. The results of 
the Monte-Carlo simulations in the elliptical case (Appendix~\ref{ss:filtering} 
and Table~\ref{t:filtering}) imply that these sources, although most of them 
are faint with a peak flux density lower than 150 mJy/beam, are not 
significantly affected by the spatial filtering due to the sky noise removal, 
with less than 15$\%$ loss of peak flux density and size. 

The fitting done on the control sample of artificial, circular, Gaussian 
sources shows that the accuracy of the sizes derived for sources with a 
peak flux equal to $\sim 5\sigma$ is roughly $20\%$ (see 
Figs.~\ref{f:filtering}c and d). With $HPBW = 21.2\arcsec$, the size accuracy 
of a weak $\sim 5\sigma$ unresolved source is $4.2\arcsec$. Therefore, faint 
sources with a size smaller
than $\sim 25.4\arcsec$ cannot be reliably deconvolved and we artificially set
their size to $25.4\arcsec$ to perform the deconvolution. As a result the 
minimum deconvolved $FWHM$ size that we can measure is $\sim 14\arcsec$ 
(2100 AU). The average deconvolved mean $FWHM$ size is 
$43\,^{+20}_{-17}\,\arcsec$, i.e. $6500\,^{+3000}_{-2600}$~AU (see 
Fig.~\ref{f:histo}b). 

\citet{Enoch08} measure average deconvolved mean sizes of $16500 \pm 6000$, 
$10700 \pm 3100$, and $7500 \pm 3500$~AU ($FWHM$) for the starless cores 
extracted with \textit{Clumpfind} from their Bolocam maps of Perseus, Serpens, 
and Ophiuchus, respectively (see Col.~9 of Table~\ref{t:allclouds}). With the 
MPIfR 19-channel bolometer array of the IRAM 30m telescope ($HPBW = 15\arcsec$ 
after smoothing), \citet{Motte98} derive significantly smaller sizes for the
starless cores in Ophiuchus based on a method similar to the one used here for 
Cha~I (multiresolution decomposition and Gaussian fitting): they find an 
average deconvolved mean size of $2100^{+1200}_{-1400}$~AU when rescaled to the 
same distance as assumed by \citet{Enoch08}. Both the differences in angular
resolution (31$\arcsec$ versus 15$\arcsec$) and in the extraction method used 
(no background subtraction versus multiresolution analysis) certainly explain 
the discrepancy of a factor of 3 between both studies. Using 
\textit{Gaussclumps} without background subtraction, \citet{Curtis10} derive
an average deconvolved mean size of $3800 \pm 200$~AU for the starless cores 
observed in Perseus by \citet{Hatchell05} with SCUBA, 
i.e. a factor of 4 times smaller than the Bolocam cores. The angular resolution
may therefore be the dominant factor influencing the derived $FWHM$ sizes.
The angular resolution of LABOCA (21$\arcsec$ after smoothing) being a bit 
larger than the SCUBA and 30~m resolutions, we can conclude that the Cha~I 
starless cores have similar physical sizes than the Perseus cores, are probably
larger than the Serpens cores (maybe by a factor of 1.5--2), and are certainly 
larger than the Ophiuchus cores (by a factor of 2--3).

No complete survey of Taurus in dust continuum emission exists. 
\citet{Kauffmann08} mapped 10 dense cores with MAMBO at the IRAM 30~m 
telescope. After smoothing to 20$\arcsec$, they identified 23 starless 
subcores. The average deconvolved mean FWHM size of this sample is 
$20600^{+4900}_{-3800}$~AU (see Table~\ref{t:allclouds}). If this 
small sample is representative for the entire population of starless cores in 
Taurus, then the Cha~I starless cores are significantly smaller (by a factor 
of 3) than the Taurus ones.

\begin{table*}
 \caption{Sensitivities of (sub)mm surveys of nearby molecular clouds and 
average properties of the detected starless cores.}
 \label{t:allclouds}
 \begin{tabular}{lccccclcccccc}
  \hline\hline
  \multicolumn{1}{c}{Cloud} & \multicolumn{1}{c}{\hspace*{-1ex} Dist.} & \multicolumn{1}{c}{\hspace*{-1ex} Freq.} & \multicolumn{1}{c}{\hspace*{-2ex} $HPBW$} & \multicolumn{1}{c}{\hspace*{-3ex} $\sigma$\tablefootmark{a}} & \multicolumn{1}{c}{\hspace*{-2ex} $\sigma_N$\tablefootmark{b}} & \multicolumn{1}{c}{\hspace*{-2ex} $\sigma_M^{7500}$\tablefootmark{\,c}} & \multicolumn{1}{c}{\hspace*{-2ex} Ref.\tablefootmark{d}} & \multicolumn{1}{c}{\hspace*{-2ex} $<FWHM>$\tablefootmark{e}} & \multicolumn{1}{c}{$<R_a>$\tablefootmark{f}} & \multicolumn{1}{c}{$<N_{\mathrm{peak}}>$\tablefootmark{g}} &  \multicolumn{1}{c}{\hspace*{-1ex} $<n_{\mathrm{7500~AU}}>$\tablefootmark{h}} & \multicolumn{1}{c}{\hspace*{-2ex} Ref.\tablefootmark{d}} \\
   & \multicolumn{1}{c}{\hspace*{-1ex} \scriptsize (pc)}  & \multicolumn{1}{c}{\hspace*{-1ex} \scriptsize (GHz)} & \multicolumn{1}{c}{\hspace*{-2ex} \scriptsize ($\arcsec$)} & \multicolumn{1}{c}{\hspace*{-3ex} \scriptsize (mJy/beam)} & \multicolumn{1}{c}{\hspace*{-2ex} \scriptsize ($10^{21}$~cm$^{-2}$)} & \multicolumn{1}{c}{\hspace*{-2ex} \scriptsize (M$_\odot$)} & & \multicolumn{1}{c}{\hspace*{-2ex} \scriptsize (1000~AU)} & & \multicolumn{1}{c}{\scriptsize (10$^{21}$~cm$^{-2}$)} & \multicolumn{1}{c}{\hspace*{-1ex} \scriptsize (10$^5$~cm$^{-3}$)} &  \\
  \multicolumn{1}{c}{(1)} & \multicolumn{1}{c}{\hspace*{-1ex} (2)} & \multicolumn{1}{c}{\hspace*{-1ex} (3)} & \multicolumn{1}{c}{\hspace*{-2ex} (4)} & \multicolumn{1}{c}{\hspace*{-3ex} (5)} & \multicolumn{1}{c}{\hspace*{-2ex} (6)} & \multicolumn{1}{c}{\hspace*{-2ex} (7)} & \multicolumn{1}{c}{\hspace*{-2ex} (8)} & \multicolumn{1}{c}{\hspace*{-2ex} (9)} & \multicolumn{1}{c}{(10)} & \multicolumn{1}{c}{(11)} & \multicolumn{1}{c}{\hspace*{-1ex} (12)} & \multicolumn{1}{c}{\hspace*{-2ex} (13)}  \\
  \hline
  \noalign{\smallskip}
  Chamaeleon~I & \hspace*{-1ex} 150 & \hspace*{-1ex} 345 & \hspace*{-2ex} 21 & \hspace*{-3ex} 12  & \hspace*{-2ex} 1.1 & \hspace*{-2ex} 0.012 & \hspace*{-2ex} 1 & \hspace*{-2ex} $6.5^{+3.0}_{-2.6}$ & $2.1^{+0.8}_{-0.6}$  & $9^{+5}_{-4}$   & \hspace*{-1ex} 1.0$^{+1.1}_{-0.5}$  & \hspace*{-2ex} 1 \\
  \noalign{\smallskip}
  Ophiuchus    & \hspace*{-1ex} 125 & \hspace*{-1ex} 268 & \hspace*{-2ex} 31 & \hspace*{-3ex} 27  & \hspace*{-2ex} 2.4 & \hspace*{-2ex} 0.035 & \hspace*{-2ex} 2 & \hspace*{-2ex} $7.5 \pm 3.5$      & $1.3 \pm 0.2$      & 60             & \hspace*{-1ex} 5.7--6.5            & \hspace*{-2ex} 8 \\
  \noalign{\smallskip}
               &                    & \hspace*{-1ex} 240 & \hspace*{-2ex} 13 & \hspace*{-3ex}  8  & \hspace*{-2ex} 5.9 & \hspace*{-2ex} 0.035 & \hspace*{-2ex} 3 & \hspace*{-2ex} $2.1^{+1.2}_{-1.4}$  & $1.3^{+0.7}_{-0.3}$ & $26^{+23}_{-13}$ &                                   & \hspace*{-2ex} 3 \\
  \noalign{\smallskip}
  Taurus       & \hspace*{-1ex} 140 & \hspace*{-1ex} 250 & \hspace*{-2ex} 11 & \hspace*{-3ex} 1.4 & \hspace*{-2ex} 1.3 & \hspace*{-2ex} 0.007 & \hspace*{-2ex} 4 & \hspace*{-2ex} $20.6^{+4.9}_{-3.8}$ & $1.9^{+0.5}_{-0.4}$ & $23^{+13}_{-11}$ & \hspace*{-1ex} 3.0                & \hspace*{-2ex} 4 \\
  \noalign{\smallskip}
  Perseus      & \hspace*{-1ex} 250 & \hspace*{-1ex} 268 & \hspace*{-2ex} 31 & \hspace*{-3ex} 15  & \hspace*{-2ex} 1.4 & \hspace*{-2ex} 0.039 & \hspace*{-2ex} 5 & \hspace*{-2ex} $16.5 \pm 6.0$     & $1.7 \pm 0.9$      & 38              & \hspace*{-1ex} 3.5--3.9           & \hspace*{-2ex} 8 \\
  \noalign{\smallskip}
               &                    & \hspace*{-1ex} 353 & \hspace*{-2ex} 14 & \hspace*{-3ex} 35  & \hspace*{-2ex} 6.6 & \hspace*{-2ex} 0.086 & \hspace*{-2ex} 6 & \hspace*{-2ex} $3.8 \pm 0.2$      & $1.4 \pm 0.1$      & $35 \pm 3$      &                                   & \hspace*{-2ex} 9 \\
  \noalign{\smallskip}
  Serpens      & \hspace*{-1ex} 260 & \hspace*{-1ex} 268 & \hspace*{-2ex} 31 & \hspace*{-3ex} 9.5 & \hspace*{-2ex} 0.9 & \hspace*{-2ex} 0.026 & \hspace*{-2ex} 7 & \hspace*{-2ex} $10.7 \pm 3.1$     & $1.7 \pm 0.5$      & 33              & \hspace*{-1ex} 2.8--3.2           & \hspace*{-2ex} 8 \\
  \hline
 \end{tabular}
 \tablefoot{The values in Cols.~6 and 7 were computed with the same 
 assumptions for all clouds (see Appendix~\ref{s:assumptions}). The values in 
 Cols.~9 to 12 obtained from the litterature were rescaled to match the 
 physical assumptions and size criteria adopted for our analysis of Cha~I 
 (see Appendix~\ref{ss:rescaling} for details).
 \tablefoottext{a}{Peak flux density rms sensitivity in the original beam.}
 \tablefoottext{b}{Peak column density rms sensitivity in the original beam.}
 \tablefoottext{c}{Mass rms sensitivity per aperture of diameter 7500~AU.}
 \tablefoottext{d}{References: (1) this work; (2) \citet{Young06}; (3)
  \citet{Motte98}; (4) \citet{Kauffmann08}; (5) \citet{Enoch06}; 
   (6) \citet{Hatchell05}; (7) \citet{Enoch07}; (8) \citet{Enoch08}; 
   (9) \citet{Curtis10}.}
 \tablefoottext{e}{Average deconvolved mean size of the detected starless 
   cores.}
 \tablefoottext{f}{Average aspect ratio.}
 \tablefoottext{g}{Average peak column density in a $FWHM$ size of 3180~AU, 
  corresponding to our angular resolution of 21.2$\arcsec$ at the distance of 
  Cha~I.}
 \tablefoottext{h}{Average mean density in an aperture of diameter 7500~AU.}
}
\end{table*}

\subsubsection{Aspect ratios}
\label{sss:aspect_ratios}

The distribution of aspect ratios computed with the deconvolved $FWHM$ sizes 
is shown in Fig.~\ref{f:histo}c. Based on the Monte Carlo simulations of 
Appendix~\ref{ss:filtering}, we estimate that a faint source can reliably be
considered as intrinsically elongated when its aspect ratio is higher than 
1.4. 17$\%$ of the sources are below this threshold while 83$\%$ can be 
considered as elongated. The average aspect ratio is $2.1\,^{+0.8}_{-0.6}$.
It is similar to the ones measured in Serpens and Taurus, somewhat larger than 
in Perseus, and significantly larger than in Ophiuchus (see Col.~10 of 
Table~\ref{t:allclouds}).

\subsubsection{Column densities}
\label{sss:starless_coldens}

The average peak H$_2$ column density of the starless sources in Cha~I 
is $9\,^{+5}_{-4} \times 10^{21}$~cm$^{-2}$ (Fig.~\ref{f:histo}d). 
This is 4 times lower than the average peak column density of the
starless cores in Perseus and Serpens, and 3 to 7 times lower than in 
Ophiuchus (see Col.~11 of Table~\ref{t:allclouds}, and 
Appendix~\ref{ss:rescaling} for details). It appears to be significantly 
smaller than in Taurus too (by a factor of 2.6), but since the 
Taurus sample is not complete and the source extraction methods differ, this 
may not be significant.

\begin{figure*}
\centerline{\resizebox{1.00\hsize}{!}{\includegraphics[angle=270]{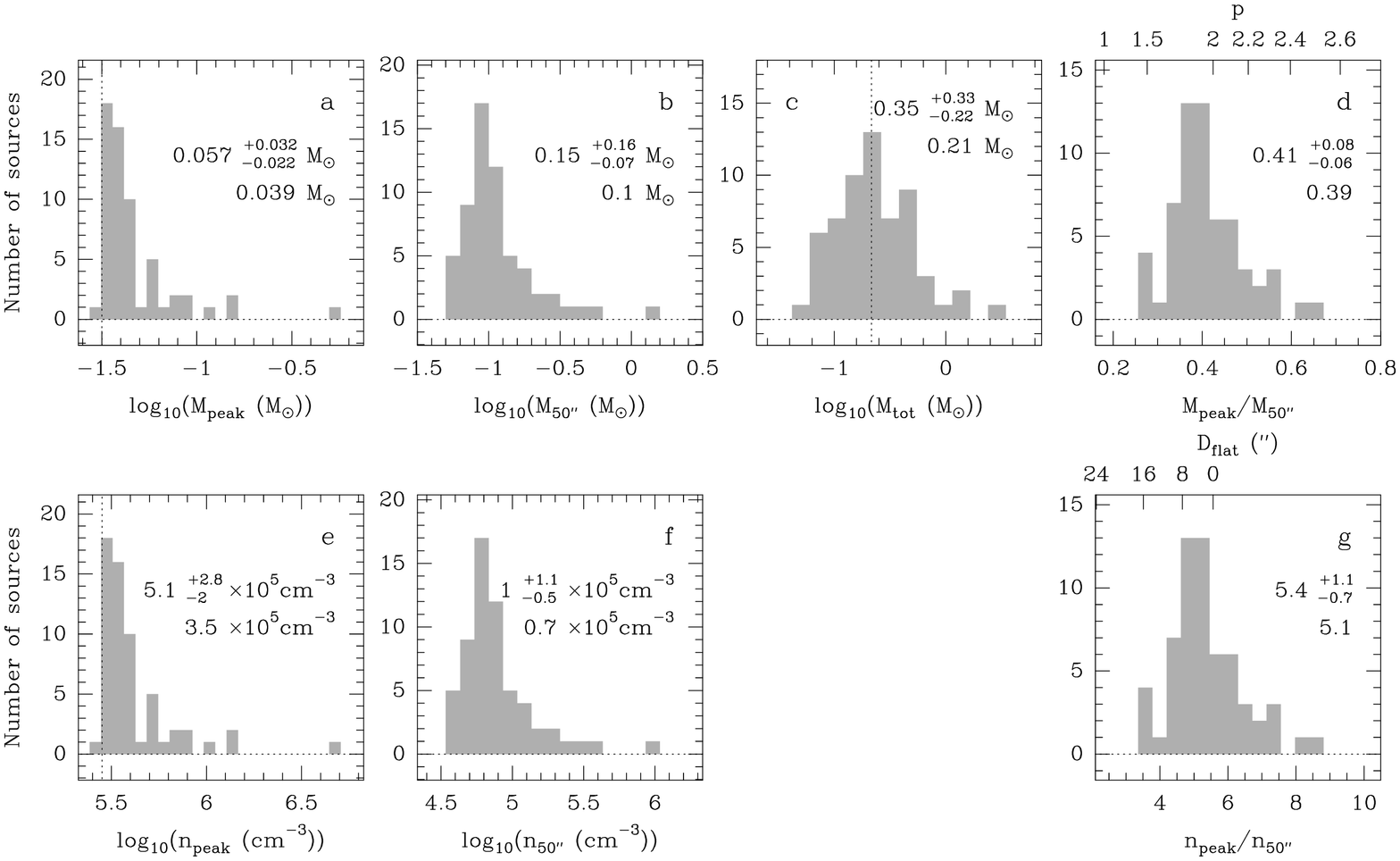}}}
\caption{Distribution of masses and densities obtained for the 60 starless (or 
Class 0) sources found with \textit{Gaussclumps} in the \textit{sum} map of 
Cha~I at scale 5. The mean, standard deviation, and median of the 
distribution are given in each panel. The asymmetric standard deviation defines
the range containing 68$\%$ of the sample.
\textbf{a} Peak mass of the fitted Gaussian.
\textbf{b} Mass within an aperture of diameter $50\arcsec$.
\textbf{c} Total mass of the fitted Gaussian. The dotted line indicates the 
estimated completeness limit at 90$\%$ for Gaussian sources corresponding to a 
6.3$\sigma$ peak detection limit for the average source size. 
\textbf{d} Mass concentration, ratio of the peak mass to the mass within an 
aperture of diameter $50\arcsec$. 
\textbf{e} Peak density.
\textbf{f} Mean density within an aperture of diameter $50\arcsec$.
\textbf{g} Density contrast, ratio of peak density to mean density.
In panels \textbf{a} and \textbf{e}, the dotted line indicates the 
$5\sigma$ sensitivity limit.
The upper axis of panel \textbf{d}, which can also be used for panel 
\textbf{g}, shows the power-law exponent $p$ derived assuming that the sources 
have a density profile proportional to $r^{-p}$ and a uniform dust temperature. 
Alternately, the upper axis of panel \textbf{g}, which can also be used for 
panel \textbf{d}, deals with the case where the density is uniform within a 
diameter $D_{\mathrm{flat}}$ and proportional to $r^{-2}$ outside, still with 
the assumption of a uniform temperature. See Sect.~\ref{sss:starless_masses}
for the limitations of these upper axes.
}
\label{f:histom}
\end{figure*}

\subsubsection{Masses and densities}
\label{sss:starless_masses}

The distribution of masses and free-particle densities listed 
in Table~\ref{t:starless} are displayed in Fig.~\ref{f:histom}.
The $5\sigma$ sensitivity limit used to 
extract sources with \textit{Gaussclumps} corresponds to a peak mass 
of 0.032 M$_\odot$ and a peak density of $2.8 \times 10^5$~cm$^{-3}$, computed 
for a diameter of 21.2$\arcsec$ (3200~AU). The median of the peak mass 
distribution is 0.039~M$_\odot$, implying a median peak density of 
$3.5 \times 10^5$~cm$^{-3}$ (Figs.~\ref{f:histom}a and e). We give the mass 
integrated within an aperture of diameter 50$\arcsec$ (7500~AU) in Col.~10 of 
Table~\ref{t:starless}. This aperture is well adapted to our sample for three 
reasons: it corresponds to the average mean, undeconvolved $FWHM$ size (see 
Sect.~\ref{sss:starless_sizes} and Fig.~\ref{f:histo}a), it is not affected by 
the spatial filtering due to the sky noise removal (see 
Appendix~\ref{ss:filtering} and Table~\ref{t:filtering}), and it is still 
preserved in the \textit{sum} map at scale 5 (see Appendix~\ref{s:mrmed} and 
Table~\ref{t:mrmed}). The median of the mass integrated within this aperture 
is 0.10~M$_\odot$, corresponding to a median mean density of 
$7 \times 10^4$~cm$^{-3}$ (Figs.~\ref{f:histom}b and f).

The cores in Cha~I are about 3--4 times less dense within an aperture 
of diameter 7500~AU than those in Perseus and Serpens, 6 times less dense 
than those in Ophiuchus, and  3 times less dense than those in Taurus (see 
Col.~12 of Table~\ref{t:allclouds}, and Appendix~\ref{sss:rescaling_dens} for 
details). However, both studies of \citet{Enoch08} and \citet{Kauffmann08}
do not filter the extended emission as we do with our multiresolution 
decomposition to isolate the starless cores from the environment in which they
are embedded. This may bias the densities in both studies to somewhat higher 
values compared to Cha~I. In addition, the sensitivities of the different 
surveys in terms of mass (and mean density) within the aperture of 7500~AU are 
not the same (see Col.~7 of Table~\ref{t:allclouds}). Our survey is more 
sensitive than those reported by \citet{Enoch08} by a factor of 2 to 3, but 
less sensitive than the one of \citet{Kauffmann08} by a factor of nearly 2. 
This means that the surveys of Perseus, Serpens, and Ophiuchus may be biased 
towards higher masses and mean densities and that they may have missed a 
population of less dense starless cores similar to the one found in Cha~I. 
In the case of Taurus, the bias of the source sample of \citet{Kauffmann08} 
prevents any conclusion concerning the possible existence of a population of 
less dense starless cores.

Figure~\ref{f:histom}c shows the distribution of total masses computed 
from the Gaussian fits (Col.~9 of Table~\ref{t:starless}). The 
completeness limit at 90$\%$ is estimated from a peak flux detection threshold 
at 6.3$\sigma$ for the average size of the source sample 
($FWHM = 49\arcsec$)\footnote{For a Gaussian distribution of mean value $m$ 
and standard deviation $\sigma$, the relative population below $m-1.28\sigma$ 
represents 10$\%$. Therefore our peak flux detection threshold at 5$\sigma$ 
implies a 90$\%$ completeness limit at $6.3\sigma$, with $\sigma$ the rms 
noise  level in the \textit{sum} map at scale 5.}. It corresponds to a total 
mass of 0.22~M$_\odot$.  The median total mass is very close (0.21~M$_\odot$), 
which implies that only 50$\%$ of the detected sources are above the estimated 
$90\%$ completeness limit. For comparison, the $90\%$ completeness limit of 
the starless core sample of \citet{Enoch08} in Perseus is $\sim 0.9$~M$_\odot$, 
i.e. 1.2~M$_\odot$ when rescaled to the same temperature and dust opacity as 
we use here (12~K and $\beta = 1.85$). Our completeness limit 
is similar to that obtained by \citet{Konyves10} for their 11 deg$^2$ 
sensitive continuum survey of the Aquila Rift cloud complex (distance 260~pc) 
with \textit{Herschel}: assuming a mass distribution proportional to the 
radius, they estimated their sample of 541 starless cores to be 75 and 85$\%$ 
complete above masses of 0.2 and 0.3 M$_\odot$, respectively. 
Their assumed dust opacity law yields $\kappa_{870} = 0.012$~cm$^2$~g$^{-1}$,
i.e. only 20$\%$ different from our assumption so the completeness limits can 
be directly compared.

We estimate the mass concentration of the sources from the ratio of the peak 
mass to the mass within an aperture of 50$\arcsec$ (Col.~11 of 
Table~\ref{t:starless}) which is insensitive to the spatial filtering due to
the data reduction (see Table~\ref{t:filtering}). A similar property is the 
density contrast measured as the ratio of the peak density to the mean density 
within this aperture (Col.~14 of Table~\ref{t:starless}). The statistical rms 
uncertainties on the peak mass and the mass within 50$\arcsec$ are 0.006 and 
0.011 M$_\odot$, respectively, which means a relative uncertainty of up to 
20$\%$ for the weakest sources. The distributions of both ratios are shown in 
Figs.~\ref{f:histom}d and g and their rms uncertainties\footnote{The relative 
uncertainty of the ratio is equal to the square root of the quadratic sum 
of the relative uncertainties of its two terms, i.e. we assume both terms are 
uncorrelated.} are given in parentheses in Cols.~11 and 14 of 
Table~\ref{t:starless}. The two outliers with the largest ratios are also 
those with the highest relative uncertainty (about $30\%$).
The upper axis of Fig.~\ref{f:histom}d, which can also be used for 
Fig.~\ref{f:histom}g, displays the exponent of the density profile under the 
assumptions that the sources are spherically symmetric with a power-law 
density profile, i.e. $\rho \propto r^{-p}$, and that the dust temperature is 
uniform. The median mass concentration and density contrast are 0.39 and 5.1, 
respectively. This corresponds to $p \sim 1.9$, suggesting that most sources 
are significantly centrally-peaked. It is very close to the exponent of the
singular isothermal sphere ($p = 2$).
As a caveat, we should mention that the relation used to derive $p$ is valid 
only for sources with a mass between 21.2$\arcsec$ and 50$\arcsec$ that is 
measurable and extends up to 50$\arcsec$. In practice, most sources have 
$C_M < 50\%$ with a relative uncertainty less than $20\%$ (see Col.~11 of 
Table~\ref{t:starless}), meaning that for those sources, the mass between both 
diameters is measured with a reasonable accuracy. The few sources with 
$C_M > 50\%$ ($13\%$ of the sample), especially those with $C_M > 60\%$, are 
also those with the highest relative uncertainties ($\sim 30\%$ for the 
latter). The relation to derive $p$ may not be valid for these few sources, or 
they are simply too weak to give a reliable estimate. As a result, these 
sources may bias the mean and median values of $p$ to slightly higher values, 
but since they are not numerous, the bias is certainly small\footnote{If we 
remove the sources with $C_M > 60\%$ and  $C_M > 50\%$, the mean of $C_M$ is 
only reduced by $2\%$ and $6\%$, respectively, and the median decreases by less 
than $1\%$.}.

The upper axis of Fig.~\ref{f:histom}g, which can also be used for 
Fig.~\ref{f:histom}d, deals with an alternate case where the sources have a 
constant density within a diameter $D_{\mathrm{flat}}$ and a density 
proportional to $r^{-2}$ outside, still with the assumption of a uniform 
temperature. Under these assumptions, the measurements are consistent with a 
flat inner region of diameter 16$\arcsec$ \textit{at most} (2400~AU) for a few 
sources, but most sources have $D_{\mathrm{flat}} < 10\arcsec$ (1500~AU), or 
cannot be described with such a density profile.

\subsubsection{Mass versus size}
\label{sss:starless_massvssize}

\begin{figure*}
\centerline{\resizebox{1.0\hsize}{!}{\includegraphics[angle=270]{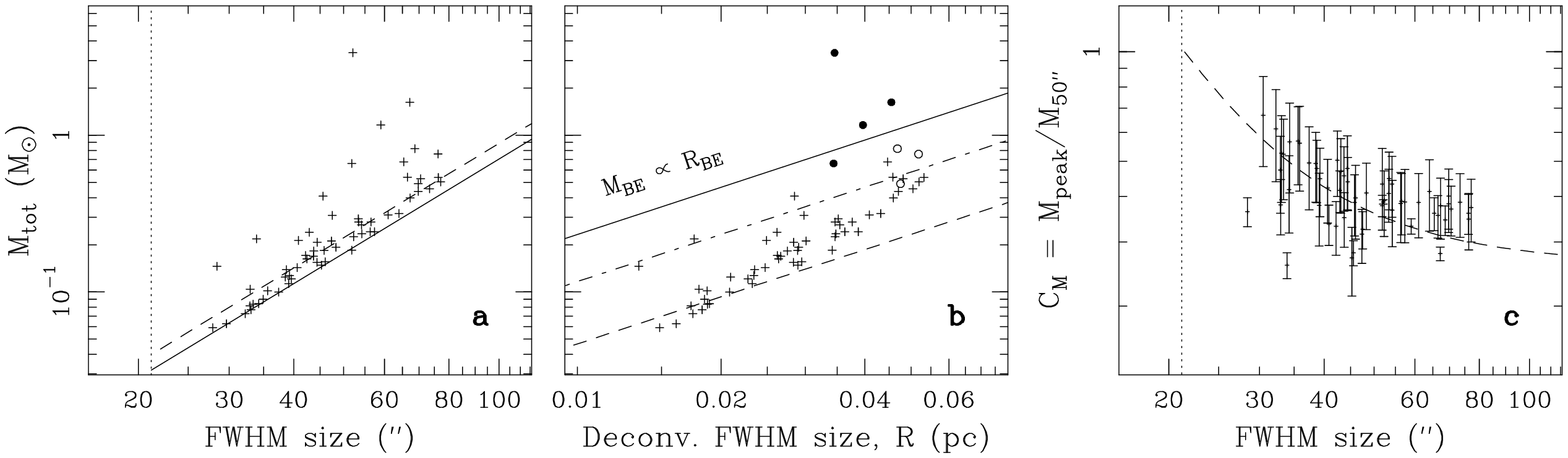}}}
\caption{\textbf{a} Total mass versus mean $FWHM$ size for the 60 starless (or 
Class 0) sources found with \textit{Gaussclumps} in the \textit{sum} map of 
Cha~I at scale 5. The angular resolution (21.2$\arcsec$) is marked by the 
dotted line. The solid line ($M \propto FWHM^2$) is the 5$\sigma$ peak 
sensitivity limit for Gaussian sources. The dashed line shows the 6.3$\sigma$ 
peak sensitivity limit which corresponds to a completeness limit of 90$\%$ for 
Gaussian sources. 
\textbf{b} Total mass versus mean deconvolved $FWHM$ size. Sizes smaller than
25.4$\arcsec$ were set to 25.4$\arcsec$ before deconvolution (see note b of 
Table~\ref{t:starless}). The solid line shows the relation 
$M = 2.4 \, Ra_{\mathrm{s}}^2/G$ that characterizes critical Bonnor-Ebert 
spheres 
(see Sect.~\ref{sss:starless_massvssize}). The dash-dotted and dashed lines 
show the location of this relation when divided by 2 and 5, respectively.
The sources with a mass larger than the critical Bonnor-Ebert mass
$M_{\mathrm{BE,}P_{\mathrm{ext}}}$ estimated from the ambient cloud pressure are 
shown with a filled circle. Those with 
$0.5 \, M_{\mathrm{BE,}P_{\mathrm{ext}}} < M < M_{\mathrm{BE,}P_{\mathrm{ext}}}$ 
are shown with a circle.
\textbf{c} Mass concentration versus mean $FWHM$ size. The dashed line is the 
expectation for a circular Gaussian flux density distribution.}
\label{f:massvssize}
\end{figure*}

The distribution of total masses versus source sizes derived from the Gaussian 
fits is shown in Fig.~\ref{f:massvssize}a. About 50$\%$ of the sources are 
located between the 5$\sigma$ detection limit (full line) and the estimated 
$90\%$ completeness limit (dashed line), suggesting that we most likely miss a 
significant number of sources with a low peak column density. 
Figure~\ref{f:massvssize}b shows a similar diagram for the deconvolved source
size. If we assume that the deconvolved $FWHM$ size is a good estimate of the
external \textit{radius} of each source, then we can compare this distribution 
to the critical Bonnor-Ebert mass that characterizes the limit above which 
the hydrostatic equilibrium of an isothermal sphere with thermal support only 
is gravitationally unstable. This relation 
$M_{\mathrm{BE}}(R) = 2.4 \, Ra_{\mathrm{s}}^2/G$ \citep[][]{Bonnor56}, with 
$M_{\mathrm{BE}}(R)$ the total mass, $R$ the external radius, 
$a_{\mathrm{s}}$ the 
sound speed, and $G$ the gravitational constant, is drawn for a temperature of 
12~K as a solid line in Fig.~\ref{f:massvssize}b. Only three sources 
(Cha1-C1 to C3) are located above this critical mass limit. If we account for 
a factor of 2 uncertainty on the mass, then seven additional sources could 
also fall above this limit provided the mass is underestimated (Cha1-C4 to C7, 
C9, C11, and C13, see dash-dotted line in Fig.~\ref{f:massvssize}b). Most 
sources, however, have a mass lower than the critical Bonnor-Ebert mass by a 
factor of 2 to 5. The uncertainty on the temperature (see 
Appendix~\ref{ss:temperature}) does not influence these results much since, 
even in the unlikely case of the \textit{bulk} of the mass being at a 
temperature of 7~K, the measured masses would move upwards relative to the 
critical Bonnor-Ebert mass limit by a factor of 1.9 only, because the latter 
is also temperature dependent.

Following the analysis of \citet{Konyves10} for the Aquila starless cores, we 
can also estimate the critical Bonnor-Ebert mass with the equation 
$M_{\mathrm{BE}}(P_{\mathrm{ext}}) = 1.18 \, a_{\mathrm{s}}^4 \,
G^{-\frac{3}{2}} \, P^{-\frac{1}{2}}_{\mathrm{ext}}$
\citep{Bonnor56}. The external pressure is estimated with the equation 
$P_{\mathrm{ext}} = 0.88 \, G \, (\mu m_{\mathrm{H}} N_{\mathrm{cl}})^2$, 
with $\mu$ the mean molecular weight per free particle, $m_{\mathrm{H}}$ the
mass of hydrogen, and $N_{\mathrm{cl}}$ the column density of the local ambient 
cloud in which the sources are embedded 
\citep[][]{McKee03}. We use the extinction listed in Table~\ref{t:starless} to
estimate this background column density with the conversion factors given in 
Appendix~\ref{ss:avtoN}. The sources with a mass larger than 
$M_{\mathrm{BE}}(P_{\mathrm{ext}})$ are the same as for $M_{\mathrm{BE}}(R)$, plus 
Cha1-C4. The agreement between both estimates of $M_{\mathrm{BE}}$
suggests that our estimates of the external radius and external pressure 
are consistent. If we account again for a possible factor of 2 uncertainty on 
the mass measurement, only three additional sources would fall above the 
critical limit based on the pressure (Cha1-C7, C11, and C18). In summary,
four sources are likely above the critical Bonnor-Ebert limit (Cha1-C1 to C4),
and seven additional ones may also be if their mass is underestimated by a 
factor of 2 (Cha1-C5 to C7, C9, C11, C13, and C18). The implications of this
analysis will be discussed in Sect.~\ref{s:discussion}.

The mass concentration $C_M$ is plotted versus source size in 
Fig.~\ref{f:massvssize}c. $C_M$ is actually equal to the ratio of the 
peak flux to the flux integrated within the aperture of diameter 50$\arcsec$.
When the sources do not overlap, this ratio is nearly independent of the 
Gaussian fitting since the second and third 
stiffness parameters of \textit{Gaussclumps} were set to 1, i.e. biasing 
\textit{Gaussclumps} to keep the fitted peak amplitude close to the observed 
one and the fitted center position close to the position of the observed peak.
The dashed line shows the expected ratio if the (not deconvolved) sources were 
exactly Gaussian and circular and allows to estimate the departure of the 
sources from being Gaussian within 50$\arcsec$. 
Most sources have a mass concentration consistent with the Gaussian 
expectation, but many of them have a significant uncertainty on $C_M$ that
prevents a more accurate analysis. The two obvious outliers toward the lower 
left are sources Cha1-C6 and C8, which have strong neighbors significantly 
contaminating their flux within 50$\arcsec$ (sources Cha1-S1 and S4, 
respectively).

\begin{figure}
\centerline{\resizebox{1.00\hsize}{!}{\includegraphics[angle=270]{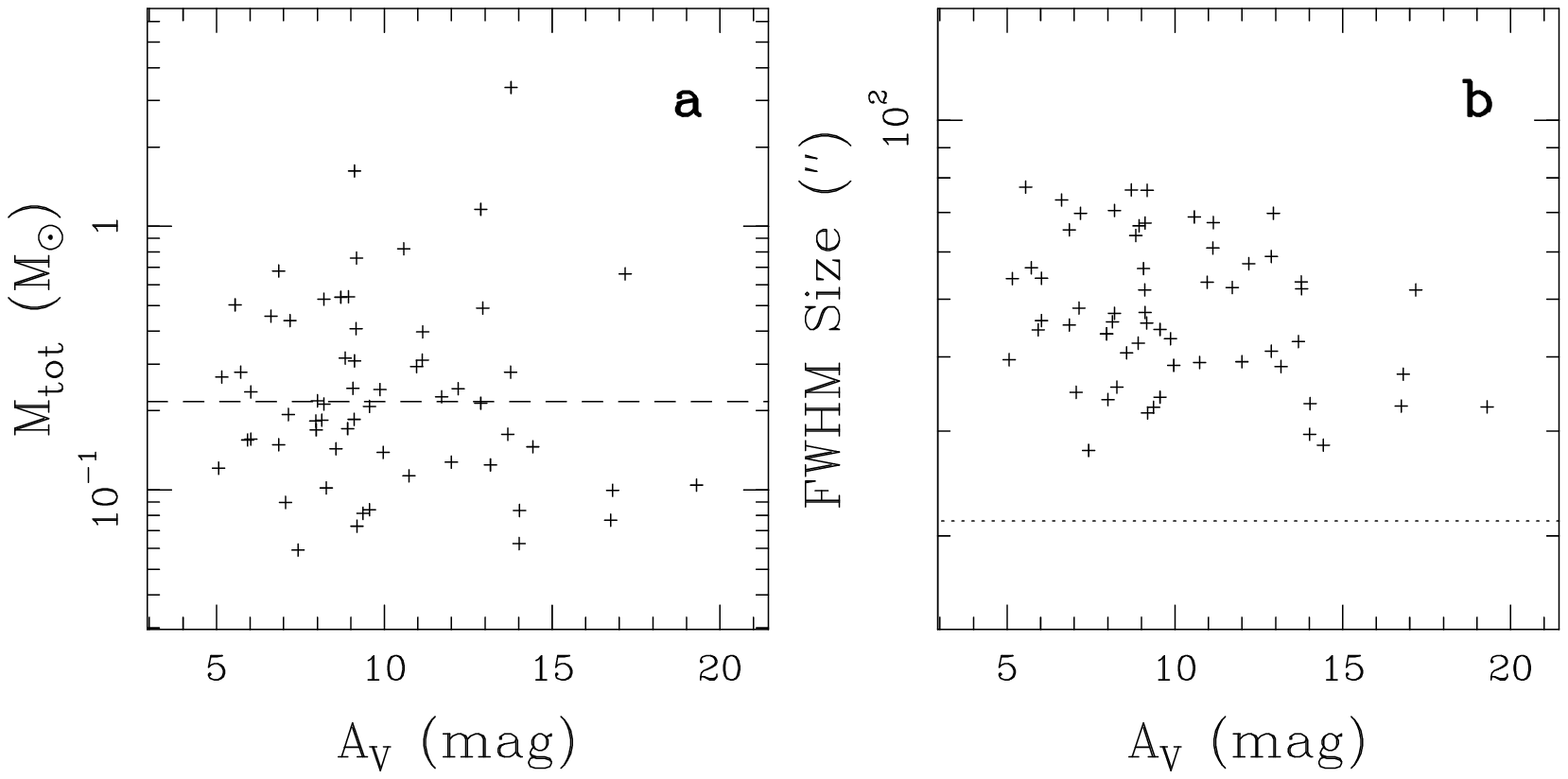}}}
\caption{\textbf{a} Total mass versus visual extinction A$_V$ for the 60 
starless (or Class 0) sources found with \textit{Gaussclumps} in the 
\textit{sum} map of Cha~I at scale 5. The dashed line shows the estimated 
90$\%$ completeness limit (0.22 M$_\odot$).
\textbf{b} FWHM size versus visual extinction. The angular resolution 
(21.2$\arcsec$) is marked by the dotted line.}
\label{f:massvsav}
\end{figure}

There is no obvious correlation between the total mass or FWHM size of the 
sources and the visual extinction of the environment in which they are 
embedded (see Fig.~\ref{f:massvsav}). A similar conclusion was drawn by 
\citet{Sadavoy10} for the five nearby molecular clouds Ophiuchus, Taurus, 
Perseus, Serpens, and Orion based on SCUBA data.

\subsubsection{Core mass distribution (CMD)}
\label{sss:cmd}

\begin{figure}
\centerline{\resizebox{1.00\hsize}{!}{\includegraphics[angle=270]{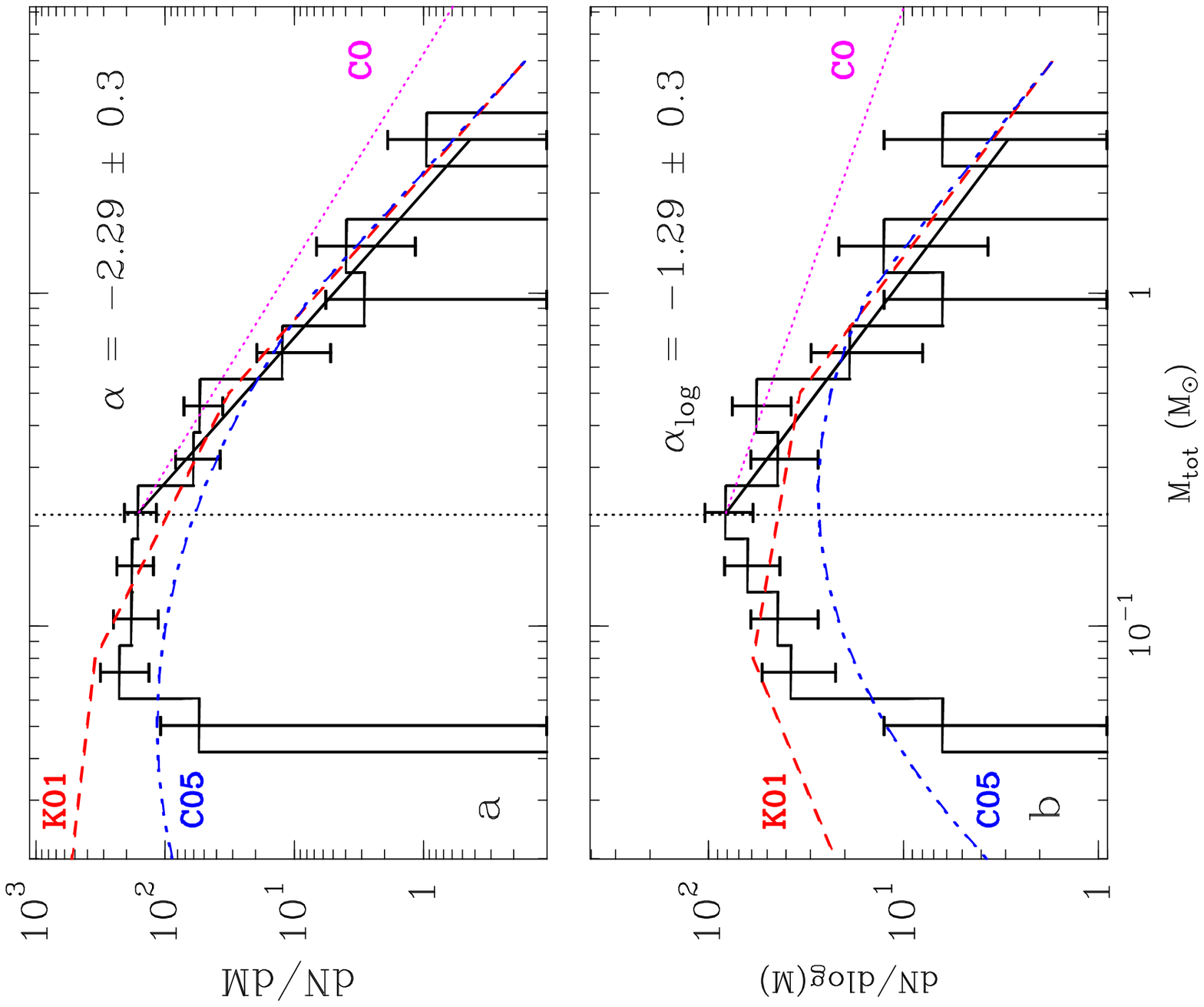}}}
\caption{Mass distribution $\mathrm{d}N/\mathrm{d}M$ (\textbf{a}) and 
$\mathrm{d}N/\mathrm{d}\log(M)$ (\textbf{b}) of the 60 starless (or Class 0) 
sources. The error bars represent the Poisson noise (in $\sqrt{N}$). The 
vertical dotted line is the estimated 90$\%$ completeness limit. The thick 
solid line is the best power-law fit performed on the mass bins above the 
completeness limit. The best fit exponent, $\alpha$ and $\alpha_{\mathrm{log}}$, 
respectively, is given in the upper right corner of each panel. The IMF of 
single stars corrected for binaries \citep[][, K01]{Kroupa01} and the IMF of 
multiple systems \citep[][, C05]{Chabrier05} are shown in dashed (red) and 
dot-dashed (blue) lines, respectively. They are both vertically shifted to 
the same number at 5~M$_\odot$. The dotted (purple) curve is the 
typical mass spectrum of CO clumps \citep[][]{Blitz93,Kramer98}.}
\label{f:cmd}
\end{figure}

The mass distribution of the 60 starless (or Class 0) sources is shown in
Fig.~\ref{f:cmd}, in both forms $\mathrm{d}N/\mathrm{d}M$ (a) and 
$\mathrm{d}N/\mathrm{d}\log(M)$ (b). Its shape in Fig.~\ref{f:cmd}a looks very 
similar to the shape of the mass distribution found in other star forming 
regions with a power-law-like behavior at the high-mass end and a flattening 
toward the low-mass end. In our case, the flattening occurs below the 
estimated $90\%$ completeness limit (0.22~M$_\odot$) and may not be
significant. Above this limit, the distribution is consistent with a power-law.
The exponent of the best power-law fit ($\alpha = -2.29 \pm 0.30$ for 
$\mathrm{d}N/\mathrm{d}M$, $\alpha_{\mathrm{log}} = -1.29 \pm 0.30$ for 
$\mathrm{d}N/\mathrm{d}\log(M)$) is very close to the value of
\citet{Salpeter55} that characterizes the high-mass end of the stellar initial 
mass function ($\alpha = -2.35$) and steeper than the exponent of the typical 
mass spectrum of CO clumps \citep[$\alpha = -1.6$, see][]{Blitz93,Kramer98}. 
Such a result was also obtained for the population of starless cores in, e.g., 
Ophiuchus \citep[$\alpha \sim -2.5$,][]{Motte98}, 
Serpens, Perseus \citep[$\alpha = -2.3 \pm 0.4$,][]{Enoch08}, 
the Pipe nebula \citep[][]{Alves07}, 
Taurus \citep[$\alpha_{\mathrm{log}} = -1.2 \pm 0.1$\footnote{Given the small 
size of the Taurus sample (69 starless sources), the uncertainty seems 
underestimated compared to the other clouds.},][]{Sadavoy10}, 
and the Aquila Rift 
\citep[$\alpha_{\mathrm{log}} = -1.5 \pm 0.2$,][]{Andre10a,Konyves10}.

The mass distribution of the Cha~I starless sources does not show any 
significant flattening down to the estimated $90\%$ completeness limit 
(0.22~M$_\odot$), although the signal-to-noise ratio may not be sufficient to
draw a firm conclusion. This seems to contrast with the mass distributions of 
Ophiuchus and Aquila which flatten below 0.4 and 1.0~M$_\odot$, a factor 
$\sim 4$ and $\sim 3$ above their completeness limits, respectively 
\citep[][]{Motte98,Konyves10}. This flattening translates into a 
$\mathrm{d}N/\mathrm{d}\log(M)$ curve with a significant maximum around 
0.5--0.6~M$_\odot$ for Aquila while, in Ophiuchus, it keeps increasing with a 
reduced exponent ($\alpha_{\mathrm{log}} \sim -0.5$) down to the completeness 
limit \citep[$\sim 0.1$ M$_\odot$,][]{Motte98}. The 
$\mathrm{d}N/\mathrm{d}\log(M)$ curve in Cha~I reaches a maximum at a mass of 
0.2 M$_\odot$ (Fig.~\ref{f:cmd}b), significantly lower than in Aquila. 
\citet{Sadavoy10} found a maximum around 0.16~M$_\odot$ for Taurus, which 
\textit{a priori} looks similar to our result for Cha~I. However, their 
analysis is based on masses computed within a 850~$\mu$m contour of 
90~mJy/14$\arcsec$-beam, meaning that they miss a large fraction of the total 
mass of each core. It is very likely that the mass corresponding to the true 
maximum is higher than 0.16~M$_\odot$ by a factor of a few, even maybe an 
order of magnitude. For comparison, the average mass measured by 
\citet{Kauffmann08} within an aperture of diameter 8400~AU, which represents 
only a fraction of the total mass (roughly 50$\%$), is already a factor of 3 
higher for their sample of 28 starless ``peaks'' (see 
Sect.~\ref{sss:starless_masses}). The discrepancy with the mass distribution 
obtained by \citet{Onishi02} based on H$^{13}$CO$^+$ observations is even 
larger, since their distribution $\mathrm{d}N/\mathrm{d}M$ flattens around 
2~$M_\odot$. Therefore, the true maximum of the $\mathrm{d}N/\mathrm{d}\log(M)$ 
curve in Taurus is certainly at a mass significantly higher than in Cha~I.

\subsection{YSO candidates}
\label{ss:pointsources}

\begin{table*}
 \caption{
 Characteristics of YSOs extracted with \textit{Gaussclumps} in the 870~$\mu$m continuum map of Cha~I filtered up to scale 5.
 }
 \label{t:ysos}
 \centering
 \begin{tabular}{ccccccccccccc}
 \hline\hline
 \multicolumn{1}{c}{Name} & \multicolumn{1}{c}{\hspace*{-2ex} ${N_{\mathrm{gcl}}}$\tablefootmark{a}} & \multicolumn{1}{c}{$FWHM$\tablefootmark{b}} & \multicolumn{1}{c}{$FWHM$\tablefootmark{b}} & \multicolumn{1}{c}{$SNR$\tablefootmark{c}} & \multicolumn{1}{c}{\hspace*{-2ex} ${N_{\mathrm{peak}}}$\tablefootmark{d}} & \multicolumn{1}{c}{\hspace*{-2ex} ${A_V}$\tablefootmark{e}} & \multicolumn{1}{c}{${M_{\mathrm{peak}}}$\tablefootmark{f}} & \multicolumn{1}{c}{${M_{\mathrm{tot}}}$\tablefootmark{f}} & \multicolumn{1}{c}{${M_{50\arcsec}}$\tablefootmark{f}} & \multicolumn{1}{c}{${C_M}$\tablefootmark{g}} & \multicolumn{1}{c}{Class\tablefootmark{h}} & \multicolumn{1}{c}{Ref.\tablefootmark{h}} \\ 
  & \hspace*{-2ex} & \multicolumn{1}{c}{($\arcsec \times \arcsec$)} & \multicolumn{1}{c}{(1000 AU)$^2$} & & \multicolumn{1}{c}{\hspace*{-2ex} ($10^{21}$ cm$^{-2}$)} & \multicolumn{1}{c}{\hspace*{-2ex} (mag)} & \multicolumn{1}{c}{(M$_\odot$)} & \multicolumn{1}{c}{(M$_\odot$)} & \multicolumn{1}{c}{(M$_\odot$)} &  \multicolumn{1}{c}{($\%$)} & & \\ 
 \multicolumn{1}{c}{(1)} & \multicolumn{1}{c}{\hspace*{-2ex} (2)} & \multicolumn{1}{c}{(3)} & \multicolumn{1}{c}{(4)} & \multicolumn{1}{c}{(5)} & \multicolumn{1}{c}{\hspace*{-2ex} (6)} & \multicolumn{1}{c}{\hspace*{-2ex} (7)} & \multicolumn{1}{c}{(8)} & \multicolumn{1}{c}{(9)} & \multicolumn{1}{c}{(10)} & \multicolumn{1}{c}{(11)} & \multicolumn{1}{c}{(12)} & \multicolumn{1}{c}{(13)} \\ 
 \hline
 Cha1-S1 & \hspace*{-2ex} 1 & 23.5 $\times$ 22.9 &  1.6 $\times$  1.3  &  215 & \hspace*{-2ex}  33 & \hspace*{-2ex}    7 &  0.20 &  0.24 &  0.23 &  86( 1)&   II    & 1 \\  Cha1-S2 & \hspace*{-2ex} 2 & 26.6 $\times$ 26.6 &  2.5 $\times$  2.5  &  101 & \hspace*{-2ex}  15 & \hspace*{-2ex}    9 &  0.092 &  0.14 &  0.13 &  69( 1)&    I    & 1 \\  Cha1-S3 & \hspace*{-2ex} 3 & 25.1 $\times$ 23.0 &  2.1 $\times$  1.4  &   97 & \hspace*{-2ex}  15 & \hspace*{-2ex}   11 &  0.089 &  0.11 &  0.11 &  79( 1)&   II    & 1 \\  Cha1-S4 & \hspace*{-2ex} 5 & 57.8 $\times$ 31.0 &  8.3 $\times$  3.5  &   51 & \hspace*{-2ex} 7.7 & \hspace*{-2ex}   14 &  0.046 &  0.18 &  0.097 &  48( 1)&    I    & 1 \\  Cha1-S5 & \hspace*{-2ex} 7 & 22.7 $\times$ 21.3 &  1.5 $\times$  1.0  &   24 & \hspace*{-2ex} 3.7 & \hspace*{-2ex}  2.7 &  0.022 &  0.024 &  0.024 &  95( 7)&  II  & 2 \\  Cha1-S6 & \hspace*{-2ex} 10 & 21.4 $\times$ 21.2 &  1.2 $\times$  1.1  &   16 & \hspace*{-2ex} 2.5 & \hspace*{-2ex}   11 &  0.015 &  0.015 &  0.013 & 112(15)&  II  & 1 \\  Cha1-S7 & \hspace*{-2ex} 12 & 21.9 $\times$ 21.2 &  1.5 $\times$  1.2  &   14 & \hspace*{-2ex} 2.1 & \hspace*{-2ex}    6 &  0.013 &  0.013 &  0.013 &  99(14)&  Flat   & 1 \\  Cha1-S8 & \hspace*{-2ex} 17 & 22.0 $\times$ 21.2 &  1.7 $\times$  1.4  &   11 & \hspace*{-2ex} 1.7 & \hspace*{-2ex}  3.1 &  0.010 &  0.011 & 0.0098 & 107(19)&   II    & 3 \\  Cha1-S9 & \hspace*{-2ex} 18 & 21.8 $\times$ 21.2 &  1.6 $\times$  1.4  &   11 & \hspace*{-2ex} 1.6 & \hspace*{-2ex}    7 & 0.0098 &  0.010 & 0.0095 & 103(19)&   II    & 1 \\  Cha1-S10 & \hspace*{-2ex} 20 & 21.6 $\times$ 21.2 &  1.6 $\times$  1.5  &  9.9 & \hspace*{-2ex} 1.5 & \hspace*{-2ex}   14 & 0.0091 & 0.0093 &  0.012 &  73(12)&  Flat   & 1 \\  Cha1-S11 & \hspace*{-2ex} 22 & 21.4 $\times$ 21.2 &  1.6 $\times$  1.5  &  9.5 & \hspace*{-2ex} 1.4 & \hspace*{-2ex}  3.2 & 0.0087 & 0.0088 & 0.0078 & 112(25)&    I    & 1 \\  Cha1-S12 & \hspace*{-2ex} 28 & 21.4 $\times$ 21.2 &  1.8 $\times$  1.7  &  7.7 & \hspace*{-2ex} 1.2 & \hspace*{-2ex}    8 & 0.0071 & 0.0072 & 0.0071 &  99(25)&   II    & 1 \\  Cha1-S13 & \hspace*{-2ex} 31 & 35.8 $\times$ 28.5 &  5.2 $\times$  3.6  &  7.7 & \hspace*{-2ex} 1.2 & \hspace*{-2ex}    7 & 0.0071 &  0.016 &  0.014 &  52( 9)&   II    & 1 \\  Cha1-S14 & \hspace*{-2ex} 62 & 21.5 $\times$ 21.2 &  2.0 $\times$  1.9  &  5.9 & \hspace*{-2ex} 0.89 & \hspace*{-2ex}   11 & 0.0054 & 0.0054 &  0.020 &  27( 5)&   II    & 1 \\  Cha1-S15 & \hspace*{-2ex} 63 & 54.5 $\times$ 24.3 &  9.1 $\times$  2.9  &  5.6 & \hspace*{-2ex} 0.85 & \hspace*{-2ex}    8 & 0.0052 &  0.015 & 0.0097 &  53(13)&   II    & 4 \\  Cha1-S16 & \hspace*{-2ex} 77 & 24.8 $\times$ 21.2 &  3.1 $\times$  2.1  &  5.1 & \hspace*{-2ex} 0.78 & \hspace*{-2ex}    6 & 0.0047 & 0.0055 & 0.0053 &  89(31)&   II    & 1 \\  \hline
 \end{tabular}
 \tablefoot{
 \tablefoottext{a}{Numbering of \textit{Gaussclumps} sources like in Table~\ref{t:id_gcl_simbad}.}
 \tablefoottext{b}{Size of the fitted Gaussian and deconvolved source size. The deconvolved size is computed from the fitted size multiplied by ($1+1/SNR$), with $SNR$ the peak signal-to-noise ratio, and therefore is an upper limit in most cases.}
 \tablefoottext{c}{Peak signal-to-noise ratio of the fitted Gaussian.}
 \tablefoottext{d}{Peak H$_2$ column density computed assuming a dust temperature of 20~K and a dust opacity of 0.03~cm$^2$~g$^{-1}$. The statistical rms uncertainty is $1.5 \times 10^{20}$~cm$^{-2}$.}
 \tablefoottext{e}{Visual extinction derived from 2MASS.}
 \tablefoottext{f}{Mass in the central beam ($HPBW = 21.2\arcsec$), total mass derived from the Gaussian fit, and mass computed from the flux measured in an aperture of 50$\arcsec$ in diameter. The statistical rms uncertainties of $M_{\mathrm{peak}}$ and $M_{50\arcsec}$ are 0.0009 and 0.0015 M$_\odot$, respectively.}
 \tablefoottext{g}{Mass concentration $m_{\mathrm{peak}}/m_{50\arcsec}$. The statistical rms uncertainty is given in parentheses.}
 \tablefoottext{h}{Infrared class of the nearest YSO listed in Col.~11 of Table~\ref{t:id_gcl_simbad} and reference: (1) \citet{Luhman08a}; (2) \citet{Furlan09}; (3) \citet{Henning93}; (4) \citet{Luhman08b}.}
 }
\end{table*}

The sources extracted with \textit{Gaussclumps} in the \textit{sum} map at 
scale 5 and associated with known YSOs found in the SIMBAD database are 
listed in Table~\ref{t:ysos}. Their possible SIMBAD associations are listed 
in Table~\ref{t:id_gcl_simbad}. Most of these 16 sources are barely 
resolved. Their deconvolved sizes are computed from the fitted sizes after 
multiplying the latter by $(1+1/SNR)$, with $SNR$ the peak signal-to-noise 
ratio. In this way, the deconvolved sizes are about $1\sigma$ upper limits on 
the true physical sizes of the barely resolved sources. The peak masses, 
total masses, and masses within an aperture of diameter 50$\arcsec$ are 
computed with the same assumptions as for the starless sources, except for the 
dust temperature and mass opacity that we here take as 20~K and 
0.03~cm$^2$~g$^{-1}$, respectively (see Appendix~\ref{s:assumptions}). The
mass concentration $C_M$ is higher 
than $70\%$ for 11 sources ($69\%$), consistent with those sources being 
nearly point-like and dominating their environment within a diameter of 
50$\arcsec$. The remaining 5 sources are contaminated by nearby sources (case 
of Cha1-S2 and S4) or were qualified as ``candidate'' (Sc) in 
Table~\ref{t:id_gcl_simbad} because they are significantly embedded in a 
larger-scale dense core and are difficult to extract from their 
environment (Cha1-S13 to S15).

The same analysis was performed for the 5 additional compact sources found in 
Sect.~\ref{ss:additional} based on a SIMBAD association. Their deconvolved 
sizes, peak column density, and masses are given in Table~\ref{t:motherysos}.
They all have a concentration parameter consistent with 
their being nearly point-like, but the uncertainties are large because they
are faint and close to the detection limit.

\begin{table*}
 \caption{
 Characteristics of the additional compact sources listed in Table~\ref{t:otherysos}.
 }
 \label{t:motherysos}
 \centering
 \begin{tabular}{ccccccccccccc}
 \hline\hline
  \multicolumn{1}{c}{Name} & \multicolumn{1}{c}{\hspace*{-2ex} ${N_{\mathrm{add}}}$\tablefootmark{a}} & \multicolumn{1}{c}{$FWHM$\tablefootmark{b}} & \multicolumn{1}{c}{$FWHM$\tablefootmark{b}} & \multicolumn{1}{c}{$SNR$\tablefootmark{c}} & \multicolumn{1}{c}{\hspace*{-2ex} ${N_{\mathrm{peak}}}$\tablefootmark{d}} & \multicolumn{1}{c}{\hspace*{-2ex} ${A_V}$\tablefootmark{e}} & \multicolumn{1}{c}{${M_{\mathrm{peak}}}$\tablefootmark{f}} & \multicolumn{1}{c}{${M_{\mathrm{tot}}}$\tablefootmark{f}} & \multicolumn{1}{c}{${M_{50\arcsec}}$\tablefootmark{f}} & \multicolumn{1}{c}{${C_M}$\tablefootmark{g}} & \multicolumn{1}{c}{Class\tablefootmark{h}} & \multicolumn{1}{c}{Ref.\tablefootmark{h}} \\ 
  & \hspace*{-2ex} & \multicolumn{1}{c}{($\arcsec \times \arcsec$)} & \multicolumn{1}{c}{(1000 AU)$^2$} & & \multicolumn{1}{c}{\hspace*{-2ex} ($10^{21}$ cm$^{-2}$)} & \multicolumn{1}{c}{\hspace*{-2ex} (mag)} & \multicolumn{1}{c}{(M$_\odot$)} & \multicolumn{1}{c}{(M$_\odot$)} & \multicolumn{1}{c}{(M$_\odot$)} &  \multicolumn{1}{c}{($\%$)} & & \\ 
 \multicolumn{1}{c}{(1)} & \multicolumn{1}{c}{\hspace*{-2ex} (2)} & \multicolumn{1}{c}{(3)} & \multicolumn{1}{c}{(4)} & \multicolumn{1}{c}{(5)} & \multicolumn{1}{c}{\hspace*{-2ex} (6)} & \multicolumn{1}{c}{\hspace*{-2ex} (7)} & \multicolumn{1}{c}{(8)} & \multicolumn{1}{c}{(9)} & \multicolumn{1}{c}{(10)} & \multicolumn{1}{c}{(11)} & \multicolumn{1}{c}{(12)} & \multicolumn{1}{c}{(13)} \\ 
 \hline
 Cha1-S17 & \hspace*{-2ex} 1 & 24.4 $\times$ 20.7 &  3.1 $\times$  2.1  &  4.6 & \hspace*{-2ex} 0.69 & \hspace*{-2ex}  2.7 & 0.0042 & 0.0047 & 0.0048 &  87(34)&   II    & 2 \\  Cha1-S18 & \hspace*{-2ex} 2 & 56.9 $\times$ 18.6 & 11.3 $\times$  2.1  &  2.7 & \hspace*{-2ex} 0.40 & \hspace*{-2ex}  2.3 & 0.0024 & 0.0057 & 0.0039 &  63(35)&   II    & 2 \\  Cha1-S19 & \hspace*{-2ex} 3 & 23.2 $\times$ 18.4 &  3.0 $\times$  1.4  &  3.9 & \hspace*{-2ex} 0.59 & \hspace*{-2ex}    6 & 0.0036 & 0.0034 & 0.0030 & 118(66)&   II    & 1 \\  Cha1-S20 & \hspace*{-2ex} 4 & 20.4 $\times$ 16.6 &  1.9 $\times$  0.0  &  4.9 & \hspace*{-2ex} 0.74 & \hspace*{-2ex}  2.9 & 0.0045 & 0.0034 & 0.0030 & 149(82)&   II    & 1 \\  Cha1-S21 & \hspace*{-2ex} 5 & 18.6 $\times$ 16.3 &  1.0 $\times$  0.0  &  5.1 & \hspace*{-2ex} 0.77 & \hspace*{-2ex}  2.2 & 0.0047 & 0.0031 & 0.0022 & 211(152)&   II    & 1 \\  \hline
 \end{tabular}
 \tablefoot{
 \tablefoottext{a}{Numbering of additional sources like in Table~\ref{t:otherysos}.}
 \tablefoottext{b}{Size of the fitted Gaussian and deconvolved source size. The deconvolved size is computed from the fitted size multiplied by ($1+1/SNR$), with $SNR$ the peak signal-to-noise ratio, and therefore is an upper limit in most cases. A zero value means that the uncertainty was still underestimated.}
 \tablefoottext{c}{Peak signal-to-noise ratio of the fitted Gaussian.}
 \tablefoottext{d}{Peak H$_2$ column density computed assuming a dust temperature of 20~K and a dust opacity of 0.03~cm$^2$~g$^{-1}$. The statistical rms uncertainty is $1.5 \times 10^{20}$~cm$^{-2}$.}
 \tablefoottext{e}{Visual extinction derived from 2MASS.}
 \tablefoottext{f}{Mass in the central beam ($HPBW = 21.2\arcsec$), total mass derived from the Gaussian fit, and mass computed from the flux measured in an aperture of 50$\arcsec$ in diameter. The statistical rms uncertainties of $M_{\mathrm{peak}}$ and $M_{50\arcsec}$ are 0.0009 and 0.0015 M$_\odot$, respectively.}
 \tablefoottext{g}{Mass concentration $m_{\mathrm{peak}}/m_{50\arcsec}$. The statistical rms uncertainty is given in parentheses.}
 \tablefoottext{h}{Infrared class of the nearest YSO listed in Col. 10 of Table~\ref{t:otherysos} and reference: (1) \citet{Luhman08a}; (2) \citet{Furlan09}.}
 }
\end{table*}

In total, 21 compact sources associated with a YSO in the SIMBAD database are 
detected above the $3.5\sigma$ level (42 mJy/21.2$\arcsec$-beam) in the 
870~$\mu$m continuum map of Cha~I. This represents about 10$\%$ of the known
YSO members of Cha~I \citep[237 YSOs, see][]{Luhman08c}. The infrared class 
of the nearest YSO associated to each compact source is given 
in Tables~\ref{t:ysos} and \ref{t:motherysos}. There are 3 Class~I, 
2 ``Flat'', and 16 Class~II sources. This represents a 870~$\mu$m detection 
rate of 75$\%$, 20$\%$, and 17$\%$ for the respective infrared classes, based 
on the census of \citet{Luhman08c} (see Table~\ref{t:census}). Two of the
detected Class~II sources are transitional disks, i.e. objects for which the 
spectral energy distribution indicates central clearings or gaps in the dust 
distribution \citep[Cha-S5 and S6, see][]{Kim09}. The missing Class~I YSO not 
listed in Tables~\ref{t:ysos} and \ref{t:motherysos} 
\citep[2M~J11092855-7633281 in][]{Luhman08c} is located about 8$\arcsec$ east 
of Cha-C12. Given that Cha-C12 is extended and does 
not show any peak at the position of the YSO, we did not consider the 
870~$\mu$m emission as being primarily associated with the YSO itself. We can 
however not exclude that part of this emission comes from the disk and/or the 
residual envelope of this YSO.

\subsection{Large scale structures}
\label{ss:largescale}

\begin{table*}
 \caption{
 Large-scale filamentary structures in Cha~I probed with LABOCA at 870~$\mu$m (see Fig.~\ref{f:mrmed}b).
 }
 \label{t:clf_ls}
 \centering
 \begin{tabular}{ccccccccccccccccc}
 \hline\hline
 \multicolumn{1}{c}{$N_{\mathrm{fil}}$} & \multicolumn{1}{c}{\hspace*{-2ex} R.A.\tablefootmark{a}} & \multicolumn{1}{c}{\hspace*{-1ex} Decl.\tablefootmark{a}} & \multicolumn{1}{c}{$f_{\mathrm{peak}}$\tablefootmark{a}} & \multicolumn{1}{c}{\hspace*{-2ex} $f_{\mathrm{mean}}$\tablefootmark{b}} & \multicolumn{1}{c}{\hspace*{-2ex} $f_{\mathrm{tot}}$\tablefootmark{b}} & \multicolumn{1}{c}{$A$\tablefootmark{c}} & \multicolumn{1}{c}{$\theta_{\mathrm{maj}}$\tablefootmark{d}} & \multicolumn{1}{c}{\hspace*{-2ex} $\theta_{\mathrm{min}}$\tablefootmark{d}} & \multicolumn{1}{c}{\hspace*{-1ex} $R_{\mathrm{a}}$\tablefootmark{e}} & \multicolumn{1}{c}{\hspace*{-1ex} ${N_{\mathrm{peak}}}$\tablefootmark{f}} & \multicolumn{1}{c}{\hspace*{-3ex} ${N_{\mathrm{mean}}}$\tablefootmark{f}} & \multicolumn{1}{c}{\hspace*{-2ex} $M$\tablefootmark{g}} & \multicolumn{1}{c}{$a_{\mathrm{maj}}$\tablefootmark{h}} & \multicolumn{1}{c}{\hspace*{-1ex} $a_{\mathrm{min}}$\tablefootmark{h}} & \multicolumn{1}{c}{${M_{\mathrm{tot}}}$\tablefootmark{i}} & \multicolumn{1}{c}{\hspace*{-2ex} ${M_{\mathrm{lin}}}$\tablefootmark{j}} \\ 
  & \multicolumn{1}{c}{\hspace*{-2ex} \scriptsize (J2000)} & \multicolumn{1}{c}{\hspace*{-1ex} \scriptsize (J2000)} & \multicolumn{2}{c}{\scriptsize (mJy/21$\arcsec$-beam)} & \multicolumn{1}{c}{\hspace*{-2ex} \scriptsize (Jy)} & \multicolumn{1}{c}{\scriptsize ($\arcmin \times \arcmin$)} & \multicolumn{1}{c}{\scriptsize ($\arcmin$)} & \multicolumn{1}{c}{\hspace*{-2ex} \scriptsize ($\arcmin$)} & & \multicolumn{2}{c}{\hspace*{-1ex} \scriptsize{($10^{21}$ cm$^{-2}$)}} & \multicolumn{1}{c}{\hspace*{-2ex} \scriptsize{(M$_\odot$)}} & \multicolumn{1}{c}{\scriptsize (pc)} & \multicolumn{1}{c}{\hspace*{-1ex} \scriptsize (pc)} & \multicolumn{1}{c}{\scriptsize{(M$_\odot$)}} & \multicolumn{1}{c}{\hspace*{-2ex} \scriptsize (M$_\odot/$pc)} \\ 
 \multicolumn{1}{c}{(1)} & \multicolumn{1}{c}{\hspace*{-2ex} (2)} & \multicolumn{1}{c}{\hspace*{-1ex} (3)} & \multicolumn{1}{c}{(4)} & \multicolumn{1}{c}{\hspace*{-2ex} (5)} & \multicolumn{1}{c}{\hspace*{-2ex} (6)} & \multicolumn{1}{c}{(7)} & \multicolumn{1}{c}{(8)} & \multicolumn{1}{c}{\hspace*{-2ex} (9)} & \multicolumn{1}{c}{\hspace*{-1ex} (10)} & \multicolumn{1}{c}{\hspace*{-1ex} (11)} & \multicolumn{1}{c}{\hspace*{-3ex} (12)} & \multicolumn{1}{c}{\hspace*{-2ex} (13)} & \multicolumn{1}{c}{(14)} & \multicolumn{1}{c}{\hspace*{-1ex} (15)} & \multicolumn{1}{c}{\hspace*{-2ex}(16)} & \multicolumn{1}{c}{\hspace*{-2ex}(17)} \\ 
 \hline
  1 & \hspace*{-2ex} 11:06:38.57 & \hspace*{-1ex} -77:23:48.4 &   200 & \hspace*{-2ex}    55 & \hspace*{-2ex}  18.2 &  46.6 &  13.0 & \hspace*{-2ex}   4.2 & \hspace*{-1ex}   3.1 & \hspace*{-1ex} 17.5 & \hspace*{-3ex}  4.9 & \hspace*{-2ex}   9.7 &  0.57 & \hspace*{-1ex}  0.18 & \hspace*{-2ex} 19.0  & \hspace*{-2ex}   33 \\ 
   2 & \hspace*{-2ex} 11:10:00.10 & \hspace*{-1ex} -76:35:37.2 &   116 & \hspace*{-2ex}    36 & \hspace*{-2ex}  20.2 &  79.4 &  22.2 & \hspace*{-2ex}   3.9 & \hspace*{-1ex}   5.7 & \hspace*{-1ex} 10.2 & \hspace*{-3ex}  3.2 & \hspace*{-2ex}  10.7 &  0.97 & \hspace*{-1ex}  0.17 & \hspace*{-2ex} 16.0  & \hspace*{-2ex}   17 \\ 
   3 & \hspace*{-2ex} 11:07:46.92 & \hspace*{-1ex} -77:38:04.9 &    73 & \hspace*{-2ex}    24 & \hspace*{-2ex}   7.8 &  45.2 &  10.5 & \hspace*{-2ex}   4.9 & \hspace*{-1ex}   2.1 & \hspace*{-1ex}  6.4 & \hspace*{-3ex}  2.1 & \hspace*{-2ex}   4.1 &  0.46 & \hspace*{-1ex}  0.22 & \hspace*{-2ex}  6.6  & \hspace*{-2ex}   14 \\ 
   4\tablefootmark{k} & \hspace*{-2ex} 11:02:27.42 & \hspace*{-1ex} -77:38:59.6 &    51 & \hspace*{-2ex}    23 & \hspace*{-2ex}  10.8 &  67.0 &  10.2 & \hspace*{-2ex}   4.4 & \hspace*{-1ex}   2.3 & \hspace*{-1ex}  4.5 & \hspace*{-3ex}  2.0 & \hspace*{-2ex}   5.7 &  0.45 & \hspace*{-1ex}  0.19 & \hspace*{-2ex}  9.1  & \hspace*{-2ex}    8 \\ 
    4\tablefootmark{k} & \hspace*{-2ex} -- & \hspace*{-1ex} -- & -- & \hspace*{-2ex} -- & \hspace*{-2ex} -- & --  &  16.5 & \hspace*{-2ex}   2.3 & \hspace*{-1ex}   7.3 & \hspace*{-1ex} -- & \hspace*{-3ex} -- & \hspace*{-2ex} --  &  0.72 & \hspace*{-1ex}  0.10 & \hspace*{-2ex} -- & \hspace*{-2ex} -- \\ 
   5 & \hspace*{-2ex} 11:04:22.92 & \hspace*{-1ex} -77:46:59.4 &    44 & \hspace*{-2ex}    21 & \hspace*{-2ex}   2.2 &  14.5 &   5.9 & \hspace*{-2ex}   2.7 & \hspace*{-1ex}   2.1 & \hspace*{-1ex}  3.9 & \hspace*{-3ex}  1.8 & \hspace*{-2ex}   1.1 &  0.26 & \hspace*{-1ex}  0.12 & \hspace*{-2ex}  2.6  & \hspace*{-2ex}   10 \\ 
  \hline 
 \end{tabular}
 \tablefoot{
 \tablefoottext{a}{Peak position and flux density measured with \textit{Clumpfind}.}
 \tablefoottext{b}{Mean and total flux densities inside the contour at 10 mJy/21$\arcsec$-beam (about $3.3\sigma$) -- hereafter called contour C.}
 \tablefoottext{c}{Angular area of contour C.}
 \tablefoottext{d}{Major and minor angular axes of contour C measured by hand.}
 \tablefoottext{e}{Aspect ratio.}
 \tablefoottext{f}{Peak and mean H$_2$ column density.}
 \tablefoottext{g}{Mass within contour C in smoothed map at scale 5.}
 \tablefoottext{h}{Physical major and minor axes of contour C.}
 \tablefoottext{i}{Mass within contour C in full LABOCA map.}
 \tablefoottext{j}{Mass per unit length $M_{\mathrm{tot}}/a_{\mathrm{maj}}$.}
 \tablefoottext{k}{This structure has a T shape and consists of two filaments that were identified as a single structure by \textit{Clumpfind}. The major and minor axes were measured for each filament separately, but all other parameters correspond to the entire structure.}
 }
 \end{table*}

The smoothed map at scale 5 clearly shows the two main filamentary structures 
in fields Cha-Center and Cha-North already mentioned in 
Sect.~\ref{ss:labocamap} (see Fig.~\ref{f:mrmed}b). It also reveals two 
fainter filaments that seem to be connected with a \textit{T}-shape in the 
south-western part of field Cha-Center. Two other faint, elongated, 
large-scale structures are present in fields Cha-Center and Cha-South, 
respectively. These filamentary structures were already noticed by 
\citet{Haikala05} in their C$^{18}$O~1--0 map where they appear to be broader,
certainly because their map is more sensitive to lower densities than our
870~$\mu$m dust emission map. These authors mentioned that the C$^{18}$O 
filaments are oriented at roughly right angles to the magnetic field (see their 
Fig.~8).

The properties of the filamentary structures detected with LABOCA were 
obtained with \textit{Clumpfind}. Since we 
are here interested in the largest structures, only one contour level at 
10~~mJy/21$\arcsec$-beam (about 3.3$\sigma$ in the smoothed map) was defined
to run \textit{Clumpfind}. In this way, these filaments were not splitted 
into sub-structures. \textit{Clumpfind} extracted 11 structures. Only
those with a peak flux density higher than 40~mJy/beam were selected.
Their properties are listed in Table~\ref{t:clf_ls} and they are labeled in
Fig.~\ref{f:mrmed}b accordingly. The major and minor sizes given in 
Table~\ref{t:clf_ls} correspond to the full extent of each filament measured
by hand using the contour at 10 mJy/beam.

The length of the filaments varies between 6$\arcmin$ and 22$\arcmin$ (0.25 
and 1 pc) and their width between 2.3$\arcmin$ and 4.9$\arcmin$ (0.1 
and 0.2 pc). Their aspect ratio ranges from 2 to 7. The Monte Carlo 
simulations performed in Appendix~\ref{ss:filtering} (see 
Table~\ref{t:filtering}) show that weak elliptical sources with a minor FWHM 
larger than 2.0$\arcmin$, i.e. a full extent along the minor axis larger than 
$\sim 4.0\arcmin$, lost more than 15$\%$ of their peak flux density and size 
because of the correlated noise removal of the data reduction. Filaments 1, 3, 
and the widest filament of the T-shaped structure 4 may therefore be affected 
by this spatial filtering and their intrinsic widths may be somewhat larger 
than measured here. On the other hand, the measured width of filaments 2, 5, 
and of the thinnest filament of structure 4 likely corresponds to 
their actual minor size. The total mass of these filaments was measured in the 
full map (Fig.~\ref{f:labocamap}) within the contour defined above in the 
smoothed map at scale 5. They are listed in Col.~16 of Table~\ref{t:clf_ls}, 
as well as the corresponding mass per unit length along the major axis in 
Col.~17.

With a total flux in the smoothed map at scale 5 of 59 Jy, corresponding to a 
mass of 31 M$_\odot$, these filaments contribute to 52$\%$ of the total 
870~$\mu$m flux measured in Cha~I. Their mean H$_2$ column density 
is relatively low, varying from 2 to $5 \times 10^{21}$ cm$^{-2}$, equivalent 
to a visual extinction $A_V$ from 2 to 5 mag. When the mass included in the
smaller scales is added (i.e. in the \textit{sum} map at scale 5), the total 
mass of these filaments becomes 53~M$_\odot$, i.e. about 88$\%$ of the total 
mass traced with LABOCA in Cha~I.

\subsection{Comparison with extinction and C$^{18}$O 1-0 emission}
\label{ss:comp_avc18o}

\subsubsection{Correlation between continuum emission and $A_V$}
\label{sss:av_cont}

\begin{figure}
\centerline{\resizebox{1.00\hsize}{!}{\includegraphics[angle=270]{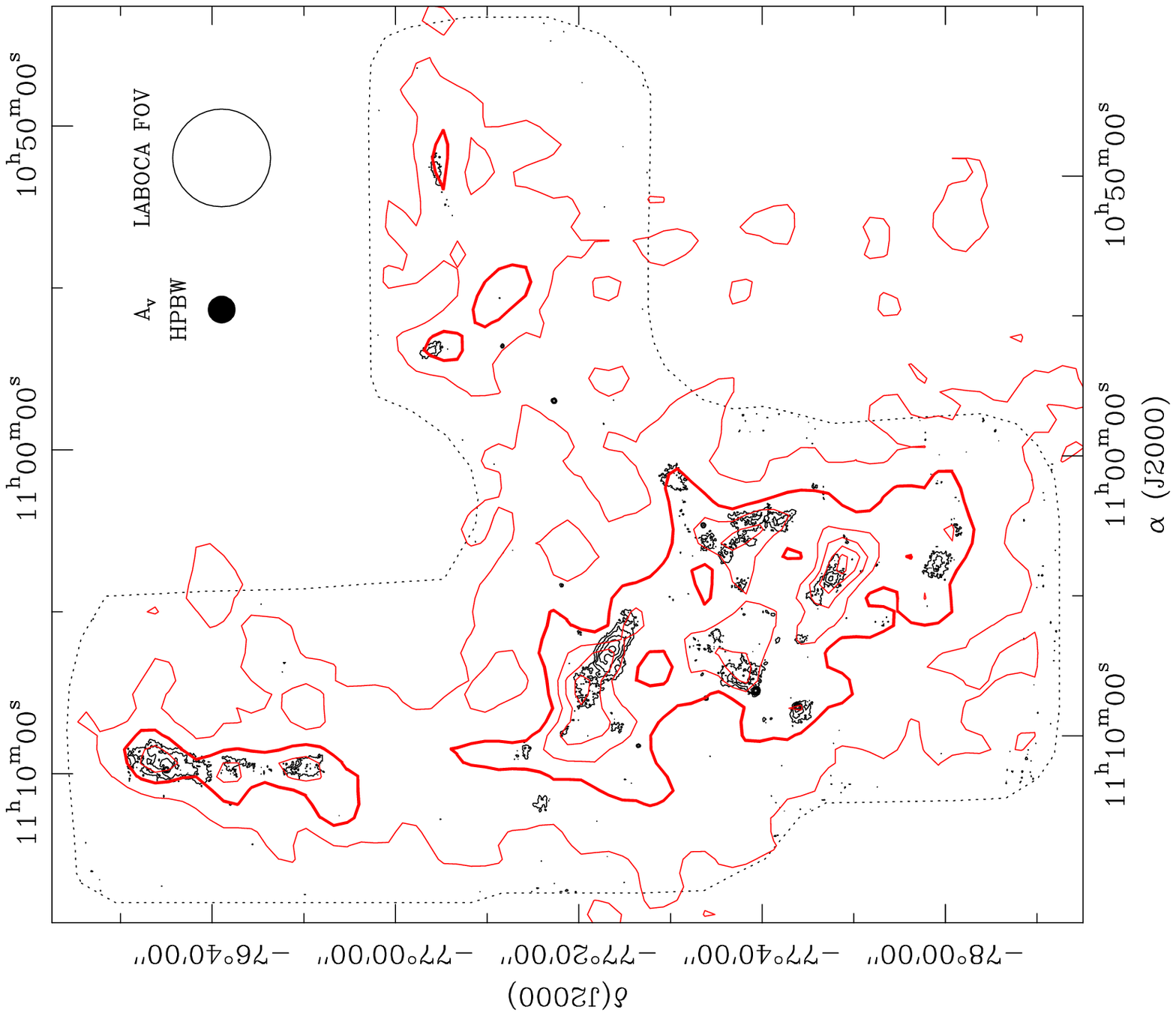}}}
\caption{Extinction map of Fig.~\ref{f:av} (red contours) overlaid on the 
870~$\mu$m continuum emission map of Cha~I (black contours). The contour levels 
of the extinction map increase from 3 to 18~mag by step of 3~mag. The thicker 
red contour corresponds to $A_V = 6$~mag. The contour levels of the 870~$\mu$m 
map are the same as in Fig.~\ref{f:labocamap}. The dotted line 
delimits the field mapped at 870~$\mu$m. The field of view of LABOCA and the
angular resolution of the extinction map are shown in the upper right corner.}
\label{f:avcont}
\end{figure}

The 870~$\mu$m dust continuum emission map of Cha~I and the extinction map 
derived from 2MASS are overlaid in Fig.~\ref{f:avcont}. The overall 
correspondence is relatively good, especially in field 
Cha-North. In the other fields, the extinction peaks are somewhat
shifted compared to the 870~$\mu$m peaks, but these shifts ($< 2\arcmin$) are 
smaller than the angular resolution of the extinction map and may not be 
significant. The only exception is close to Cha-MMS1 (C1) in field 
Cha-Center, where the extinction map peaks about 5$\arcmin$ north-east of the 
870~$\mu$m peak. This offset still remains when the 870~$\mu$m map is 
smoothed to 3$\arcmin$,  showing that it does not result from the different 
angular resolutions of the original data. An explanation could be that the 
extinction map does not trace well the H$_2$ column density in this 
high-extinction 
region. The average intrinsic colors of the small number of background stars 
detected in each resolution element of this part of the map may deviate 
significantly from the assumed average intrinsic colors (Sect.~\ref{ss:2mass}). 
Another contributing effect could be the spatial filtering of the 
870~$\mu$m data. The extinction map (and the C$^{18}$O~1--0 map, see 
Sect.~\ref{sss:c18o}) suggests a structure more extended toward the north-east
than traced by LABOCA. This missing extended structure could bias the peak 
position of the 3$\arcmin$-smoothed 870~$\mu$m map toward the south-west.

The $4\sigma$ H$_2$ column density sensitivity limit of the 
870~$\mu$m map is $4.2 \times 10^{21}$~cm$^{-2}$, which corresponds to 
$A_V \sim 4.5$~mag. Most of the continuum emission detected at 
870~$\mu$m is above the contour level $A_V = 9$~mag, and most extended 
regions with $4.5 < A_V < 9$~mag traced by the extinction map are not detected
at 870~$\mu$m. This is due to the spatial filtering related to the 
sky noise removal (see Appendix~\ref{ss:filtering}) since these low-extinction 
regions have sizes on the order of 10$\arcmin$--20$\arcmin$, larger than the 
field of view of LABOCA, especially in fields Cha-Center and Cha-South. On the 
other hand, there are also a few 870~$\mu$m clumps detected between 4.5 and 
9~mag that are not seen in the extinction map, most likely because of its poor 
angular resolution (e.g. Cha1-C13 and C22 in the southern part and Cha1-C20 
and C45 in the eastern part). A few YSO candidates are even found below 3~mag 
(e.g. Cha1-S5, S17, and S18 in field Cha-West, Cha1-S20 and S21 in field 
Cha-North).

\subsubsection{Correlation between C$^{18}$O 1--0 emission and $A_V$}
\label{sss:c18o}

\begin{figure}
\centerline{\resizebox{\hsize}{!}{\includegraphics[angle=270]{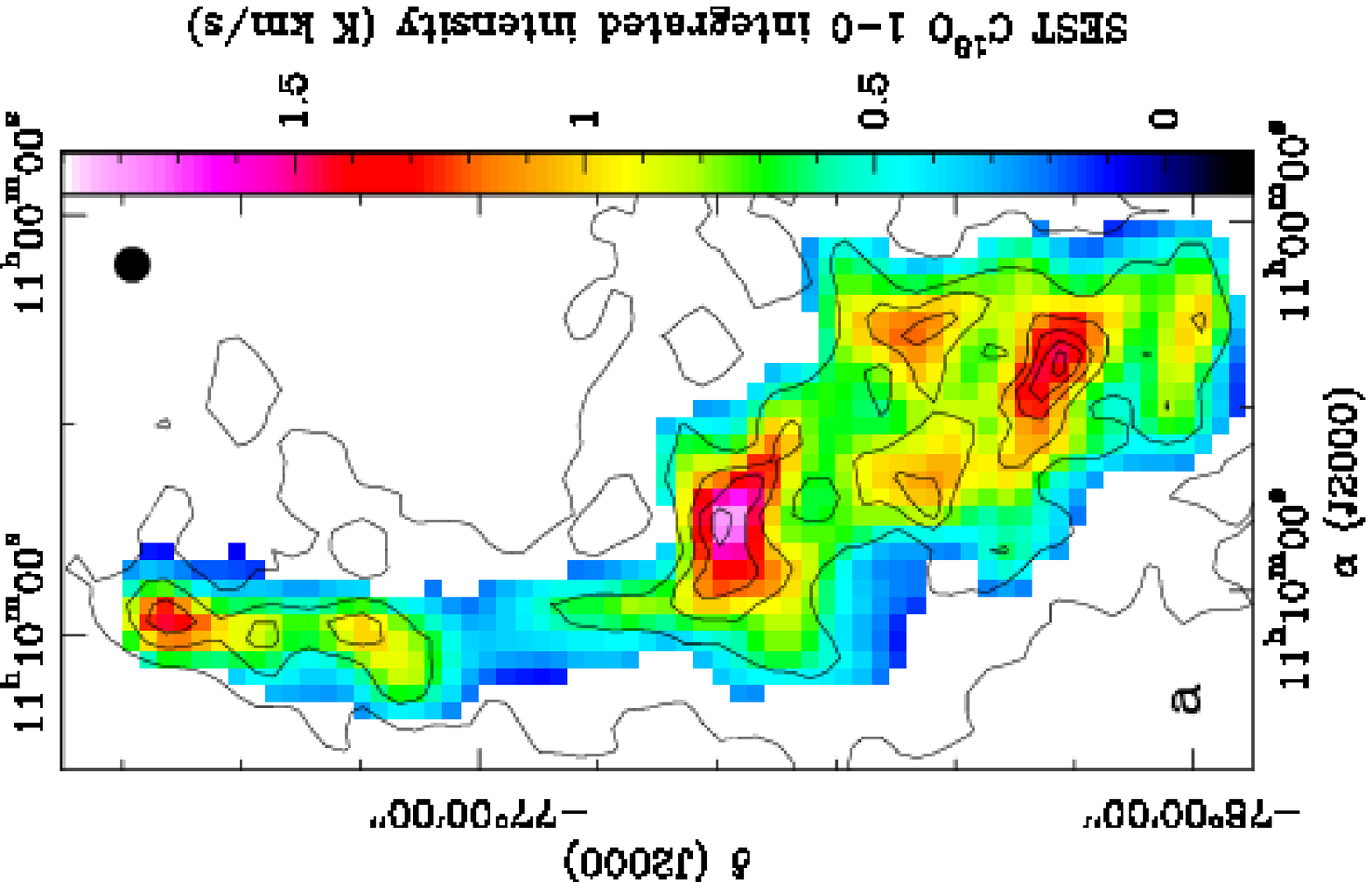}\hspace*{3ex}\includegraphics[angle=270]{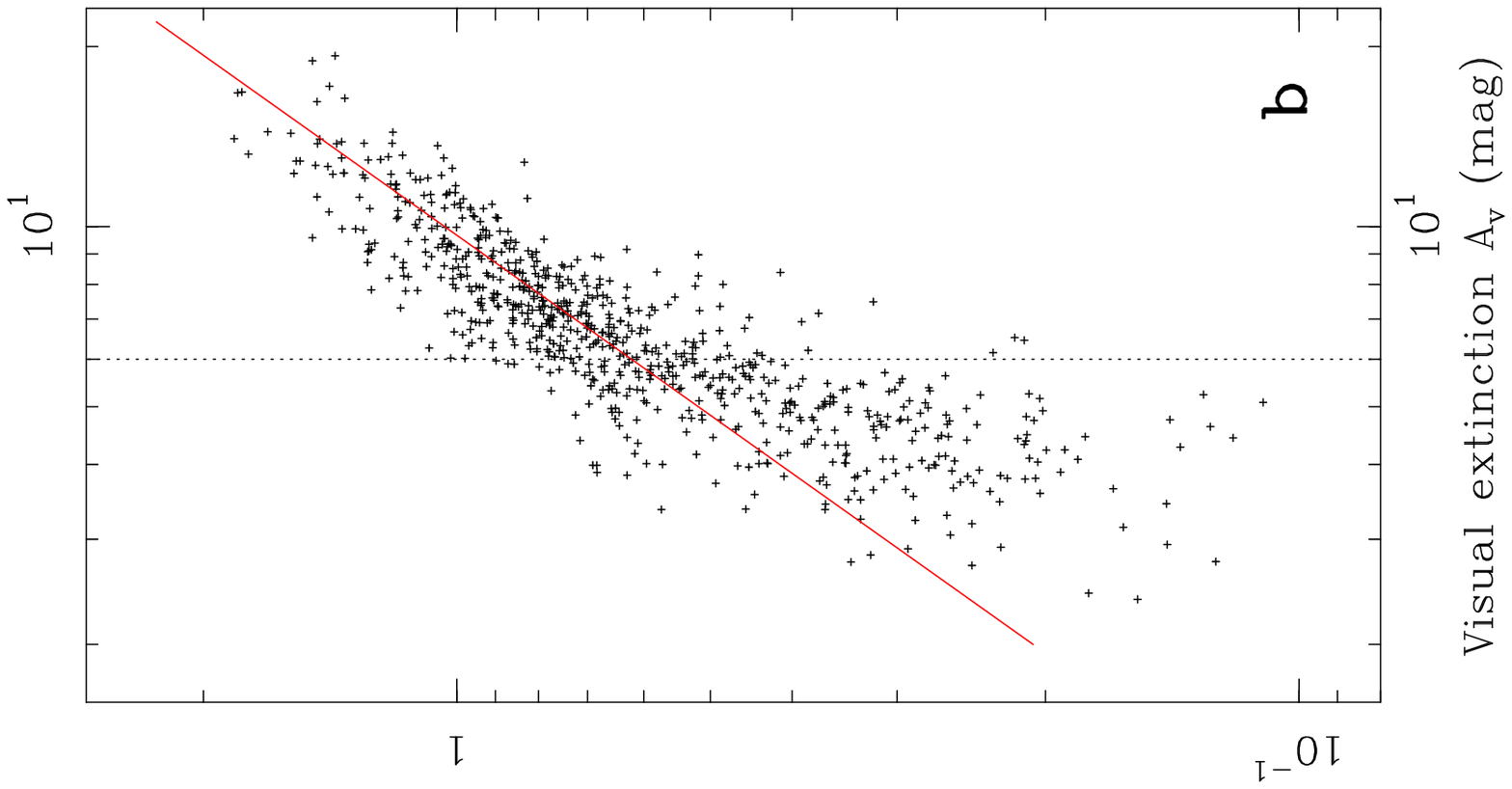}}}
\caption{\textbf{a} Integrated intensity map of C$^{18}$O 1--0 emission 
detected with the SEST by \citet{Haikala05} in $T_a^\star$ scale, smoothed to 
an effective angular resolution of 3$\arcmin$ (filled circle). The extinction 
map of Fig.~\ref{f:av} is overlaid as contours that increase from 3 to 
18~mag by step of 3 mag.
\textbf{b} Correlation between the C$^{18}$O 1--0 integrated intensity and the
visual extinction. Only data points with a C$^{18}$O 1--0 integrated intensity 
higher than 0.1 K~km~s$^{-1}$ ($>3\sigma$ for 90$\%$ of the map) and 
$A_V > 2$~mag are plotted. The dotted line marks the extinction $A_V = 6$~mag 
above which a linear fit with a fixed slope of 1 (in log-log space) was 
performed. The red line shows the result of this fit. Note that all visual 
extinction values where derived assuming the standard extinction law and have 
to be multiplied by 0.85 for $A_V$ above $\sim 6$~mag (see 
Appendix~\ref{ss:avtoN}).}
\label{f:c18oav}
\end{figure}

The extinction map derived from 2MASS is compared with the C$^{18}$O~1--0 map 
obtained by \citet{Haikala05} with the SEST in Fig.~\ref{f:c18oav}a. We 
smoothed the C$^{18}$O~1--0 map to an effective angular resolution of 
3$\arcmin$ and a pixel size of 1.5$\arcmin$, to match those 
of the extinction map. The correlation between the two tracers looks very 
good above $A_V = 6$~mag where the C$^{18}$O~1--0 
integrated intensity is proportional to the visual extinction in good 
approximation (see Fig.~\ref{f:c18oav}b). The high-extinction end 
($A_V > 15$~mag) is however not well constrained.

At low extinction ($A_V < 6$ mag), the C$^{18}$O~1--0 integrated intensity 
drops steeply with decreasing extinction. This sharp decrease is most 
likely due to photodissociation. The photodissociation models of 
\citet{Visser09} show such a steep C$^{18}$O column density variation in the 
range of extinction they investigated (slope $\sim 2$ to 7 in a log-log diagram
for $A_V < 5$~mag, depending on the density and the UV field). The extension of 
their model to $A_V \sim 15$~mag (Visser 2010, 
\textit{priv. comm.}) shows that for a density 
$n_H = 300$~cm$^{-3}$, the turnover occurs at $A_V \sim 4-5$~mag for a 
UV-field scaling factor of $1-10$ and a gas temperature of 15~K, while it 
occurs at $A_V \sim 2-3$~mag for $n_H = 1000$~cm$^{-3}$ and at 
$A_V \sim 7-8$~mag for $n_H = 100$~cm$^{-3}$. The density of free particles 
corresponding to the median extinction in Cha~I is $\sim 350$~cm$^{-3}$ (see 
Sect.~\ref{ss:cha1masses}), i.e. $n_H \sim 590$~cm$^{-3}$. This density, which 
should characterize the density close to the edges of the cloud better than the
mean density, should imply a turnover between $A_V \sim 2-3$ and $4-5$ mag 
according to the models of \citeauthor{Visser09}. A reason why the turnover in 
Cha~I occurs at a higher level ($A_V \sim 6$~mag\footnote{Since this 
visual extinction is about the threshold above which $R_V$ in Cha~I is 
thought to have a value higher than the standard one (see 
Appendix~\ref{ss:avtoN}), we may have to apply a correction to this visual 
extinction that was computed assuming the standard extinction law. With 
$R_V = 5.5$, the true visual extinction would become $A_V \sim 5$~mag instead 
of 6, still high compared to the model predictions.}) may be that the 
low-density medium is inhomogeneous and the UV radiation can penetrate deeper 
in the cloud than predicted by the homogeneous model of Visser et al..

It follows from the previous paragraphs that the C$^{18}$O~1--0 emission does 
not trace well the cloud mass at $A_V < 6$~mag but that it is well correlated 
with the extinction between $A_V = 6$ and $\sim$15~mag and should 
thus be a reliable tracer of mass in this range of extinction. We would then 
expect C$^{18}$O~1--0 to trace about the same mass as the extinction map above 
$A_V = 6$~mag (220~M$_\odot$, see Sect.~\ref{ss:cha1masses}). 
\citet{Haikala05} derived a mass of 230~M$_\odot$ from the C$^{18}$O~1--0 map 
shown in Fig.~\ref{f:c18ocont}a. This map does not contain field Cha-West. 
This field was covered by \citet{Mizuno99} and its total C$^{18}$O~1--0 
integrated intensity represents an additional contribution of 15$\%$. The mass 
traced by C$^{18}$O~1--0 in Cha~I is thus about 260~M$_\odot$. Given the 
uncertainties on the C$^{18}$O abundance, the excitation temperature, and the 
$A_V$-to-mass conversion, this is in very good agreement 
with the mass derived 
from the extinction map above $A_V = 6$~mag.

\subsubsection{Comparison of continuum emission with C$^{18}$O 1--0 emission}
\label{sss:contc18o}

\begin{figure*}
\centerline{\resizebox{1.00\hsize}{!}{\includegraphics[angle=270]{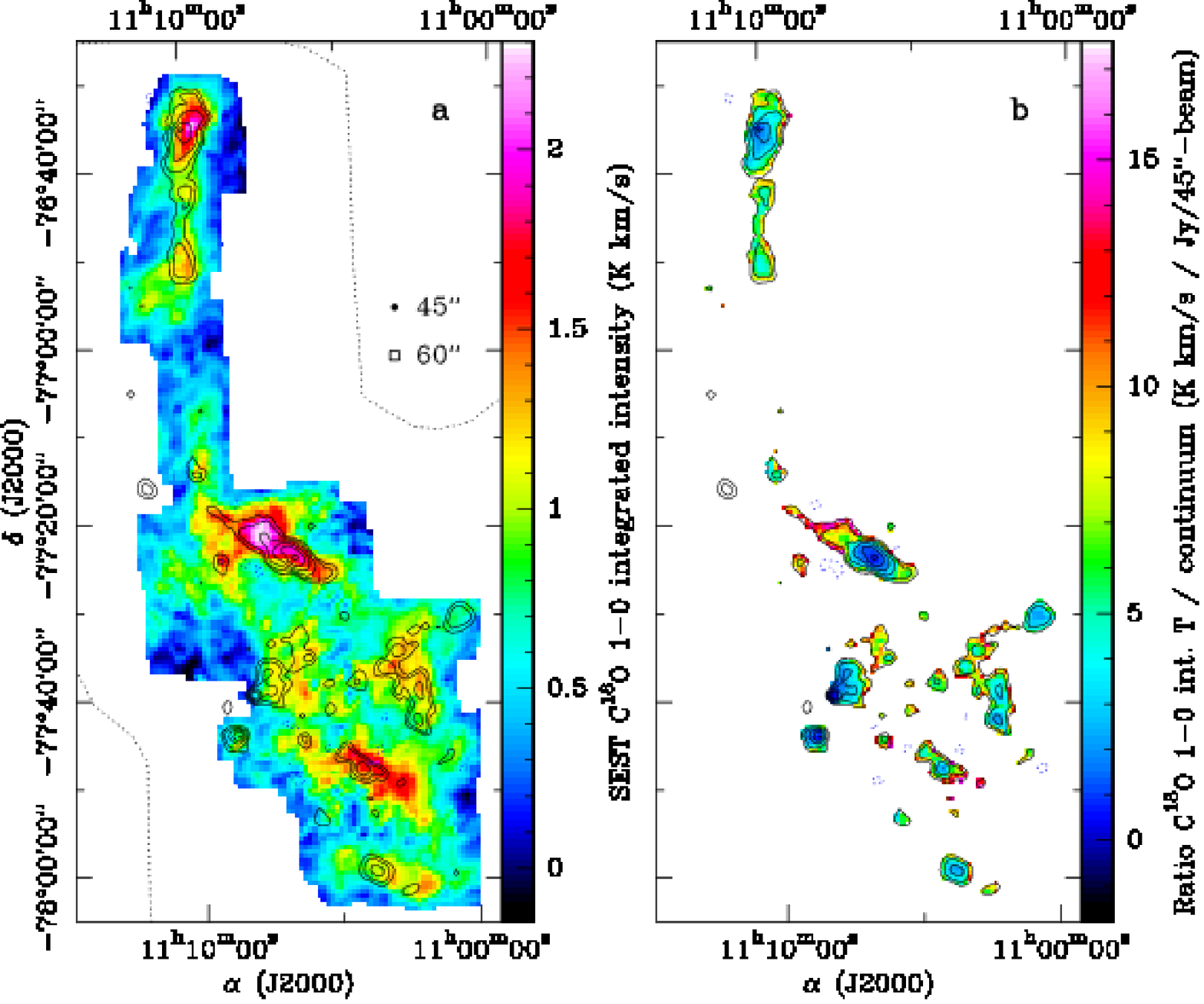}\hspace*{5ex}\includegraphics[angle=270]{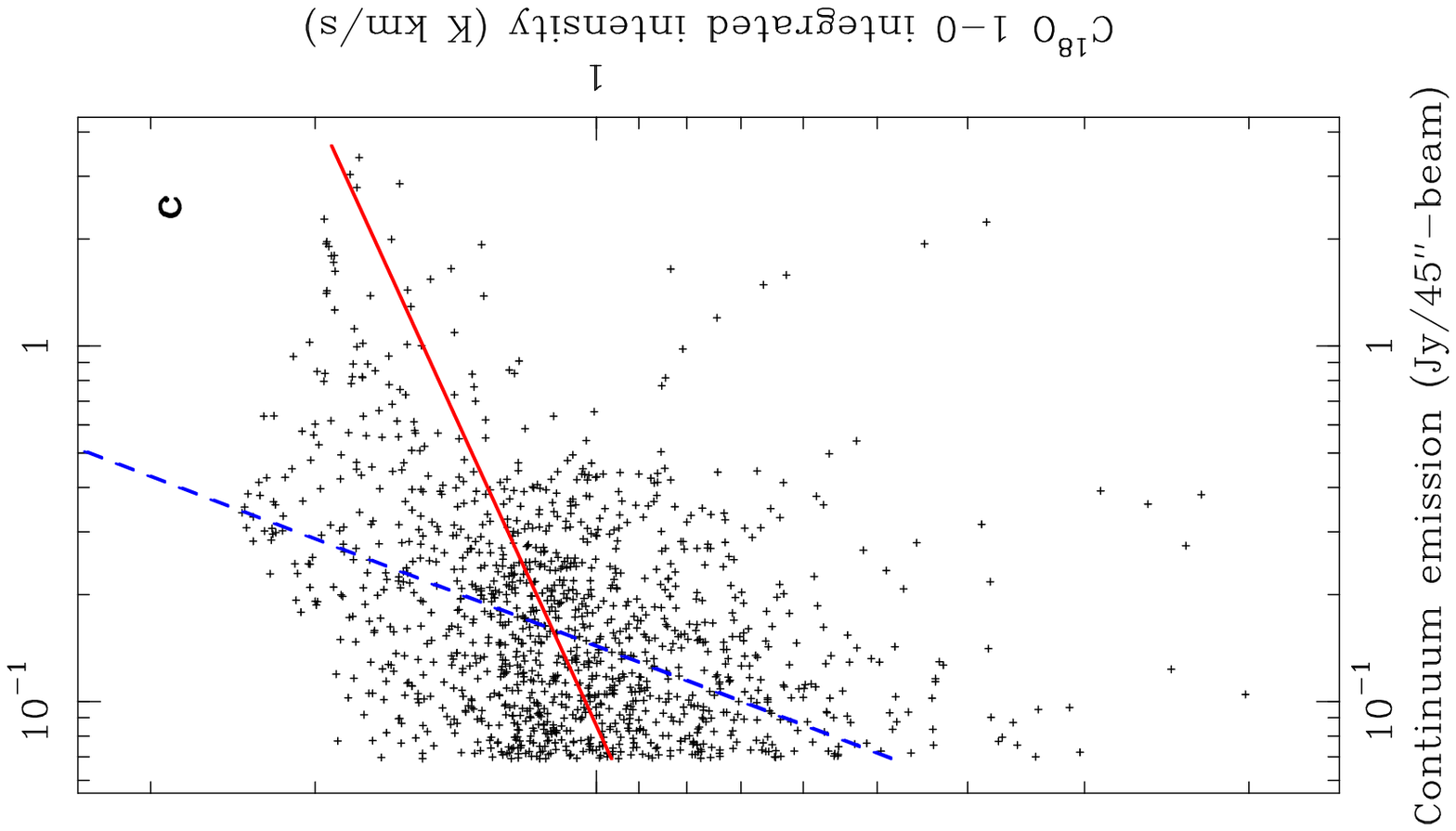}}}
\caption{\textbf{a} Integrated intensity map of C$^{18}$O 1--0 emission 
detected with the SEST by \citet{Haikala05} in $T_a^\star$ scale. The 
870~$\mu$m continuum map smoothed to 45$\arcsec$ is overlaid as contours. The 
contours are  $-f$ (dashed), $f$, $2f$, $4f$, $8f$, $16f$, and $32f$, with 
$f = 4\sigma$ and $\sigma$ the rms noise level (23~mJy/45$\arcsec$-beam). The 
dotted line delimits the field mapped at 870~$\mu$m. The filled circle shows 
the SEST angular resolution, and the square the sampling of the C$^{18}$O map.
\textbf{b} Ratio of C$^{18}$O 1--0 integrated intensity to 870~$\mu$m 
continuum flux density with a 870~$\mu$m clip at $4\sigma$. The contours are 
the same as in \textbf{a}. \textbf{c} Correlation between the C$^{18}$O~1--0 
integrated intensity and the 870~$\mu$m continuum emission. Only data points
with a C$^{18}$O~1--0 integrated intensity higher than 0.2 K~km~s$^{-1}$ 
($> 3\sigma$ for 90$\%$ of the map) and a 870~$\mu$m flux density higher than
69 mJy/45$\arcsec$-beam ($3\sigma$) are plotted. The red line shows the result 
of a linear fit over the full range. The blue dashed line has a 
slope of 1 in log-log space and its vertical position is arbitrary.}
\label{f:c18ocont}
\end{figure*}

The 870~$\mu$m dust continuum emission is much less correlated with
the C$^{18}$O~1--0 emission than the 2MASS extinction, as shown in 
Fig.~\ref{f:c18ocont}a. We smoothed the 870~$\mu$m continuum map to an 
effective angular resolution of 45$\arcsec$ and a pixel size of 30$\arcsec$ to 
match the C$^{18}$O map resolution and pixel size\footnote{But beware that the
C$^{18}$O map is highly undersampled since it was observed with a step of
60$\arcsec$.}. The ratio of C$^{18}$O~1--0 integrated intensity to 870~$\mu$m 
peak flux density is shown in Fig.~\ref{f:c18ocont}b in color scale. It varies 
by more than one order of magnitude, being the lowest at the 870~$\mu$m peaks. 
If we assume that the dust temperature and the excitation temperature of 
C$^{18}$O~1--0 are the same, the gas-to-dust mass ratio is uniform, and the 
line is optically thin \citep[see][]{Haikala05}, then the ratio 
plotted in Fig.~\ref{f:c18ocont}b should roughly trace the CO abundance. Its 
variations point to significant CO depletion toward the center of the 
870~$\mu$m sources\footnote{The temperature dependences of 
the C$^{18}$O and H$_2$ column densities when derived from the C$^{18}$O~1--0 
integrated intensity and the 870~$\mu$m flux density are such that their ratio
decreases by a factor 5 when the temperature drops from 12 to 6~K. 
Therefore, if the temperature is not uniform but rather decreases toward the 
center of each core, the drop of CO abundance toward the center of each 
core would be even more severe than derived above with the assumption of a 
uniform temperature.}.
Figure~\ref{f:c18ocont}c displays the relation between the
C$^{18}$O~1--0 integrated intensity and the 870~$\mu$m flux density for all 
pixels above $3\sigma$. Again, this plot shows the poor correlation between
both tracers. In addition, a linear fit in log-log space yields a slope of 
$\sim 0.2$, far from what would be expected if C$^{18}$O~1--0 were tracing the 
870~$\mu$m continuum emission well (see the red versus blue lines 
in Fig.~\ref{f:c18ocont}c). 

However, these figures are \textit{a priori} biased because of the spatial 
filtering of the 870~$\mu$m data that does not affect the C$^{18}$O~1--0 data.
The mass traced by the 870~$\mu$m map represents only $\sim 30\%$ of the mass 
derived from C$^{18}$O~1--0 (see Sect.~\ref{ss:cha1masses}). Since the 
estimate of the latter does not take into account depletion in regions of high 
column density, the discrepancy between both tracers must be even worse than a 
factor of 3. A proper correction would be to filter the C$^{18}$O~1--0 data in 
the same way as the 870~$\mu$m data are filtered. This is not an easy task 
and, because of the undersampling of the C$^{18}$O~1--0 data, may not be fully 
reliable either.
At this stage of the analysis, we cannot exclude that a significant 
part of the spatial variations seen in Fig.~\ref{f:c18ocont}b are due to the 
filtering of the continuum data rather than solely to variations of the 
C$^{18}$O abundance.

%

\section{Discussion}
\label{s:discussion}

\subsection{Evolutionary state of Cha-MMS1 and Cha-MMS2}
\label{ss:chamms}

Cha-MMS1 and Cha-MMS2 were the only two dense cores known in Cha~I before our 
LABOCA survey. \citet{Reipurth96} discovered them at 1.3~mm with the SEST and 
suggested that both could be Class~0 protostars based on their tentative
association with known outflows. In Cha-North, the peak of the 870~$\mu$m 
emission detected with LABOCA (source S3) corresponds to the source Cha-MMS2 
detected by \citet{Reipurth96} within 7$\arcsec$, which is certainly within 
the errors of the SEST measurements that were performed with a single-channel 
bolometer. Our source S3 is clearly associated with the T~Tauri star 
\object{WW~Cha} 
(coincidence within 1.2$\arcsec$, see Table~\ref{t:id_gcl_simbad}). As a 
result, Cha-MMS2 is associated with WW~Cha and is certainly not a Class~0 
protostar \citep[see also][ who came to the same conclusion]{Lommen09}.

\citet{Belloche06} discovered a faint \textit{Spitzer} source at 24 and 
70~$\mu$m located 5--7$\arcsec$ to the east of Cha-MMS1 (source 
$N_{\mathrm{sp}} = 1$ in Table~\ref{t:starless_spitzer}). They also reported
a \textit{Spitzer} non-detection at 8~$\mu$m 
\citep[see also Fig.~8 of][]{Luhman08c}, at a $3\sigma$ level $\sim 9$ 
and 7 times lower than the peak fluxes of the Taurus Class~0 protostars 
\object{IRAM~04191+1522} and \object{L1521F}, respectively. They did not find 
any evidence for an
outflow associated with Cha-MMS1 based on CO 3--2 observations done with APEX, 
and derived a deuterium fractionation of 11$\%$ for N$_2$H$^+$, a value 
typical for evolved prestellar cores and young protostars. Based on 
these two 
results, they concluded that Cha-MMS1 could be at the stage of the first 
hydrostatic core, i.e. just before the dissociation of H$_2$ and the second 
collapse. Our LABOCA observations refine the position of Cha-MMS1. The 
870~$\mu$m emission peaks at a position 1.3$\arcsec$ away from the 
\textit{Spitzer} source, which is well within the positional accuracy of our 
measurement. This strengthens even further the association of Cha-MMS1 with 
the faint \textit{Spitzer} source.

Recently, \citet{Dunham08} found a very tight correlation between the 
\textit{Spitzer} 70~$\mu$m flux and the internal luminosity of the 
low-luminosity YSOs of the c2d legacy project. They define the internal 
luminosity as the sum of the luminosities of the central protostar and its 
circumstellar disk, if present. 
With a 70~$\mu$m flux of $\sim 200$~mJy \citep[][]{Belloche06}, this 
correlation implies that Cha-MMS1 has a very low internal luminosity of 
$\sim0.015$~L$_\odot$. Luminosities ranging between $10^{-4}$ and 
0.1~L$_\odot$ have been theoretically predicted at the stage of the first 
core, depending on the collapse model and the amount of rotation 
\citep[][]{Saigo06,Omukai07}. The faint luminosity of Cha-MMS1 looks therefore 
consistent with the luminosity of a first core. It is about 2 and 3 times 
fainter than the internal luminosities of the young Class~0 protostars L1521F 
and IRAM~04191+1522, respectively \citep[][]{Dunham08}. Since IRAM~04191+1522 
drives a prominent bipolar outflow \citep[][]{Andre99} and L1521F 
may also drive a very compact one \citep[][]{Shinnaga09}, the absence of 
evidence for an outflow driven by Cha-MMS1 is an additional argument 
suggesting that it is in a different stage than these two young protostars,
i.e. that it may be at the stage of the first core. 

We note that \citet{Chen10} proposed \object{L1448~IRS2E} as a candidate first 
hydrostatic core based on its low bolometric luminosity 
($L_{\mathrm{bol}}< 0.1$~L$_\odot$, with no \textit{Spitzer} detection at 
70~$\mu$m and below) and their \textit{detection of a compact outflow}. 
Magnetohydrodynamical simulations do predict that a first core may drive an 
outflow but they predict that it should be slow 
\citep[a couple of km~s$^{-1}$ at most, see, e.g.,][]{Tomisaka02,Machida08}. 
The outflow of L1448~IRS2E has velocities about one order of magnitude higher 
than the predictions of \citet{Machida08}, more consistent with velocities 
predicted for the second outflow produced once the protostar (second core) is 
formed. It is therefore somewhat unclear whether L1448~IRS2E can really be at 
the stage of the first core. 

On the other hand, the absence of evidence for an 
outflow powered by Cha-MMS1 does not contradict these MHD simulations since a 
compact, slow outflow is \textit{a priori} not ruled out in this source. We 
thus confirm the suggestion of \citet{Belloche06} that Cha-MMS1 is a candidate 
first hydrostatic core.

\subsection{Nature of the LABOCA starless sources}
\label{ss:nature}

In the past decade, the mass distribution of starless dense cores in 
many nearby molecular clouds as traced mainly by (sub)mm continuum emission 
(see references in Sect.~\ref{sss:cmd}) was found to have the same power-law 
behaviour as the stellar IMF toward the high-mass end, while the mass 
distribution of lower-density clumps traced with CO has a much flatter shape.
This was put forward as a strong argument suggesting that the IMF in these 
low-mass-star forming clouds is essentially set at the early stage of 
cloud fragmentation already. The consistency of the high-mass end of the mass 
distribution of the Cha~I starless sources with the stellar IMF suggests that 
they may be \textit{prestellar}\footnote{A prestellar core is usually 
defined as a starless core that is gravitationally bound 
\citep[][]{Andre00,Ward-Thompson07}.} in nature, i.e. form stars in the future.

However, the distribution deviates significantly from the stellar IMF at the 
low-mass end. This CMD is compared to the stellar IMF of \citet{Kroupa01} and 
the IMF of multiple systems of \citet{Chabrier05} in Fig.~\ref{f:cmd}. Since 
LABOCA does not probe the scales at which members of a (future) multiple 
system can be seen individually, only the comparison to the IMF of multiple 
systems should make sense. While the Cha~I mass distribution looks consistent 
with the IMF of multiple systems above $\sim 0.3$~M$_\odot$, it deviates 
significantly below this threshold, the Cha~I starless sources being more 
numerous at the low-mass end than expected from the IMF. Since the 
completeness limit of the Cha~I survey is $\sim 0.2$~M$_\odot$, the 
discrepancy is \textit{a priori} even worse at very low mass. Such a deviation 
has already been noticed in Ophiuchus by \citet{Andre07} who concluded that a 
fraction of the lowest-mass starless sources may not be prestellar and that 
the link between the CMD and the stellar IMF is less robust at the low-mass 
end than at the high-mass end \citep[see also][]{Andre09}. 
Our LABOCA survey of Cha~I supports this conclusion. Follow-up 
heterodyne observations with MOPRA and APEX to measure the fraction of 
gravitationally bound sources are underway and should help better understanding
the low-mass end of the mass distribution in Cha~I.

Only four sources -- among them Cha1-C1, i.e. Cha-MMS1, that is not starless 
(see Sect.~\ref{ss:chamms}) -- have a mass larger than the critical 
Bonnor-Ebert value, and seven additional ones may also belong to this group if 
their mass is underestimated by a factor of 2 (see 
Sect.~\ref{sss:starless_massvssize}). As a result, at most 17$\%$ of the 
starless sources appear to be gravitationally unstable and thus likely 
\textit{prestellar}, provided their turbulent and magnetic supports are small 
compared to their thermal support or will vanish if they are not. This 
fraction is significantly smaller than in Ophiuchus and Aquila. 
\citet{Andre10a} estimate that more than $60\%$ of the 541 starless sources 
detected with \textit{Herschel} in Aquila, with a completeness limit similar 
to ours (see Sect.~\ref{sss:starless_masses}), are candidate prestellar cores
(with $M_{\mathrm{BE}}/M_{\mathrm{tot}} < 2$). 
Similarly, about 70$\%$ of the starless sources detected by 
\citet{Motte98} in Ophiuchus with a completeness limit somewhat better than 
ours have $M_{\mathrm{BE}}(R)/M_{\mathrm{tot}} < 2$
(see definition of $M_{\mathrm{BE}}(R)$ in 
Sect.~\ref{sss:starless_massvssize})\footnote{The percentage quoted 
here assumes $R = FWHM$ and a distance of 125~pc.}. In Taurus, more than 
50$\%$ of the 
starless cores identified by \citet{Kauffmann08} are gravitationally unstable,
but this sample is biased and inhomogeneous so a direct comparison to Cha~I is
more uncertain.

The very low fraction of candidate prestellar cores among the starless cores 
detected in Cha~I suggests that we may be witnessing the end of star formation 
in this cloud or, on the contrary, that these starless cores are in a less 
advanced stage of evolution compared to those in Aquila, Ophiuchus, and 
Taurus. The latter suggestion is less likely because we already assessed 
the gravitational state of the cores assuming that they have no turbulent 
and magnetic support. A reduction of these support mechanisms via turbulence 
dissipation or ambipolar diffusion could in principle lead to gravitational 
instability and eventually gravitational collapse, but only if the cores are 
above the (purely thermal) critical Bonnor-Ebert mass limit already. A 
reduction of the thermal support via cooling could in principle also lead to a 
gravitational instability. In Sect.~\ref{sss:starless_massvssize}, we 
estimated that even if the current temperature of the cores is 7~K instead of 
12~K, this would not change the current fraction of unstable cores 
significantly. If the cores are currently at 12~K even in their center (i.e. 
the masses we measure are correct) and are still cooling down, then the 
fraction of unstable cores will not change much either, because the 
Bonnor-Ebert critical mass will decrease by less than a factor of 2 while most 
cores are at least a factor of 3 below the current critical mass limit. It is 
thus unlikely that most starless cores that appear to be currently stable will 
become unstable in the future, unless they experience a strong 
external perturbation.

\subsection{Star formation efficiency}
\label{ss:efficiency}

\subsubsection{Local star formation efficiency}
\label{sss:localeff}

Apart from the deviation at low mass, the mass scale of the Cha~I CMD 
agrees quite well with that of the IMF of multiple systems in 
Fig.~\ref{f:cmd}. Taken at face value, this would suggest that the 
star formation efficiency at the core level in Cha~I will be as high as 
$\sim 100\%$ for all prestellar sources that populate the high mass end above 
$\sim 0.3$~M$_\odot$. For comparison, \citet{Andre10a} found an efficiency 
$\epsilon_{\mathrm{core}} = M_\star/M_{\mathrm{core}} \sim$~20--40\% in Aquila
and \citet{Alves07} obtained $\epsilon_{\mathrm{core}} = 30 \pm 10\%$ for the 
Pipe nebula. \citet{Enoch08} could only set a lower limit of
25\% for their combined sample of Perseus, Serpens, and Ophiuchus sources, 
because of their poor completeness limit (0.8~M$_\odot$).

\citet{Luhman07} computes the specific IMF of the stellar population of Cha~I
and find a peak at 0.1--0.15~M$_\odot$. If we take this turnover mass as 
reference, then the (apparent) turnover of the Cha~I CMD at the completeness 
limit still suggests a local star formation efficiency 
$\epsilon_{\mathrm{core}} >$~50--70\%. Such a very high efficiency for 
Cha~I sounds \textit{a priori} unlikely. One explanation could be that the 
contamination by gravitationally unbound, starless sources (see 
Sect.~\ref{ss:nature}) may extend further to higher masses and mask the 
maximum of the true \textit{prestellar} CMD. Indeed, \citet{Konyves10} found 
that the peak of the Aquila CMD shifted from 0.6 to 0.9~M$_\odot$ when they 
removed from their sample the sources that they estimated not to be 
\textit{prestellar} ($\sim 35\%$). On the other hand, the majority of the Pipe 
nebula sources appear to be gravitationally unbound \citep[][]{Lada08} and 
their CMD still closely resembles the IMF in shape \citep[][]{Alves07}. 
Therefore it is unclear by how much the Cha~I CMD could be affected by 
contamination from unbound, starless sources. The on-going follow-up 
heterodyne observations should help answer this question.

\subsubsection{Global star formation efficiency}
\label{sss:globaleff}

At the scale of the cloud, \citet{Luhman07} estimate a current 
star formation efficiency, defined as 
$\mathrm{SFE} = \frac{M_\star}{M_\star+M_{\mathrm{cloud}}}$, of about 10$\%$, 
based on the cloud mass derived from CO ($\sim$~1000~M$_\odot$, see 
Sect.~\ref{ss:cha1masses}) and the mass $\sim$~100~$M_\odot$ of the 215 young 
stars and brown dwarfs known to belong to Cha~I at that time. This star 
formation efficiency is a factor of 1.5 to 3 larger than the SFEs of the five 
nearby molecular clouds targeted by the c2d \textit{Spitzer} legacy project 
\citep[between 3.0 and 6.3$\%$, see Table 4 of][]{Evans09}. It is even more 
than one order of magnitude higher than the SFE of the Taurus molecular cloud
\citep[0.6$\%$\footnote{This SFE was estimated with a cloud mass that includes 
very low column-density material. If we take the mass of the regions with 
$A_V >3$~mag (see Table~\ref{t:clouddensities}), then the SFE would be 
$\sim 4$ times larger.}, see][]{Goldsmith08}. Assuming a constant rate of star 
formation, \citet{Evans09} suggest that the c2d clouds could reach a final 
SFE of 15 to 30$\%$ given the estimated typical lifetime of a molecular cloud. 
The SFE of Cha~I is close to this limit, this being even more true if the rate 
of star formation in the c2d clouds decreases with time like in Cha~I (see 
Sect.~\ref{ss:lifetimes}), which would reduce the expected final 
SFE. We note that \citet{Bonnell10} find an overall efficiency of star 
formation of $\sim 15\%$ in their numerical simulation of a turbulent, 
molecular cloud that undergoes star formation.
It is thus very likely that Cha~I is much closer to the end of the star 
formation process than the other clouds mentioned above.

\subsection{Lifetimes}
\label{ss:lifetimes}

The median age of the YSO population of Cha~I is $\sim 2$~Myr 
\citep[][]{Luhman08c}\footnote{If the further distance is 
confirmed for Cha~I (196~pc, see Appendix~\ref{ss:distance}), the median age 
would be reduced to $\sim 1$~Myr \citep[][]{Knude10}.}. With the census of 
starless cores and YSOs presented
in Table~\ref{t:census}, we can statistically estimate the lifetimes of each
category if we assume that all prestellar cores evolve into Class 0/Class I
protostars and then into Class II/Class III YSOs, and that star formation has 
proceeded at a constant rate over time. Five to 15 YSOs out of 204 are 
protostars, implying a typical lifetime of $0.5 - 1.5 \times 10^5$~yr for the 
protostars in Cha~I. Only one object is a Class 0 protostar (or first core), 
hence a very short lifetime of $10^4$~yr for the Class 0 phase. Given that a 
fraction of the starless cores may not be prestellar in nature (see 
Sect.~\ref{ss:nature}), we can only derive an upper limit of 
$6 \times 10^5$~yr to the duration of the prestellar phase in Cha~I.

\begin{table}
 \caption{Census of starless cores and YSOs in Cha~I.}
 \label{t:census}
 \centering
 \begin{tabular}{lc}
 \hline\hline
 \multicolumn{1}{c}{Class} & \multicolumn{1}{c}{Number} \\
 \hline
 Starless cores          & 59\tablefootmark{a} \\
 Class 0 (or first core) &  1\tablefootmark{b} \\
 I                       &  4\tablefootmark{c} \\
 Flat                    & 10\tablefootmark{c} \\
 II                      & 94\tablefootmark{c} \\
 III                     & 95\tablefootmark{c} \\
 \hline
 \end{tabular}
 \tablefoot{
 \tablefoottext{a}{This work.}
 \tablefoottext{b}{Cha-MMS1 \citep[][]{Belloche06}.}
 \tablefoottext{c}{From \citet{Luhman08a}, over a large (but not complete) 
 part of Cha~I.}
 }
\end{table}

The median peak density of the Cha~I starless cores is 
$3.5 \times 10^5$~cm$^{-3}$ and their median mean density within 50$\arcsec$
is $7 \times 10^4$~cm$^{-3}$ (Figs.~\ref{f:histom}e and f). This median mean
density corresponds to a prestellar lifetime of $\sim 4 \times 10^5$~yr
according to the empirical relation of \citet{Ward-Thompson07}.
If the fraction of non-prestellar sources among the Cha~I starless sources is 
on the order of $30\%$ or less, then the agreement between this prestellar 
lifetime and the lifetime derived from the statistical counts in the previous
paragraph suggests that the assumption of a star formation rate 
constant over time until now is valid in Cha~I. It is also consistent with the 
prestellar lifetime of $4.5 \pm 0.8 \times 10^5$~yr estimated by 
\citet{Enoch08} for their combined sample of Perseus, Serpens, and Ophiuchus
\citep[see also][ for Perseus alone]{Jorgensen07}. The very short lifetime of 
$10^4$~yr derived here for the Class 0 phase is similar to the one found in 
Ophiuchus \citep[$1-3 \times 10^4$~yr, see][]{Andre00}, but about one order
of magnitude shorter than in Perseus \citep[e.g.][]{Hatchell07}.

On the other hand, if, e.g., $90\%$ of the Cha~I starless sources turn out not 
to be prestellar in nature, then the unrealistically low statistical lifetime 
that would be inferred for the prestellar sources ($6 \times 10^4$~yr) would 
rather point to a decrease of the star formation rate with time and confirm 
the trend pointed out by \citet{Luhman07}. This could also explain why so few 
Class 0 sources (only one) are found in Cha~I, if the Class 0 lifetime derived 
in Perseus rather than in Ophiuchus applies to Cha~I. In any case, an 
\textit{increase} of the star formation rate until the present time 
\citep[][, see Sect.~\ref{s:intro}]{Palla00} is ruled out for Cha~I.

\subsection{Filaments}
\label{ss:filaments}

The \textit{Herschel} satellite has recently revealed an impressive network
of parsec-scale filamentary structures in the nearby molecular clouds Aquila 
and Polaris \citep[e.g.][]{Andre10a,Menshchikov10}. The few filaments detected 
in Cha~I  with LABOCA (see Sect.~\ref{ss:largescale}) and in C$^{18}$O 1--0
\citep[][]{Haikala05} are strong evidence for the existence of such a network 
in this cloud, of which we only see the tip. These filaments have a 
deconvolved width of $0.16 \pm 0.05$~pc on average (33000~AU), measured at a
level of 10~mJy/21.2$\arcsec$-beam. This width may be somewhat underestimated 
because of the spatial filtering due to the correlated noise removal 
(see Sect.~\ref{ss:largescale}). It is similar to the typical 
width of the Perseus filaments as measured by \citet{Hatchell05} with SCUBA 
($FWHM \sim 1\arcmin$, i.e. 15000~AU at 250~pc). However, the Cha~I filaments 
are less massive, with a mass per unit length ranging between 8 and 
33~M$_\odot$~pc$^{-1}$ (see Table~\ref{t:clf_ls}) while it ranges 
between 47 and 115~M$_\odot$~pc$^{-1}$ for the Perseus filaments 
\citep[][]{Hatchell05}. 

The mass per unit length of filaments is an essential parameter to evaluate 
their stability. In Aquila, most ($> 60\%$) candidate \textit{prestellar} 
cores were found in (thermally) supercritical filaments defined by 
$M_{\mathrm{lin}} > M_{\mathrm{lin, crit}}^{\mathrm{unmagn.}} = 
2\,a_{\mathrm{s}}^2\,/G$ \citep[see][, and references therein]{Andre10a}. Only 
one filament in Cha~I ($N_{\mathrm{fil}} = 1$) has a mass per unit length 
larger than the critical value (19~M$_\odot$~pc$^{-1}$ at 12~K). It contains 
the single Class~0 protostar (Cha1-C1) and two of the three candidate 
prestellar cores (Cha1-C2 and C3), while only one of the four other 
(subcritical) filaments contains a candidate prestellar core (Cha1-C4 in 
filament 5). Even if these numbers are much too small to derive any robust 
conclusion, they seem to be consistent with the results obtained by 
\citet{Andre10a} for Aquila.

However, the total velocity dispersion as probed by \hbox{C$^{18}$O~1--0} in 
Cha~I is significantly larger than the purely thermal dispersion. The 
C$^{18}$O~1--0 linewidth ($FWHM$) is typically on the order of 
0.5--0.8~km~s$^{-1}$ \citep[see Table~1 of][]{Haikala05}, yielding a 
non-thermal velocity dispersion $\sigma_{\mathrm{NT}} = 0.20$--0.38~km~s$^{-1}$ 
and an effective sound speed $a_{\mathrm{eff}} = 0.29$--0.43~km~s$^{-1}$ at 12~K. 
We can compute the mean free-particle density $n_{\mathrm{mean}}$ of the 
filaments with the equation $n_{\mathrm{mean}} = 
\frac{N_{\mathrm{mean}}}{a_{\mathrm{min}}} \times \frac{\mu_{\mathrm{H}_2}}{\mu}$, 
with $N_{\mathrm{mean}}$ and $a_{\mathrm{min}}$ given in Table~\ref{t:clf_ls} and 
$\mu_{\mathrm{H}_2}$ and $\mu$ defined in Appendix~\ref{ss:mu}. These mean 
densities are in the range \hbox{3.7--$10 \times 10^{3}$~cm$^{-3}$}. Since the
critical density of the C$^{18}$O~1--0 line at 10~K is about 
$3 \times 10^{3}$~cm$^{-3}$, 
this transition traces the bulk of the filaments very well and its 
linewidth should reliably trace their total velocity dispersion, although some 
contamination from the ambient cloud cannot be completely excluded. If we 
assume that C$^{18}$O~1--0 traces only the filaments, then the critical mass 
per unit length becomes the virial mass per unit length 
$2\,a_{\mathrm{eff}}^2\,/G$ \citep[][]{Fiege00} and ranges from 39 to 
86~M$_\odot$~pc$^{-1}$. The Cha~I filaments have a mass per unit length lower 
than this critical value by a factor of 2 at least (the velocity dispersion is 
the highest in filament $N_{\mathrm{fil}} = 1$). Unless they are compressed by 
a significant toroidal magnetic field \citep[][]{Fiege00} or the C$^{18}$O~1--0 
linewidth overestimates their total velocity dispersion, they do not seem to 
be supercritical and will most likely neither collapse nor fragment in the 
future. This would be consistent with our conclusion that the process of star 
formation is nearly over in Cha~I.

\subsection{The end of star formation in Cha~I\,?}
\label{ss:end}

\begin{table*}
 \caption{
 Mass, area, and average surface density of a sample of nearby molecular clouds, measured over the regions with $A_V > 3$~mag, $3 < A_V < 6$~mag, and $A_V > 6$~mag.
 }
 \label{t:clouddensities}
 \centering
 \begin{tabular}{lcrrcrccrcrccc}
 \hline\hline
  & & \multicolumn{3}{c}{$A_V > 3$~mag} & \multicolumn{3}{c}{$3 < A_V < 6$~mag} & \multicolumn{3}{c}{$A_V > 6$~mag} & & & \\ 
 \multicolumn{1}{c}{Cloud} & \multicolumn{1}{c}{Distance} & \multicolumn{1}{c}{Mass} & \multicolumn{1}{c}{Area} & \multicolumn{1}{c}{$\Sigma$\tablefootmark{a}} & \multicolumn{1}{c}{Mass} & \multicolumn{1}{c}{Area} & \multicolumn{1}{c}{$\Sigma$\tablefootmark{a}} & \multicolumn{1}{c}{Mass} & \multicolumn{1}{c}{Area} & \multicolumn{1}{c}{$\Sigma$\tablefootmark{a}} & \multicolumn{1}{c}{$R_M$\tablefootmark{b}} & \multicolumn{1}{c}{$R_A$\tablefootmark{b}} & \multicolumn{1}{c}{$R_\Sigma$\tablefootmark{b}}\\ 
  & \multicolumn{1}{c}{\scriptsize (pc)} & \multicolumn{1}{c}{\scriptsize (M$_\odot$)} & \multicolumn{1}{c}{\scriptsize (deg$^2$)} & \multicolumn{1}{c}{\scriptsize (M$_\odot$~pc$^{-2}$)} & \multicolumn{1}{c}{\scriptsize (M$_\odot$)} & \multicolumn{1}{c}{\scriptsize (deg$^2$)} & \multicolumn{1}{c}{\scriptsize (M$_\odot$~pc$^{-2}$)} & \multicolumn{1}{c}{\scriptsize (M$_\odot$)} & \multicolumn{1}{c}{\scriptsize (deg$^2$)} & \multicolumn{1}{c}{\scriptsize (M$_\odot$~pc$^{-2}$)} & & & \\ 
 \multicolumn{1}{c}{(1)} & \multicolumn{1}{c}{(2)} & \multicolumn{1}{c}{(3)} & \multicolumn{1}{c}{(4)} & \multicolumn{1}{c}{(5)} & \multicolumn{1}{c}{(6)} & \multicolumn{1}{c}{(7)} & \multicolumn{1}{c}{(8)} & \multicolumn{1}{c}{(9)} & \multicolumn{1}{c}{(10)} & \multicolumn{1}{c}{(11)} & \multicolumn{1}{c}{(12)} & \multicolumn{1}{c}{(13)} & \multicolumn{1}{c}{(14)} \\ 
 \hline
Chamaeleon I &  150 &    720 &  1.2 &     88 &    500 &  0.9 &     81 &    220 & 0.30 &    109 &   0.45 &   0.33 &    1.4 \\ Ophiuchus &  125 &   3340 & 10.1 &     69 &   2350 &  8.3 &     60 &    990 & 1.87 &    111 &   0.42 &   0.23 &    1.9 \\ Taurus &  140 &   6120 & 15.1 &     68 &   4320 & 12.5 &     58 &   1800 & 2.59 &    116 &   0.42 &   0.21 &    2.0 \\ Perseus &  250 &   6140 &  4.8 &     68 &   4510 &  3.9 &     61 &   1620 & 0.87 &     98 &   0.36 &   0.22 &    1.6 \\ Serpens &  260 &   7980 &  5.5 &     70 &   6300 &  4.5 &     68 &   1680 & 1.00 &     82 &   0.27 &   0.22 &    1.2 \\ Orion &  450 &  63320 & 15.0 &     68 &  44180 & 11.8 &     61 &  19140 & 3.23 &     96 &   0.43 &   0.27 &    1.6 \\  \hline
 \end{tabular}
 \tablefoot{The masses and areas for all clouds but Chamaeleon~I were provided by S. Sadavoy (\textit{priv. comm.}), based on \citet{Sadavoy10}. The masses for $A_V > 6$~mag, initially calibrated for $R_V = 3.1$, are here recalibrated for $R_V = 5.5$ (see Appendix~\ref{ss:avtoN}). The masses for $A_V > 3$~mag are corrected accordingly. The distances are the same as in \citet{Sadavoy10}.
 \tablefoottext{a}{Average surface density, i.e. mass-to-area ratio.}
 \tablefoottext{b}{Mass, area, and surface-density ratios between the regions with $A_V>6$~mag} and $3<A_V<6$~mag.
}
 \end{table*}

The low fraction of candidate prestellar cores among the population of 
starless cores detected with LABOCA, the small number of Class 0 protostars, 
the high global star formation efficiency, the decrease of the star formation 
rate with time, and the low mass per unit length of the detected filaments 
all suggest that we may be witnessing the end of the star formation process in 
Cha~I. While the conclusion of \citet{Luhman07} that the star formation rate 
in Cha~I has declined for several Myr already questions the scenario of 
\citet{Palla00} in which star formation is a threshold phenomenon and 
accelerates as a result of cloud contraction (see Sect.~\ref{s:intro}), our 
findings raise new questions about the end of star formation. \citet{Palla00} 
argue that the truncation period, i.e. the period over which the 
star formation rate drops, must be very brief (1 Myr at most) and that 
the falloff should be related to the dispersal of the parent cloud gas. In 
low-mass star forming regions, they consider that this dispersal can only 
result from the impact of molecular outflows. Even if the total current mass 
efflux due to molecular outflows in the particular case of Taurus is three 
orders of magnitude lower than the one needed to disperse the cloud in such a 
short time, they argue that the (potentially) accelerating nature of star 
formation should significantly increase the mass efflux in the future and 
that, if star formation is really a threshold phenomenon, its falloff can 
occur before the actual cloud dispersal. 

To look for signs of cloud dispersal in Cha~I, we compare its mass, area, and 
surface density to those of a sample of nearby molecular clouds in 
Table~\ref{t:clouddensities}. We list the surface densities for the low- 
($3 < A_V < 6$~mag) and high-extinction ($A_V > 6$~mag) regions in Cols.~8 and 
11, respectively. The limit $A_V =6$~mag approximately corresponds to the 
extinction threshold for star formation mentioned in Sects.~\ref{s:intro} and 
\ref{sss:starless_extinction}. If all clouds have similar depths, then the 
surface density scales with the average density and could be a good indication 
for cloud dispersal. On the one hand, Cha~I has a surface density above 
$A_V = 6$~mag very similar to the ones of the other clouds, suggesting that the 
average density of its high-extinction region is the same as in the other 
clouds, provided its depth is similar to the ones of these clouds. Only 
Serpens has a significantly lower surface density compared to the rest of the 
sample. On the other hand, the surface density of Cha~I (and Serpens to a 
lesser extent too) for $3 < A_V < 6$~mag is significantly higher than in the 
other clouds. As a result, it stands out, along with Serpens, with a 
surface-density ratio between the regions with $A_V > 6$~mag and 
$3 < A_V < 6$~mag somewhat lower than in the other clouds. Since this ratio 
does not depend on the depth of the cloud, it could be a more reliable tracer 
of cloud dispersal and could indicate that Cha~I and Serpens are in the 
process of being dispersed. However, these two clouds do not seem to differ 
significantly from other nearby star-forming clouds in terms of probability 
density functions of column density \citep[][]{Kainulainen09}. Therefore the 
signs for cloud dispersal in Cha~I based on the extinction map are very 
tenuous. Further clues to explain why Cha~I could be at the end of the 
star formation process should be searched for.
Finally, we note that we cannot exclude that the star formation 
activity in Cha~I could be revived in the future if the cloud would undergo a 
significant external compression

%

\section{Conclusions}
\label{s:conclusions}

We performed a deep, unbiased, 870~$\mu$m dust continuum survey for starless 
and protostellar cores in Chamaeleon~I with the bolometer array 
LABOCA at APEX. The resulting 1.6~deg$^2$ map was compared with maps of dust 
extinction and C$^{18}$O 1--0 emission. It was decomposed into multiple 
scales to disentangle the emission of compact or moderately extended sources 
from large-scale structures. The analysis was performed by carefully taking 
into account the spatial filtering properties of the data reduction process. 
Our main results and conclusions are the following:

\begin{enumerate}
  \item The standard reduction process of LABOCA data with BoA filters the 
  emission more extended than $\sim$ 2.5--4$\arcmin$ ($FWHM$), depending on 
  its flux density. Elongated sources are better recovered than circular ones
  along their major axis, up to 6--9$\arcmin$. The iterative scheme requires 
  about 21 iterations to converge.
  \item The mass detected with LABOCA (61~M$_\odot$) represents only
  \hbox{6--8$\%$} of the cloud mass traced by the dust extinction and 
  CO, and 
  about $30\%$ of the mass traced by C$^{18}$O. Most (88$\%$) of this 
  870~$\mu$m mass is located in five filaments with typical length and width 
  of 0.6 and 0.16~pc, respectively.
  \item The C$^{18}$O 1--0 emission is a reliable tracer of mass in the 
  visual extinction range 6 to 15~mag. C$^{18}$O appears to be 
  photodissociated below 6~mag.
  \item 81 sources were extracted from the 870~$\mu$m map, 59 of which are 
  starless based on the absence of \textit{Spitzer} association, one is a 
  candidate first hydrostatic core (Cha-MMS1), and 21 are associated with 
  more evolved young stellar objects already known in the SIMBAD database.
  \item Like in other nearby molecular clouds, the starless cores in Cha~I are 
  only found above an extinction threshold $A_V = 5$~mag and their 
  distribution of extinction peaks at $A_V \sim 8$~mag (7~mag with 
  $R_V = 5.5$). Alternatively there may be no strict threshold but just a very 
  low probability of finding a core at low extinction.
  \item The 90$\%$ completeness limit of our 870~$\mu$m survey is 
  0.22~M$_\odot$ for starless cores. Only 50$\%$ of the detected starless 
  cores are above this limit, suggesting that we may miss a significant 
  fraction of the existing starless cores.
  \item The starless cores in Cha~I are less dense than those in 
  Perseus, Serpens, Ophiuchus, and Taurus by a factor of a few on average.
  This may result from the better sensitivity of the Cha~I survey and/or the
  multi-scale analysis of the Cha~I data.
  \item The CO abundance seems to decrease toward the peak of the embedded 
  starless cores, as expected for \textit{bona-fide} prestellar cores, but the
  spatial filtering of the 870~$\mu$m  data prevents a firm conclusion.
  \item On the one hand, the high-mass end of the core mass distribution in 
  Cha~I is consistent with the stellar initial mass function, suggesting that 
  a fraction of the starless cores may be \textit{prestellar}. On the 
  other hand, the low-mass end 
  is overpopulated, and at most 17$\%$ of the starless cores have a mass 
  higher than the critical Bonnor-Ebert mass. Both results suggest that a 
  large fraction of the starless cores may not be \textit{prestellar}.
  \item The local star formation efficiency at the core level in Cha~I 
  ($> 50\%$) seems to be larger than in other molecular clouds, but this 
  conclusion is uncertain due to possible contamination by unbound, starless 
  sources.
  \item The global star formation efficiency at the scale of the cloud is 
  10$\%$. It is significantly larger than in other clouds and suggests that 
  Cha~I is much closer to the end of the star formation process.
  \item The star formation rate in Cha~I can have been constant over time 
  only if $> 70\%$ of the detected starless cores
  turn out to be prestellar. This is 
  very unlikely, which suggests that the star formation rate has decreased over 
  time, as also indicated by the very small number of Class 0 protostars. An 
  increase of the star formation rate to the present time in Cha~I is ruled out.
  \item Apart from maybe one, the filaments in which most starless 
  cores are embedded do not seem to be supercritical and will most likely 
  neither collapse nor fragment in the future.
\end{enumerate}

The low fraction of candidate prestellar cores among the population of 
starless cores detected with LABOCA, the small number of Class 0 protostars, 
the high global star formation efficiency, the decrease of the star formation 
rate with time, and the low mass per unit length of the detected filaments 
all suggest that we may be witnessing the end of the star formation process in 
Cha~I. We only find a tenuous evidence for cloud dispersal based on the
distribution of extinction. Further clues to explain why Cha~I could be at the 
end of the star formation process should be searched for.

\begin{acknowledgements}
      We thank the APEX staff for their support during the observing sessions
and Attila Kov{\'a}cs and Sven Thorwirth for carrying out part of the 
observations in 2007. We also thank Lauri Haikala for providing us information 
about his C$^{18}$O data,  Carsten Kramer for his help with the use of 
\textit{Gaussclumps}, Ruud Visser for sending us new predictions of his
photodissociation model, and Sarah Sadavoy for giving us cloud masses and 
sizes on very short notice. We are grateful to the referee for his/her 
constructive criticisms that helped to improve the quality of this article.
This research has made use of the SIMBAD database, 
operated at CDS, Strasbourg, France. BP acknowledges her support by the German
\textit{Deutsche Forschungsgemeinschaft, DFG} Emmy Noether number PA~1692/1-1.
\end{acknowledgements}

\begin{appendix}
\section{LABOCA data reduction: spatial filtering and convergence}
\label{s:reduction_appendix}

\subsection{Spatial filtering due to the correlated noise removal}
\label{ss:filtering}

\begin{figure*}
\centerline{\resizebox{1.0\hsize}{!}{\includegraphics[angle=270]{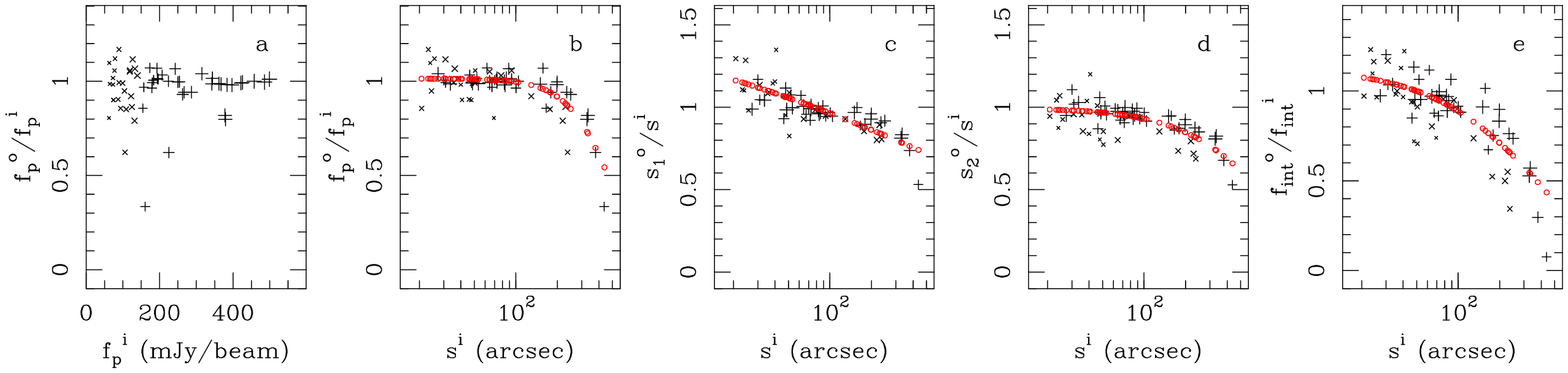}}}
\vspace*{1ex}
\centerline{\resizebox{1.0\hsize}{!}{\includegraphics[angle=270]{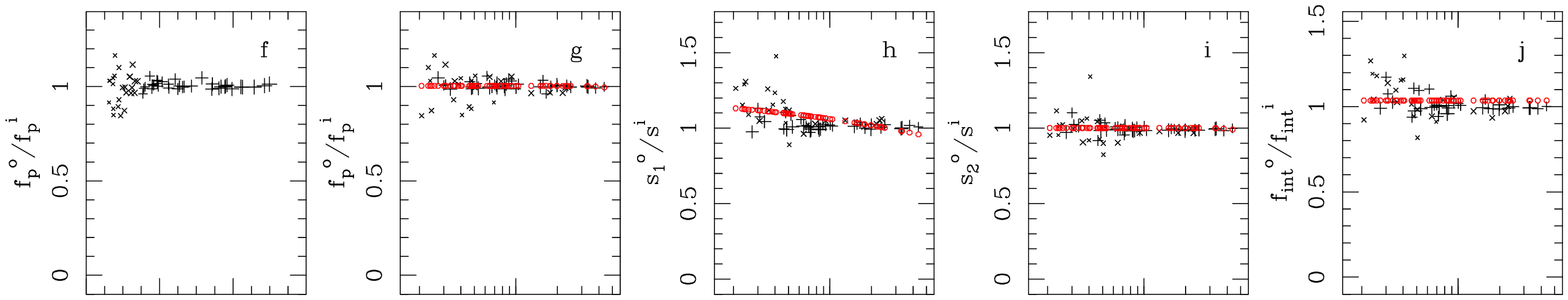}}}
\caption{Statistical properties of 60 artificial, circular, Gaussian sources 
inserted into the raw time signals before data reduction (\textbf{a} to 
\textbf{e}) or directly inserted into the final unsmoothed continuum emission 
map of Cha~I (\textbf{f} to \textbf{j}). \textbf{a} and \textbf{f}: 
ratio of output to input peak flux density as a function of input peak flux 
density. \textbf{b} and \textbf{g}: ratio of output to input peak flux density 
as a function of input size ($FWHM$). \textbf{c} and \textbf{h}: ratio of 
output major size to input size as a function of input size. \textbf{d} and 
\textbf{i}: ratio of output minor size to input size as a function of input 
size. \textbf{e} and \textbf{j}: ratio of output to input total flux as a 
function of input size. The size of the black symbols scales with the input 
size of the sources. In addition, crosses and plus symbols are for input peak 
flux densities below and above 150 mJy/19.2$\arcsec$-beam, respectively. The 
red circles in panels \textbf{b} to \textbf{e} and \textbf{g} to \textbf{j} 
show a fit to the data points using the arbitrary 3-parameter function 
$y = \log(\alpha/(1+(x/\beta)^\gamma))$.}
\label{f:filtering}
\end{figure*}

Since LABOCA is a total-power instrument used without wobbler, the standard 
data reduction relies on the correlated signal of all pixels and of smaller 
groups of pixels to estimate and remove the atmospheric and instrumental 
noise. This method acts as a high-pass filter and makes the detection of very 
extended structures extremely difficult. The strongest limiting factor is the
size of the groups of pixels connected to the same read-out cable 
\citep[about 1.5$\arcmin$ by 5$\arcmin$, see][]{Schuller09}. To analyse the 
dust continuum emission map in detail, it is therefore essential to understand 
the properties of this spatial filtering.

We performed several Monte Carlo simulations to characterize the filtering. 
First, we inserted artificial, circular, Gaussian sources in the raw time 
signals with the BoA method \textit{addSource} after calibration. The sources 
were inserted at random positions in signal-free regions of the map shown in 
Fig.~\ref{f:labocamap}. The peak flux densities and $FWHM$ sizes of the 
sources were randomly assigned in the range 50 to 500 mJy/beam and 
19.2$\arcsec$ to 8$\arcmin$, respectively. The only constraint was that the 
artificial sources should not overlap and should not be too close to the map 
edges either. The modified raw data were then reduced following exactly the 
same procedure as described in Sect.~\ref{ss:redmethod}, except that the final 
map was not smoothed and kept its original angular resolution of $19.2\arcsec$ 
(HPBW). Each artificial source was fitted with an elliptical Gaussian using 
the task GAUSS\_2D in GREG to measure its 
output peak flux density and sizes after data reduction. We also measured the 
total output flux on a circular aperture of \textit{radius} equal to $FWHM$.

The whole procedure was repeated twice, with two independent samples of 30 
artificial sources. Figures~\ref{f:filtering}a--e present the results. As a 
control check, we show in Figs.~\ref{f:filtering}f--j the results of the fits 
performed on the artificial sources directly inserted in the final unsmoothed 
continuum emission map of Cha~I. The effects of the spatial filtering 
due to the data reduction are obvious when comparing 
Figs.~\ref{f:filtering}b--e to Figs.~\ref{f:filtering}g--j. The noise produces 
a small dispersion 
in the output source properties, especially for weak sources (crosses) 
but apart from that, all curves are basically flat in the control sample
(Figs.~\ref{f:filtering}f--j). On the contrary, after data reduction, the 
peak flux densities of the sources are reduced by $\sim 15\%$ to $\sim 70\%$ 
for input sizes varying from $\sim 240\arcsec$ to $\sim 440\arcsec$. Similarly, 
the sizes of the sources are reduced by $\sim 15\%$ to $\sim 50\%$ for input 
sizes from $\sim 200\arcsec$ to $\sim 440\arcsec$. The total fluxes suffer the 
most from the spatial filtering: they are reduced by $\sim 15\%$ to 
$\sim 90\%$ for input sizes from $\sim 120\arcsec$ to $\sim 440\arcsec$. A 
careful inspection of Figs.~\ref{f:filtering}g--j shows that the sources 
with a peak flux density weaker than $\sim 150$~mJy/beam (crosses) are 
systematically below the stronger sources (plus symbols) for source sizes 
larger than $\sim 100\arcsec$. This is because the insertion of the stronger 
sources in the astronomical source model used for iterations 1 to 20 helps to 
recover more astronomical signal when their outer parts start to reach the 
thresholds mentioned in Sect.~\ref{ss:redmethod}. The spatial filtering is 
therefore less severe for strong sources than for faint sources. These results
are summarized in Table~\ref{t:filtering}.

\begin{table}
 \caption{Filtering of artificial Gaussian sources due to the 
correlated noise removal depending on the input size along the minor (major) 
axis ($FWHM$).}
 \label{t:filtering}
 \begin{tabular}{lccc}
  \hline\hline
  \multicolumn{1}{c}{Source sample} & \multicolumn{3}{c}{Size where} \\
  \cline{2-4}
  \multicolumn{1}{c}{} & \multicolumn{1}{c}{$f_p^o/f_p^i = 85\%$} & \multicolumn{1}{c}{$s_2^o/s^i = 85\%$} & \multicolumn{1}{c}{$f_{int}^o/f_{int}^i = 85\%$} \\
  \multicolumn{1}{c}{(1)} & \multicolumn{1}{c}{(2)} & \multicolumn{1}{c}{(3)} & \multicolumn{1}{c}{(4)} \\
  \hline\\[-1.5ex]
  \multicolumn{4}{l}{\textit{Circular sources:}} \\[0.3ex]
  strong sources & 280$\arcsec$ & 250$\arcsec$ & 170$\arcsec$ \\[-0.4ex]
 {\hspace*{2ex} \scriptsize ($>150$~mJy/beam)} & & & \\[0.6ex]
  weak sources   & 170$\arcsec$ & 140$\arcsec$ &  90$\arcsec$ \\[-0.4ex]
 {\hspace*{2ex} \scriptsize ($<150$~mJy/beam)} & & & \\[0.6ex]
  all sources    & 240$\arcsec$ & 200$\arcsec$ & 120$\arcsec$ \\
  \hline\\[-1.5ex]
  \multicolumn{4}{l}{\textit{Elliptical sources with aspect ratio 2.5:}} \\[0.3ex]
  strong sources & 220$\arcsec$ (550$\arcsec$) & 220$\arcsec$ (550$\arcsec$) & 130$\arcsec$ (330$\arcsec$) \\[-0.4ex]
 {\hspace*{2ex} \scriptsize ($>150$~mJy/beam)} & & & \\[0.6ex]
  weak sources   & 130$\arcsec$ (330$\arcsec$) & 120$\arcsec$ (300$\arcsec$) &  70$\arcsec$ (180$\arcsec$) \\[-0.4ex]
 {\hspace*{2ex} \scriptsize ($<150$~mJy/beam)} & & & \\[0.6ex]
  all sources    & 180$\arcsec$ (450$\arcsec$) & 160$\arcsec$ (400$\arcsec$) &  100$\arcsec$ (250$\arcsec$) \\
  \hline
 \end{tabular}
 \tablefoot{$f_p^o$ and $f_p^i$ are the output and input peak flux densities, 
$s_2^o$ and $s^i$ the output and input sizes ($FWHM$), and $f_{int}^o$ and
$f_{int}^i$ the total output and input fluxes over an aperture of 
\textit{radius} equal to the input $FWHM$. The sizes in parentheses correspond 
to the major axis of the elliptical sources.}
\end{table}

A further Monte Carlo test was performed with elliptical Gaussian sources with 
a fixed aspect ratio of 2.5. The procedure to generate these artificial 
sources was the same, with the orientation of the sources also assigned 
randomly. The peak flux densities ranged between 85 and 500~mJy/beam and the 
size along the minor axis (FWHM) between 19.2$\arcsec$ and 3.7$\arcmin$. Three 
sets of sources were analysed, providing a sample of 68 sources in total. The 
results are very similar to the circular case, but with characteristic scales 
of the filtering \textit{along the minor axis} of the elliptical sources 
20--25$\%$ smaller (see Table~\ref{t:filtering}). In return, the corresponding 
characteristic scales of the filtering \textit{along the major axis} are a 
factor of 2 larger than in the circular case. The peak flux densities are 
reduced by $\sim 15\%$ to $\sim 20\%$ for input sizes along the minor axis 
varying from $\sim 180\arcsec$ to $\sim 220\arcsec$ (i.e. sizes from 
$\sim 450\arcsec$ to $\sim 550\arcsec$ along the major axis). The aspect ratio 
is well preserved within $\sim 10\%$ after the last iteration, but the sizes 
of the sources are reduced by $\sim 15\%$ to $\sim 25\%$ for input sizes along 
the minor axis from $\sim 160\arcsec$ to $\sim 220\arcsec$ (i.e. from 
$\sim 400\arcsec$ to $\sim 550\arcsec$ along the major axis). Here again, 
sources with a peak flux density higher than $\sim$ 150~mJy/beam are better 
recovered than weaker sources that cannot or can only partly be incorporated 
into the astronomical source model used at iterations 1 to 20.

These two Monte Carlo tests clearly show that the reduction of LABOCA data 
based on correlations between pixels to reduce the atmospheric and instrumental 
noise can have a strong impact on the structure of the detected astronomical 
sources. Depending on their shape and strength, structures with (major) sizes 
larger than $\sim 140\arcsec-280\arcsec$ if circular or 
$\sim 300\arcsec-550\arcsec$ if elliptical with an aspect ratio of 2.5 are 
significantly filtered out ($>15 \%$ losses of peak flux density and size). 
Elongated structures are better recovered than circular ones. We note that the 
reduction performed for data obtained at the Caltech Submillimeter Observatory 
(CSO) with the bolometer camera Bolocam suffers from the same problem, with a 
similar characteristic scale above which losses due to spatial filtering start 
to be significant 
\citep[$\sim 120\arcsec$ for circular sources, see][]{Enoch06}.

Several attemps to attenuate the correlated noise removal by using the outer
pixels only, or to suppress the correlated noise removal and replace it 
subscan-wise with a simple first-order baseline removal in the time signals of
the OTF scans, were not successful. These alternative reduction schemes 
produced somewhat more extended emission in places where we expected some, 
but at the same time, there were much more atmospheric and instrumental noise 
residuals distributed within the map that also mimicked extended structures 
where we did not expect any, by comparison to the extinction and C$^{18}$O 1--0 
maps for instance. Attempts to reduce the data without flattening the power
spectrum at low frequency or with a lower cutoff frequency did not bring any
significant improvement either.

\subsection{Convergence of the iterative data reduction}
\label{ss:convergence}

\begin{figure}
\centerline{\resizebox{1.00\hsize}{!}{\includegraphics[angle=270]{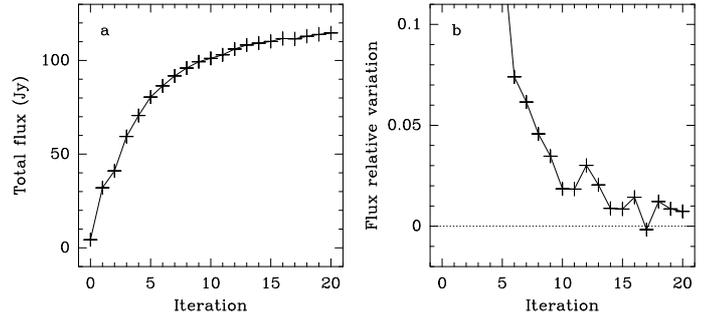}}}
\caption{\textbf{a} Convergence of the total 870~$\mu$m flux recovered in 
Cha~I by the iterative process of data reduction. \textbf{b} Relative flux 
variation between iterations $i$ and $i-1$ as a function of iteration number 
$i$.}
\label{f:total_flux}
\end{figure}

The iterative process of data reduction described in Sect.~\ref{ss:redmethod} 
allows to recover additional astronomical flux after each iteration. In this
section, we study the convergence of this method. The evolution 
of the total 870~$\mu$m flux recovered in Cha~I as a function of iteration is 
shown in Fig.~\ref{f:total_flux}a. The relative flux variation is only 0.7$\%$ 
at iteration 20, which suggests that the data reduction process indeed
converged.

\begin{figure*}[!ht]
\centerline{\resizebox{0.9\hsize}{!}{\includegraphics[angle=270]{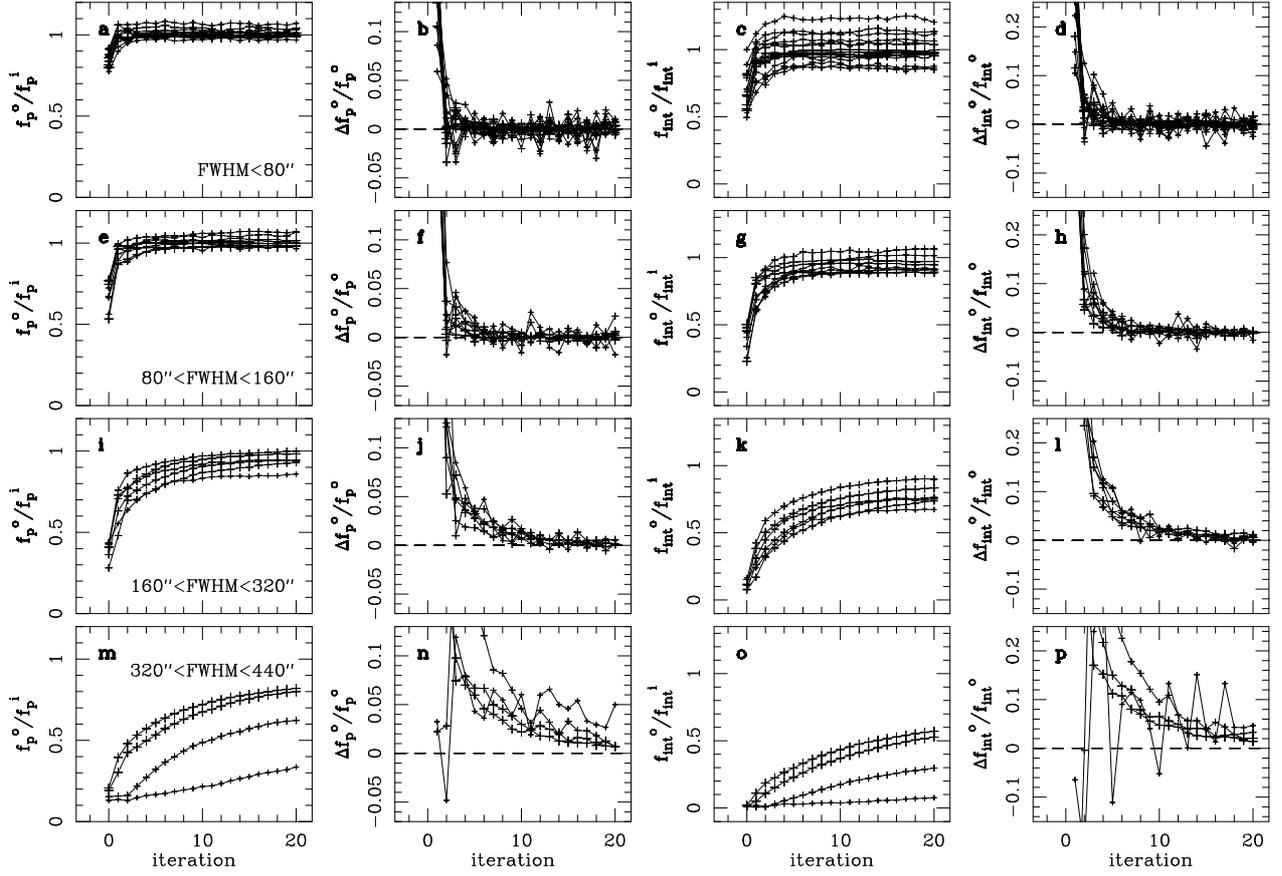}}}
\caption{Convergence of the iterative data reduction for artificial, circular, 
Gaussian sources with a peak flux density larger than 150 mJy~beam$^{-1}$. The 
first and second columns show the ratio of the output to the input peak flux 
densities, and its relative variations, respectively. The third and fourth 
columns show the same for the total flux. The different rows show sources of
different widths, as specified in each panel of the first column. The size of
the symbols increases with the input peak flux density.}
\label{f:convergence}
\end{figure*}

However, the previous estimate is based on relative variations of the 
astronomical flux from one iteration to the next one. It cannot 
completely exclude a very slow convergence. A more secure estimate can be 
derived from the set of artificial, circular, Gaussian sources described in 
Appendix~\ref{ss:filtering}. Figure~\ref{f:convergence} demonstrates that the 
number of iterations needed to reach convergence depends on the size of the 
sources. We limit the analysis to sources stronger than 150~mJy~beam$^{-1}$ 
(fainter sources show the same behaviour as far as the relative variations 
from one iteration to the next one are concerned, but the curves are noisier). 
Convergence on the peak flux density and total flux with remaining relative 
variations $< 2\%$ is reached after a few iterations (3 to 5) for sources with 
$FWHM < 160\arcsec$ (Figs.~\ref{f:convergence}b, d, f, and h). Sources with 
$160\arcsec < FWHM < 320\arcsec$ reach convergence with relative variations 
$<2\%$ at iteration 15 (Figs.~\ref{f:convergence}j and l). At iteration 
20, the peak flux densities of sources with $320\arcsec < FWHM < 380\arcsec$ 
have converged within 2$\%$ while the total fluxes may still vary by 3$\%$ 
(see Figs.~\ref{f:convergence}n and p). The convergence for the single source 
with $380\arcsec < FWHM < 440\arcsec$ is good within 5$\%$ at iteration 20. 
The results are similar for the set of elliptical sources: convergence is 
reached within $2\%$ at iteration 4--10 for sources with size along the minor 
axis $< 200\arcsec$, and within 2$\%$ at iteration 18 for sources with size 
$\sim 220\arcsec$.

This experiment shows that a significant number of iterations (about 20) are
necessary to recover as much extended emission as possible. \citet{Enoch06} 
have a similar conclusion with Bolocam for circular sources with 
$FWHM < 120\arcsec$ (convergence at iteration 5, see their Fig.~4), but they 
do not mention or show how many iterations they need to reach 
convergence\footnote{Convergence for artificial sources means here to reach a 
stable flux density, not to reach a flux density equal to the input one, which 
is never the case for large sources that will always suffer from spatial 
filtering due to the correlated noise removal.} for much larger sources.

\section{Assumptions}
\label{s:assumptions}

In this appendix, we explain the various assumptions that are used to analyse 
and interpret the Cha~I data. We also explain how the column 
densities and average densities obtained in the past for other molecular 
clouds are rescaled in the present study to compare with Cha~I 
(Table~\ref{t:allclouds}).

\subsection{Distance}
\label{ss:distance}

\citet{Whittet97} analysed the distribution of extinctions of field stars with 
distance along the line of sight of Cha~I and derived a distance of 
$150 \pm 15$~pc for this cloud. They also deduced a photometric distance for 
the B-type star HD~97300 of $152 \pm 18$~pc. They combined these results with 
the parallaxes of this and another star (HD~97048) measured with 
the \textit{Hipparcos} satellite with a larger uncertainty ($190 \pm 40$ and 
$180 \pm 20$~pc, respectively) and adopted a ``final'' distance of 
$160 \pm 15$ pc.

\citet{Knude98} used the interstellar reddening from the \textit{Hipparcos} 
and \textit{Tycho} catalogues to derive the distance to Cha~I. They found a
distance of 150 pc as revealed by a sharp increase in their plot of E(B-V) 
color excess versus distance.

However, \citet{Knude10} recently derived a further distance of $196 \pm 13$~pc
for Cha~I from a fit to an extinction vs. distance diagram. The extinction was 
obtained from a (H-K) vs. (J-H) diagram from 2MASS and the distances were 
estimated from a (J-H)$_0$ vs. $M_J$ relation based on \textit{Hipparcos}. 
Rather than considering the distance at which the first stars show a steep 
increase in extinction as the distance of the cloud, they fitted a predefined 
function to a subsample of the data. It is unclear to us whether this fitting 
procedure does give a more reliable distance estimate because the selection of 
the subsample defining the extinction jump for the fit may introduce a bias. 
Therefore, as long as this further distance has not been confirmed by another 
method, we keep the older estimate (150 pc) as the distance of Cha~I.
If the further distance turns out to be correct, then all physical sizes and 
masses derived in this article are underestimated by a factor 1.3 and 1.7, 
respectively, and the mean densities are overestimated by a factor 1.3. The 
column densities are not affected.

\subsection{Temperature}
\label{ss:temperature}

\citet{Toth00} derived a dust temperature of $12 \pm 1$~K for the 
``very cold cores'' they identified in Cha~I on somewhat large scales 
($HPBW \sim 2\arcmin$) from their analysis of the ISOPHOT Serendipity 
Survey (see cores 1 to 6 in their Table 2). The dust temperature is most 
likely not uniform and this value rather represents a mean temperature. The 
dust temperature is expected to decrease toward the center of starless dense 
cores that do not contain any central heating, even down to $\sim 5.5-7$~K as 
measured for the gas in the prestellar cores L1544 and L183 by 
\citet{Crapsi07} and \citet{Pagani07}, respectively. The modeled dust 
temperature profiles of the prestellar cores L1512, L1689B, and L1544 derived 
by \citet{Evans01} indeed vary from 12--13~K in the outer parts to 7--8~K at 
the core center. Similarly, \citet{Nutter08} measured a dust temperature 
varying from 12~K in the outer parts to 8~K in the inner parts of a filament
in the Taurus molecular cloud 1 (TMC1). Thanks to its good angular resolution
and its good sampling of the spectral energy distribution of these cores,
such temperature variations will be routinely measured in the far-infrared 
with the \textit{Herschel} satellite, which will help refining the mass and 
column density estimates \citep[see, e.g.,][]{Andre05}. For simplicity, 
we assume here a uniform dust temperature of 12~K. 

If the temperature inside the dense cores drops to, e.g., 7~K, their masses 
and column densities computed in this article could be underestimated by up to 
a factor of 3, depending on the exact shape of the temperature profile. 
However, the total mass of a core being dominated by its outer parts, it 
should not be that much underestimated.

\subsection{Conversion of visual extinction to H$_2$ column density}
\label{ss:avtoN}

The standard factor to convert a visual extinction $A_V$ to an H$_2$ column 
density is $9.4 \times 10^{20}$ cm$^{-2}$ mag$^{-1}$. It is based on the ratio 
$N_H/E(B-V) = 5.8 \times 10^{21}$ atoms cm$^{-2}$ mag$^{-1}$ measured by
\citet{Bohlin78} with data from the \textit{Copernicus} satellite, and a 
ratio of total-to-selective extinction $R_V = A_V/E(B-V) = 3.1$ corresponding 
to the standard extinction law that is representative of dust in diffuse 
regions in the local Galaxy 
\citep[see, e.g.,][ and references therein]{Draine03}.

Several measurements suggest that $R_V$ is larger toward denser 
molecular clouds. In particular, \citet{Chapman09} computed the mid-IR 
extinction law based on \textit{Spitzer} data in the three molecular clouds 
Ophiuchus, Perseus, and Serpens and found indications of an increase of $R_V$ 
from 3.1 for $A_{K_s} < 0.5$~mag to 5.5 for $A_{K_s} > 1$~mag, which they 
interpret as evidence for grain growth. Using the parameterized extinction law 
of \citet[][, their equations 1, 2a, and 2b]{Cardelli89} with $R_V = 3.1$ 
and 5.5, respectively, we find that these two thresholds correspond to 
$A_V < 4.2$~mag and $A_V > 7.2$~mag. 
\citet{Vrba84} came to a similar conclusion for Cha~I since they found
a standard extinction law toward the outer parts of the cloud but reported a
larger $R_V$ ($>4$) for field stars near the denser parts of the cloud
\citep[see also][]{Whittet94}. Therefore, we assume that the thresholds 
derived by \citet{Chapman09} also apply to Cha~I. With $R_V = 5.5$, the
factor to convert a visual extinction into an $H_2$ column density becomes
$6.9 \times 10^{20}$~cm$^{-2}$~mag$^{-1}$ 
\citep[see][ and references therein]{Evans09}. However, the visual extinction 
values presented here are derived from near-infrared measurements assuming the 
standard extinction law ($R_V = 3.1$). Compensating for this assumption, the 
``corrected'' conversion factor for $R_V = 5.5$~mag becomes 
$N_{\mathrm{H}_2}/A_V = 5.8 \times 10^{20}$~cm$^{-2}$~mag$^{-1}$ for our 
extinction map calibrated with $R_V = 3.1$. 

In summary, for $A_V$ below $\sim 6$~mag in our extinction map we assume 
$R_V = 3.1$ and the standard conversion factor 
$N_{\mathrm{H}_2}/A_V = 9.4 \times 10^{20}$~cm$^{-2}$~mag$^{-1}$. For $A_V$ above 
$\sim 6$~mag, we assume $R_V = 5.5$ and 
$N_{\mathrm{H}_2}/A_V = 5.8 \times 10^{20}$~cm$^{-2}$~mag$^{-1}$.
 
\subsection{Dust mass opacity}
\label{ss:dustopacity}

The dust mass opacity in the submm range is not well constrained and is 
typically uncertain by a factor of 2 \citep[e.g.][]{Ossenkopf94}. Since most 
sources detected in the survey of Cha~I are weak, starless cores, we adopt a 
dust mass opacity, $\kappa_{870}$, of 0.01~cm$^2$~g$^{-1}$ in most cases (and
a dust opacity index $\beta = 1.85$). It corresponds to the value recommended 
by \citet{Henning95} for ``cloud envelopes'', assuming $\beta \sim 1.5-2.0$ 
and a gas-to-dust mass ratio of 100. For higher densities 
($n_{\mathrm{H}_2} \sim 5 \times 10^6$ cm$^{-3}$) that characterize the inner 
part of a few sources in our survey, $\kappa_{870}$ is most 
likely higher, on the order of 0.02~cm$^2$~g$^{-1}$ 
\citep[][]{Ossenkopf94}. Finally, the dust mass opacity in circumstellar disks 
is expected to be even higher. We follow \citet{Beckwith90} with $\beta = 1$ 
and derive $\kappa_{870} = 0.03$~cm$^2$~g$^{-1}$ for this type of sources. In 
the article, we each time specify which dust mass opacity we use to 
compute column densities and masses. We assume in all calculations 
that the dust emission is optically thin.

\subsection{Molecular weight}
\label{ss:mu}

The mean molecular weight per hydrogen molecule and per free particle, 
$\mu_{\mathrm{H}_2}$ and $\mu$, are assumed to be 2.8 and 2.37, respectively 
\citep[see, e.g.,][]{Kauffmann08}. The former is used to convert the flux
densities to H$_2$ column densities while the latter is needed to estimate
the free-particle densities.

\subsection{Rescaling of previous results on other clouds}
\label{ss:rescaling}

\subsubsection{Column densities}
\label{sss:rescaling_coldens}

Assuming a temperature of 10~K and a dust mass opacity of 
$\kappa_{\mathrm{1.1~mm}}= 0.0114$~cm$^{2}$~g$^{-1}$,
\citet{Enoch08} derive average peak H$_2$ column densities of 1.2 and 
1.0~$\times 10^{22}$~cm$^{-2}$ in Perseus and Serpens, respectively. Their 
values have to be multiplied by 1.3 to match the same properties as we use for 
Cha~I (12~K, $\beta = 1.85$ yielding 
$\kappa_{\mathrm{268\,GHz}} = 0.0063$~cm$^2$~g$^{-1}$). In addition, both clouds 
are at larger distances than Cha~I and the Bolocam maps have a poorer angular 
resolution (31$\arcsec$) compared to our LABOCA map (21.2$\arcsec$). If the 
sources in Perseus and Serpens are not uniform in the Bolocam beam but rather 
centrally peaked with a column density scaling proportionally to the inverse 
of the beam size, then their peak column densities measured in the same 
physical size as for Cha~I would be about 2.4--2.5 times larger. The full 
rescaling factor is thus 3.1--3.5. We rescale the average column density of 
the SCUBA starless cores analysed by \citet{Curtis10} in Perseus in the same 
way (12~K and $\kappa_{\mathrm{353\,GHz}} = 0.010$~cm$^2$~g$^{-1}$), after 
correction for a small error\footnote{The solid
angle of a Gaussian beam is $\Omega = \frac{\pi}{4\,\ln(2)} \theta_{HPBW}^2$,
with $\theta_{HPBW}$ the half-power beam width 
\citep[see, e.g., Appendix~A.3 of][]{Kauffmann08}. \citet{Curtis10} seem to 
use $\Omega = \frac{\pi}{4} \theta_{HPBW}^2$.} in their 
numerical equation 5. The column density is also rescaled
proportionally to the inverse of the beam size. The rescaled values of both
studies of Perseus are consistent.

The average peak column density of the Bolocam starless cores observed 
by \citet{Enoch08} in Ophiuchus was rescaled in the same way as above. 
The average peak column density derived for the starless
cores of \citet{Motte98} after rescaling to the same physical size as for the 
Cha~I sources with an Ophiuchus distance of 125~pc is a factor of $\sim 2$ 
lower. The discrepancy between 
both studies may come from the background emission not being removed in the 
former study, from the assumption of a unique temperature for all sources in 
the former study while the second one uses different temperatures (from 12 to 
20~K), and from our assumption of a column density scaling with the inverse of 
the radius.

\citet{Kauffmann08} analysed their maps of Taurus after smoothing to 
20$\arcsec$, i.e. nearly the same spatial resolution as our Cha~I map given 
also the slightly different distances. Their column density sensitivity is 
similar to ours. Assuming the same dust properties as 
for Cha~I ($T_{\mathrm{dust}} = 12$~K and $\beta = 1.85$ yielding 
$\kappa_{\mathrm{250\,GHz}} = 0.0055$~cm$^2$~g$^{-1}$), the average column 
density in an aperture of 20$\arcsec$ derived from the flux densities listed 
in their Table~5 is $2.3\,^{+1.3}_{-1.1} \times 10^{22}$~cm$^{-2}$ for their
sample of 28 Taurus starless ``peaks''.

\subsubsection{Densities}
\label{sss:rescaling_dens}

\citet{Enoch08} obtain average densities of 1.7, 1.4, and 
$2.8 \times 10^5$~cm$^{-3}$ within an aperture of diameter $10^4$~AU for the 
starless cores in Perseus, Serpens, and Ophiuchus, respectively. We assume 
a density profile proportional to $r^{-p}$ with $p \sim$ 1.5--2, which is
typical for such cores, and rescale their densities with a temperature of 12~K 
and a dust opacity $\kappa_{\mathrm{268\,GHz}} = 0.0063$~cm$^2$~g$^{-1}$.

With the same assumptions as in Sect.~\ref{sss:rescaling_coldens}, the 
28 starless ``peaks'' of \citet{Kauffmann08} in Taurus have an average mass of 
$0.51^{+0.4}_{-0.3}$~M$_\odot$ within an aperture of diameter 8400~AU. We assume 
the same density profile as above to compute the average mass within an 
aperture of diameter 7500~AU and derive the average density.

\section{Multiresolution decomposition}
\label{s:mrmed}

As mentioned in Sect.~1.5.1 of 
\citet{Starck98}, multiresolution tools based on a wavelet transform suffer 
from several problems, in particular they create negative artefacts around the 
sources. In addition, since the emission is smoothed with the scaling 
function ($\Phi$) to produce the large-scale planes, the shape of the 
large-scale structures is not preserved and tends toward the shape of $\Phi$ 
(which, in turn, creates the negative artefacts in the small-scale planes 
mentioned above). For instance, a circularly symmetric filter will tend to 
produce roundish structures on large scale. To avoid these issues, we 
performed a multiresolution analysis based on the median transform following 
the algorithm of \citet[][, Sect.~1.5.1]{Starck98} that we implemented in 
Python. In short, the input image is first ``smoothed'' with a filter that 
consists of taking the median value on a $3\times3$-pixel grid centered on 
each pixel. The difference between the input and smoothed images yields the 
``detail'' map at scale 1 (the smallest scale). The same process is repeated 
with a $5\times5$-pixel grid using the smoothed map as input map, and the 
difference between the two images produces the detail map at scale 2. The 
process is done 7 times in total, yielding detail maps at 7 different scales.
The size of the convolution grid at scale $i$ is $n \times n$ pixels with 
$n = 2^i+1$, i.e. 18.3$\arcsec$, 30.5$\arcsec$, 54.9$\arcsec$, 
1.7$\arcmin$, 3.4$\arcmin$, 6.6$\arcmin$, and 13.1$\arcmin$ for scales 1 to 7, 
respectively. The last smoothed map contains the residuals (very large 
scale). The sum of the residual map and the 7 detail maps is strictly 
identical to the input map.

\begin{table}
 \caption{Characteristics of the multiresolution decomposition as measured on a
sample of circular and elliptical artificial Gaussian sources with a 
signal-to-noise ratio higher than 9.}
 \label{t:mrmed}
 \begin{tabular}{lccc}
  \hline\hline
  \multicolumn{1}{c}{Input size} & \multicolumn{3}{c}{Scale at which} \\
  \multicolumn{1}{c}{} & \multicolumn{1}{c}{$f_p^o/f_p^i > 40\%$} & \multicolumn{1}{c}{$f_p^o/f_p^i > 80\%$} & \multicolumn{1}{c}{$f_p^o/f_p^i \sim 100\%$} \\
  \multicolumn{1}{c}{(1)} & \multicolumn{1}{c}{(2)} & \multicolumn{1}{c}{(3)} & \multicolumn{1}{c}{(4)} \\
  \hline
  $<30\arcsec$      & 2 & 3 & 4 \\
  30--60$\arcsec$   & 3 & 4 & 5 \\
  60--120$\arcsec$  & 4 & 5 & 6 \\
  120--200$\arcsec$ & 5 & 6 & 7 \\
  200--300$\arcsec$ & 6 & 7 & 7 \\
  300--400$\arcsec$ & 6 & 7 & -- \\
  \hline
 \end{tabular}
 \tablefoot{The input size is the $FWHM$ for circular sources, and 
$\sqrt{FWHM1 \times FWHM2}$ for elliptical sources, with $FWHM1$ and $FWHM2$ 
the sizes along the major and minor axes, respectively. $f_p^o$ and $f_p^i$ 
are the output and input peak flux densities, respectively.}
\end{table}

The properties of this multiresolution decomposition were estimated with the 
samples of artificial Gaussian sources of Appendix~\ref{ss:filtering} (both 
circular and elliptical with aspect ratio 2.5) inserted into the reduced 
continuum map of Cha~I. Table~\ref{t:mrmed} presents the results for sources 
with a signal-to-noise ratio higher than 9. For each scale $i$, the artificial 
sources were fitted in the map containing the sum of the detail maps from 
scale 1 to $i$ (hereafter called \textit{sum} map at scale $i$). Compact 
sources ($FWHM < 30\arcsec$) are fully recovered at scale 4 while sources with 
$FWHM > 120\arcsec$ are strongly filtered out (losses $>60\%$). Therefore, 
scale 4 is a good scale to extract compact sources when they are embedded in 
larger scale cores ($> 120\arcsec$). Similarly, scale 5 is a good scale to 
extract sources with $FWHM < 120\arcsec$ with an accuracy better than $20\%$ 
on the peak flux density, while significantly filtering out the more extended 
background emission (losses $>60\%$ for $FWHM > 200\arcsec$). In the case of
the elliptical sources with aspect ratio 2.5, the conclusions are the same, 
the key parameter being the \hbox{geometrical} mean of the sizes along the 
major and minor axes ($\sqrt{FWHM1 \times FWHM2}$).

\end{appendix}

\listofobjects

\end{document}